\documentclass[useAMS,usegraphicx,usenatbib]{mn2e}

\usepackage{times}
\usepackage{amsmath}
\usepackage{amssymb}
\usepackage{epsfig}
\usepackage{appendix}

\begin{document}
 
\title[Can we measure the accretion efficiency of AGN?]{Can we measure the accretion efficiency of Active Galactic Nuclei?}
\author[S.I. Raimundo, A.C. Fabian, R.V. Vasudevan, P. Gandhi and J. Wu]
{\parbox[]{7.in}{S.~I. Raimundo$^{1}$\thanks{E-mail: 
sijr@ast.cam.ac.uk}, A.~C. Fabian$^{1}$, R.~V. Vasudevan$^{2}$, P.~Gandhi$^{3}$ and Jianfeng~Wu$^{4}$\\
\footnotesize
$^{1}$Institute of Astronomy, Madingley Road, Cambridge CB3 0HA\\
$^{2}$University of Maryland, College Park, MD 20742, USA \\
$^{3}$Institute of Space and Astronautical Science (ISAS), Japan Aerospace Exploration Agency, 3-1-1 Yoshinodai, chuo-ku, Sagamihara, Kanagawa 229-8510, Japan\\
$^{4}$Department of Astronomy \& Astrophysics, 525 Davey Lab, The Pennsylvania State University, University Park, PA 16802, USA}}
\maketitle
\begin{abstract} 
\noindent The accretion efficiency for individual black holes is very difficult to determine accurately. There are many factors that can influence each step of the calculation, such as the dust and host galaxy contribution to the observed luminosity, the black hole mass and more importantly, the uncertainties on the bolometric luminosity measurement. Ideally, we would measure the AGN emission at every wavelength, remove the host galaxy and dust, reconstruct the AGN spectral energy distribution and integrate to determine the intrinsic emission and the accretion rate. In reality, this is not possible due to observational limitations and our own galaxy line of sight obscuration. We have then to infer the bolometric luminosity from spectral measurements made in discontinuous wavebands and at different epochs. In this paper we tackle this issue by exploring different methods to determine the bolometric luminosity.
We first explore the trend of accretion efficiency with black hole mass ($\epsilon \propto M^{\sim 0.5}$) found in recent work by Davis \& Laor and discuss why this is most likely an artefact of the parameter space covered by their PG quasar sample. 
We then target small samples of AGN at different redshifts, luminosities and black hole masses to investigate the possible methods to calculate the accretion efficiency. For these sources we are able to determine the mass accretion rate and, with some assumptions, the accretion efficiency distributions. 
Even though we select the sources for which we are able to determine the parameters more accurately, there are still factors affecting the measurements that are hard to constrain. We suggest methods to overcome these problems based on contemporaneous multi-wavelength data measurements and specifically targeted observations for AGN in different black hole mass ranges.
\end{abstract} 


\section{Introduction}
Accretion onto a supermassive black hole is the current accepted mechanism to explain Active Galactic Nuclei (AGN) activity (\citealt{salpeter64,lynden-bell69}). This extreme process is responsible for radiative emission on a wide wavelength range, with most of the intrinsic power emitted in the optical, ultra-violet (UV) and X-ray wavebands. With multi-wavelength data it is possible to determine the AGN spectral energy distribution (SED), and investigate the physical mechanisms behind the radiative emission at each energy. 
Early work set the standard model for a thin black hole accretion disc (\citealt{shakura&sunyaev73,novikov&thorne73}), and showed that it could explain the observed emission (\citealt{shields78}), establishing a relation between the SED peak in the UV region and the thermal emission from the accretion disc.
This picture has been explored with the development of more complex models, and although there are some observational discrepancies (see \citealt{koratkar&blaes99} for a review), the thin accretion disc paradigm remains the standard adopted model.

The accretion parameters such as the rate of accreted mass ($\dot{M}$) and efficiency in converting mass to radiative energy ($\epsilon = L/(\dot{M} c^{2})$) are closely related to the energy output and play a decisive role on AGN behaviour and evolution. The study of their variation with time, environment and type of source can give insights into black hole accretion physics and constrain the current models for the evolution of AGN properties. 
\cite{soltan82} showed that it is possible to constrain the \emph{average} efficiency in converting accreted mass to emitted radiative energy (accretion efficiency hereafter), by comparing the integrated AGN luminosity with the total mass accreted onto black holes. Later work used a related method to determine the average accretion efficiency by measuring the X-ray background (\citealt{fabian&iwasawa99,elvis02}).
In studies of AGN evolution, where the redshift dependent AGN luminosity function is compared with the local black hole mass function (\citealt{small&blandford92,salucci99,marconi04,merloni04,cao07,shankar09,raimundo&fabian09}), the accretion efficiency is one of the free parameters associated with the accretion process. These can be determined on average for AGN, but when taking the observed values for Eddington ratio ($L_{\rm bol}/L_{\rm Edd}$) and obscuration into account, the possible parameter range is very wide (\citealt{raimundo&fabian09}). 
More observational results are needed to constrain the evolution of black holes and AGN properties. Direct measurements of individual accretion parameters would allow to constrain and modify the models to better take into account the variety of AGN stages and environments. 
Recent work explored the possibility of determining the mass accretion rate and accretion efficiency for individual AGN, using accretion disc theory arguments and available data on the emitted radiation.
\cite{collin02} used a simple accretion disc model to determine the mass accretion rate from the observed optical monochromatic luminosity, and explored in detail the uncertainties associated with the simple disc assumption.
\cite{bian&zhao02}, adopted a similar relation to determine mass accretion rates and accretion disc inclinations, and used this as a starting point to derive accretion properties for radio loud and radio quiet AGN (\citealt{bian&zhao03}).
More recently, \cite{davis&laor11} (DL11 hereafter), compared the simple accretion disc model with more sophisticated ones and determined individual mass accretion rates and accretion efficiencies for 80 PG quasars. Their accretion efficiency values showed an increase with increasing black hole mass, and a wide distribution along the allowed range for theoretical black hole spin predictions.
From accretion disc theory, the accretion efficiency is related with the spin of the black hole and the procedure described above provides a method to measure individual black hole spins. The relation between the two is set in first order by accretion disc theory, but note that effects such as magnetic torques can increase the amount of energy released in the disc and the relation between the spin and accretion efficiency becomes more complex (e.g. \citealt{gammie99,agol&krolik00}). The spin distribution can provide clues on the history of supermassive black holes, with low values of spin associated with black hole growth by random minor mergers and high values associated with growth by continuous gas accretion (\citealt{volonteri05, king&pringle06,berti&volonteri08}). The spin distribution determined from this method can also be compared with alternative approaches to measure black hole spin, such as spectral analysis of the X-ray Fe K$_{\alpha}$ line emission (\citealt{fabian89,laor91,brenneman&reynolds06,reynolds&fabian08}).

In this work, we discuss the uncertainties involved in the determination of accretion efficiencies and investigate the correlation between black hole mass and accretion efficiency. In Section~2, we present the main accretion disc model and how the accretion properties are determined. In Section~3, we analyse the correlation found by DL11 and describe how the sample selection effects decrease the allowed parameter space and produce the correlation found in their work. In the last sections we present several methods to determine the bolometric luminosity and the accretion efficiency of AGN based on observational data. We analyse two samples of low and higher redshift AGN and discuss our results in Section~5. Section~6 summarises our main conclusions.
Throughout this paper we adopt standard cosmological parameters of H$_0 = 71$ km s$^{-1}$ Mpc$^{-1}$, $\Omega_{\rm m} = 0.27$ and $\Omega_{\rm \Lambda} = 0.73$.
\section{Model}
\label{sec:model}
\begin{figure}
\begin{centering}
\includegraphics[width=0.8\columnwidth]{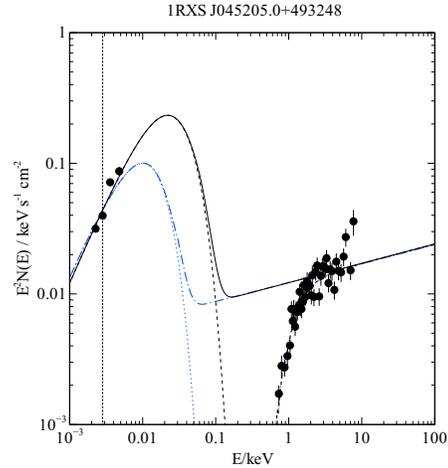}
\caption {Example of two different spins for the accretion disc model of 1RXS J045205.0+493248. The lower pair of lines (blue) represent an accretion disc with spin of $a = 0.1$, while the upper lines (black) represent a maximally rotating black hole. Solid and dash-dotted lines are the SED before absorption for each spin case, the dashed and dotted lines are the observed SED (after absorption). The vertical line shows the $B$-band observed wavelength, which is relatively insensitive to changes in spin.}
\label{spin}
\end{centering}
\end{figure}

In this section we present an overview of the model used to determine the mass accretion rate and accretion efficiency in this work.
The method makes use of the fact that the AGN spectral energy distribution at optical wavelengths is relatively insensitive to black hole spin, which allow us to establish a relation between the specific luminosity and mass accretion rate (see Fig.~\ref{spin} but also Fig.~1 of DL11).
The mass accretion rates are determined from assuming a simple accretion disc model, the measured flux at 4392 \AA\thinspace\thinspace and an estimate for the black hole mass and disc inclination.
The flux emitted by the accretion disc is the result of the integrated blackbody emission ($B_{\nu}$) at each radius $R$:
\begin{eqnarray}
F_{\nu}=\frac{2\pi\cos (i)}{D^{2}}\int_{R_{in}}^{R_{out}}B_{\nu}(T_{\rm eff}) R dR
\end{eqnarray}

Assuming isotropic emission and an effective temperature depending on the black hole mass, mass accretion rate and radius ($T_{\rm eff}\propto R^{-\frac{3}{4}}$), the relation between the luminosity at an optical frequency ($L_{\nu}$) and the mass accretion rate ($\dot{M}$) is:
\begin{eqnarray} \label{Lniu}
L_{\nu} = \frac {32}{3\pi^{2}}\left(\frac{45hG^{2}}{2c^{2}}\right)^{\frac{1}{3}}\cos\thinspace (i)\thinspace\nu^{\frac{1}{3}}M^{\frac{2}{3}}\dot{M}^{\frac{2}{3}}\int_0^{\infty}\frac{x^{\frac{5}{3}}}{e^{x}-1}dx.
\end{eqnarray}
Where $M$ is the black hole mass, $i$ the disc inclination and $x = h\nu/(k_{B}T_{eff})$. This relation is the same as the corrected version of DL11 in \cite{Laor&Davis11}.
The integral can be solved with Gamma and Zeta functions which give a numerical value of $1.93$. Changing units, we obtain:
\begin{eqnarray} \label{Mdot}
\dot{M}=1.53 \left(\frac{\nu L_{\nu}}{10^{45}\cos(i)}\right)^{3/2} \frac{10^{8}}{M} \thinspace\thinspace (M_{\odot}/yr)
\end{eqnarray}
with $M$ in $M_{\odot}$ and $\nu L_{\nu}$ (or $L_{\rm opt}$) in erg\thinspace s$^{-1}$. This equation is defined for the wavelength used in this work (4392 \AA), but the dependence in frequency is $\dot{M}\propto (\nu L_{\nu})^{3/2}\thinspace\thinspace \nu^{-2}$.
The accretion efficiency can then be determined from the mass accretion rate:
\begin{eqnarray} \label{efficiency}
\epsilon = \frac{L_{\rm bol}}{\dot{M}c^{2}}
\end{eqnarray}
but this implies having an accurate measurement of the observed bolometric luminosity ($L_{\rm bol}$). Here we decide to use the isotropic bolometric luminosity $L_{\rm bol} = 4\pi D^{2}\times F$, where $F$ is the flux. For a disc, the bolometric luminosity should be a factor of $2\times\cos(i)$ lower for a simple case and have a few extra factors for the relativistic case. We nevertheless keep this formulation throughout the paper for a clear comparison between different methods of determining $L_{\rm bol}$.
\begin{figure}
\begin{centering}
\includegraphics[width=0.8\columnwidth]{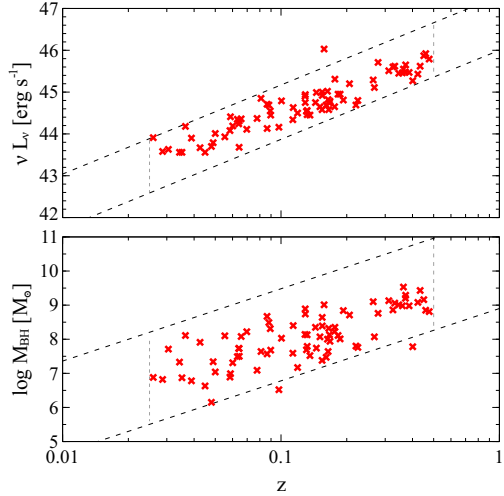}
\caption {Black hole mass and optical luminosity distribution for the sources in the DL11 PG quasar sample (red crosses). The black dashed lines represent the limits for our simulated sample. They are defined by flux and volume limitations and Eddington ratio biases. The dashed vertical grey lines mark the redshift range probed by the sample.}
\label{mass_opt_z_DL11}
\end{centering}
\end{figure}
\section{Correlations and selection effects}
\label{sec:correlations}
The approach described in the previous section was used by DL11 to determine the mass accretion rate and accretion efficiencies for a sample of 80 PG quasars. They explored complex models of accretion discs to determine the best relation between the luminosity measured at an optical wavelength (4861\thinspace\AA\thinspace\thinspace in their case) and the mass accretion rate. The bolometric luminosity for each of these quasars was determined by integrating the spectral energy distribution inferred from available data, interpolating between measured values and extrapolating to the regions where measurements are non-existent.
In this section we use the DL11 best-fit relations and explore the correlation they found between the accretion efficiency and the black hole mass. DL11 discuss the selection effects in their sample, here we analyse these effects in more detail and simulate the resulting allowed parameter range. From the discussion in the following paragraphs, it will become clear that the correlation arises from the source selection criteria. After this analysis we simulate a test sample with uniform accretion efficiency and study the effect of uncertainties in inclination and black hole mass measurements. 
\begin{figure}
\begin{centering}
\includegraphics[width=0.8\columnwidth]{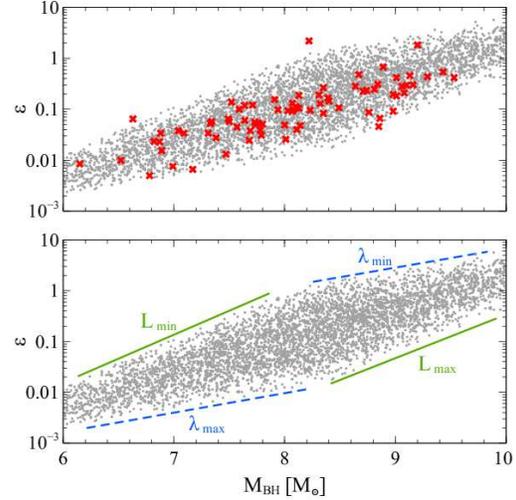}
\caption {Simulated distribution of accretion efficiency and black hole mass for AGN (grey circles) based on the flux and volume limits from Fig.~\ref{mass_opt_z_DL11} and a constant bolometric correction of 9.1. Top panel: Comparison with the DL11 observed sample (red crosses). Bottom panel: Lines indicate qualitative behaviour or our sample's distribution. The green solid lines indicate the effect of luminosity limits. Blue dashed lines indicate the general effect of two Eddington ratio limits. The parameter space possible to probe with these conditions is indicated by the distribution of grey circles.}
\label{eff_simulated}
\end{centering}
\end{figure}
\subsection{Sample selection}
The 80 PG quasars do not constitute a complete sample. Initially, these quasars were selected as part of the Palomar Bright Quasar Survey (BQS), a sub-sample of the Palomar-Green Survey (PG) of stellar-like objects with ultraviolet excess which is a volume and flux limited survey (\citealt{schmidt&green83}). The selection criteria included the presence of broad lines, a significant redshift (with the closest quasar at $z = 0.025$), and most of the emission concentrated in the inner regions of the source (star-like appearance). DL11 used the spectroscopic data on a redshift limited sample ($z < 0.5$) of the BQS from \cite{boroson&green92}. In summary, the sample is biased to luminous quasars with high Eddington ratios, and subject to flux and volume limitations. The low Eddington ratio sources are likely to be missed from this sample, due to the selection criteria based on the star-like appearance as mentioned in DL11. For the low Eddington sources, the host galaxy luminosity can dominate the emission in the optical bands and most of the radiation will not come from the central source, failing the star-like selection criterion.
The sample described above was nevertheless very suitable for the study: the multi-wavelength data available makes it easier to determine the bolometric luminosity and calculate the relevant parameters for a large sample. This approach is valid for individual objects, but one needs to consider carefully the biases when determining general trends and correlations.

From the analysis in DL11, the trends found using the more complex accretion disc models for a constant inclination of $\cos\thinspace (i) = 0.8$ were:

\begin{eqnarray} \label{DL11_mdot}
\dot{M} = 3.5\thinspace M_{\odot}\thinspace yr^{-1}  \left(\frac{M}{10^{8}}\right)^{-0.89} \left(\frac{L_{\rm opt}}{10^{45}}\right)^{1.5}
\end{eqnarray}

\begin{eqnarray} \label{DL11_eff}
\epsilon = 0.063 \left(\frac{L_{bol}}{10^{46}}\right)^{0.99} \left(\frac{M}{10^{8}}\right)^{0.89} \left(\frac{L_{\rm opt}}{10^{45}}\right)^{-1.5}
\end{eqnarray}
Assuming a bolometric correction $K_{\rm opt} = L_{\rm bol}/L_{\rm opt}$, the accretion efficiency will be:
\begin{eqnarray} \label{DL11_eff2}
\epsilon = 0.063\thinspace K^{0.99}_{opt} \left(\frac{M}{10^{8}}\right)^{0.89}\left(\frac{L_{\rm opt}}{10^{45}}\right)^{-0.51}
\end{eqnarray}
In DL11 the PG quasars are distributed along a region in the $\epsilon$ vs $M_{\rm BH}$ that corresponds to a trend of $\epsilon \propto M^{0.5}$. Here we explore this correlation keeping in mind the characteristics of the sample and its selection criteria.
In Fig.~\ref{mass_opt_z_DL11} we show the distribution of optical luminosities ($L_{\rm opt}$) and black hole masses ($M_{\rm BH}$) as a function of redshift for the DL11 sample. The redshift range is determined from the constraints discussed above, being limited to $z\thinspace\sim\thinspace$(0.025 - 0.5). 
\begin{figure}
\begin{centering}
\includegraphics[width=0.8\columnwidth]{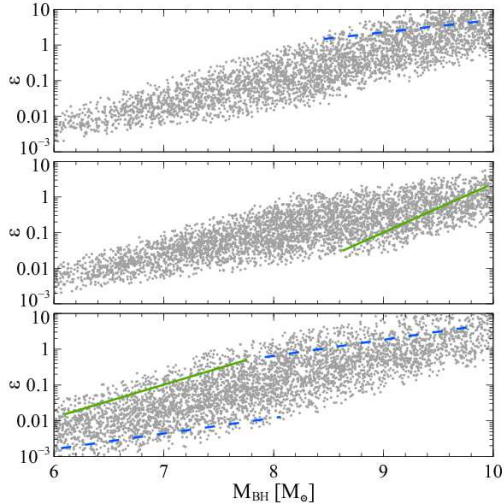}
\caption {Simulated sample in the same conditions as in Fig.~\ref{eff_simulated}, with over-plotted line limits to indicate luminosity and Eddington ratio limits. Three panels illustrate the effects of different changes. Top panel: Effect of including lower Eddington ratios, lower limit in this panel is $\lambda > 0.001$. Middle panel: Include sources at higher redshift with higher luminosity. Bottom panel: Lower the flux limit for source detection.}
\label{eff_simulated_changes}
\end{centering}
\end{figure}
We argue that these constraints, together with an almost constant bolometric correction, shape the distribution of accretion efficiencies for these sources. The correlation found between $\epsilon$ and $M_{\rm BH}$ in DL11 is caused by the narrow range of parameters that are probed by this sample and not by a specific dependence of accretion efficiency with mass. If we take the distributions in Fig.~\ref{mass_opt_z_DL11} and use Eq.~\ref{DL11_eff2} we can only probe the parameter range indicated by the grey circles in Fig.~\ref{eff_simulated}, and therefore obtain a trend between efficiency and back hole mass. To verify this argument we generate a uniform sample of AGN with properties similar to the ones in DL11. 
First we start by determining the $L_{\rm opt}$ distribution for a flux limited sample with uniform distribution within the limits given by the dashed lines in the top panel of Fig.~\ref{mass_opt_z_DL11}, black lines for the luminosity limits and grey lines for the redshift range. 
The limits we used can be understood qualitatively: at each redshift, the range of optical luminosities is determined on the lower limit side by the minimum flux for selection, and on the upper limit by both the volume limit of the survey and the shape of the luminosity function, which predicts a decrease in the number volume of sources with increasing luminosity (see for example \citealt{boyle00} for the $B$-band luminosity function).
\begin{figure}
\begin{centering}
\includegraphics[width=0.8\columnwidth]{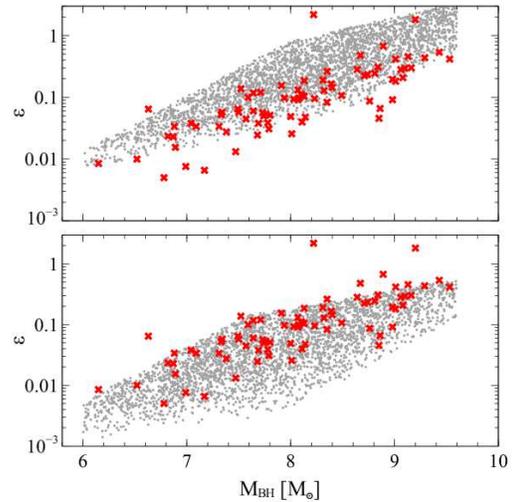}
\caption {Simulated sample (grey circles) with DL11 PG quasar sample over-plotted (red crosses) to illustrate the effect of using different bolometric corrections. Top panel: higher bolometric correction ($K_{\rm opt} = 20$). Bottom panel: Simulated distribution when considering a bolometric correction weakly dependent on $L_{\rm opt}$ \citep{marconi04}. In this case the values of accretion efficiency show a larger spread.}
\label{eff_simulated_bolcorr}
\end{centering}
\end{figure}
It is nevertheless difficult to determine the flux limit for this sample, \cite{schmidt&green83} give a $B$-band magnitude limit for each exposure field, but the values of $L_{\rm opt}$ are measured at 4861 \AA, and we would have to assume a spectral shape to determine the frequency specific flux limit. Instead, we assume a minimum $\nu F_{\nu}=3\times10^{-12}$erg cm$^{-2}$ s$^{-1}$ for every case, which is lower than the average $B$-band magnitude of 16.16 and spectral index of -0.5 (\citealt{schmidt&green83}), but reflects well the lower limit observed in DL11. For the upper limit, we did not consider the luminosity function distribution (or volume limitations), which would be beyond the scope of this paper, but assume a flux limit type border that contains almost all the sources observed in DL11. 
For the bottom panel of Fig.~\ref{mass_opt_z_DL11}, the mass distribution in DL11 corresponds to the virial estimated black hole masses and shows a broader distribution. To simulate this sample we generated redshift constrained sources within a lower and upper limit range, also represented as black and grey lines in the figure. Qualitatively, these limits can be understood as the lower and higher possible black hole masses at each redshift if we consider the luminosity limits, a bolometric correction $K_{\rm opt}$ (see discussion below) and the typical Eddington ratio of the DL11 sample ($0.04 < \lambda < 1.0$), with $\lambda = L_{\rm bol}/(1.3\times10^{38}M_{\rm BH}/M_{\odot})$. 
Using the simulated sample within the limits of Fig.~\ref{mass_opt_z_DL11} we then calculate the accretion efficiency for every source using the best fit relations from DL11 shown in Eqs.~\ref{DL11_mdot} to \ref{DL11_eff2}. 
In Fig.~\ref{eff_simulated} we present our simulated sample in grey compared with the DL11 results in the top panel, and explain qualitatively the boundary limits in the bottom panel. This simple sample shows the effect of redshift, luminosity and mass limits in limiting the parameter space probed. In the bottom panel we plot in solid green the constraints due to lower and upper luminosity limits and in dashed blue the absolute limits due to the Eddington ratio constraints. In Fig.~\ref{eff_simulated_changes} we illustrate what would happen if lower Eddington ratio objects are considered (upper panel), higher redshift sources (and consequently higher $L_{\rm opt}$) are included (middle panel) and if a lower flux limit is considered (bottom panel). We over-plot some of the lines from Fig.~\ref{eff_simulated} as a guide.
DL11 find an almost constant ratio of $L_{\rm bol}$ to $L_{\rm opt}$, in agreement with the low dependence of this ratio with bolometric luminosity (\citealt{marconi04}). In our calculations we assume a constant ratio of 9.1 which is the average found by DL11. The panels described above show that even with a constant bolometric correction, the parameter range will constrain the type of correlation one would find between the accretion efficiency and the black hole mass.
If we include a higher bolometric correction or a spread in values, the accessible parameter range would also be affected. In Fig.~\ref{eff_simulated_bolcorr} we plot how our sample (grey circles) would be distributed in the plane compared with the DL11 sources (red crosses). In the top panel we assume a higher bolometric correction ($K_{\rm opt} = 20$), and in the bottom panel, the effect of considering the $L_{\rm bol}$ dependent \cite{marconi04} bolometric correction. The first condition would cause a shift, and the second a larger spread in $\epsilon$.
\begin{figure}
\begin{centering}
\includegraphics[width=0.8\columnwidth]{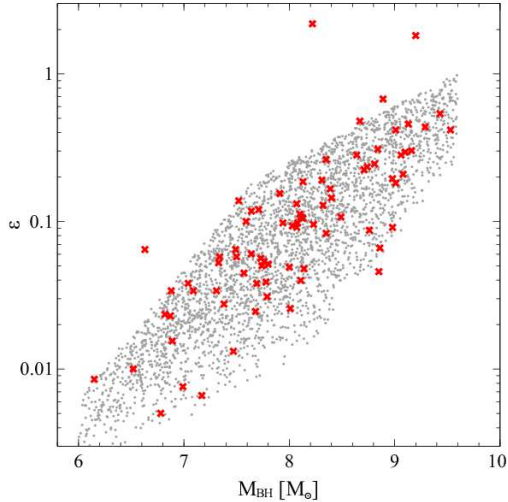}
\caption {Simulated distribution of accretion efficiency and black hole mass for a sample with a similar parameter range to DL11 (grey circles). The parameter range constrains the values of accretion efficiency and the correlations one can find. Our simulated sample shows a trend of $\epsilon \propto M^{0.56}$ similar to the $\epsilon \propto M^{0.5}$ found by DL11. For a different parameter space the correlation would be different.}
\label{eff_simulated_dl11_final}
\end{centering}
\end{figure}

The procedure described above yields a trend of efficiency with black hole mass by simply constraining the $L_{\rm opt}$, $M_{\rm BH}$ and redshift distribution to be similar to DL11 (Fig.~\ref{mass_opt_z_DL11}). In the next paragraph we add more specific constrains such as the Eddington ratio range and determine the slope determined from the simulated $\epsilon$ vs $M_{\rm BH}$.

We simulate a sample to explore the correlations one would see simply based on the parameter space available to the DL11 quasar sample. We use the constraints mentioned above for luminosity and black hole mass as a function of redshift and use a constant bolometric correction. To better represent the typically high Eddington ratio quasars in DL11, we add a $\lambda > 0.04$ cut-off but leave the possibility of $\lambda > 1.0$ to include some of the sources which, from their calculations, show an Eddington ratio above one. We also limit the mass range to the PG quasar minimum and maximum values $\log (M_{\rm BH}/M_{\odot}$) = (6.0 - 9.6) and $\log$ ($L_{\rm opt}$ [erg s$^{-1}$]) = (43.4 - 46.5). Using Eq. \ref{DL11_eff2}, we determine the accretion efficiency and plot our results in Fig.~\ref{eff_simulated_dl11_final}. We over-plot our simulated sample (grey dots) and the results by DL11 (red crosses). We can see that with a constrained distribution of $L_{\rm opt}$ and $M_{\rm BH}$ due to selection effects, we can only probe a small range of parameters. Our simulated sources show a trend $\epsilon \propto M^{0.56}$ similar to the one found in DL11 ($\epsilon \propto M^{0.5}$), illustrating that this is mainly caused by the allowed parameter space covered by the sample.
In general, if our AGN has a lower Eddington ratio and/or a lower $L_{\rm opt}$, it should populate the empty upper region of the plot. If we select sources from larger volume surveys, we should also be able to fill in the region of high mass and high luminosity accretion efficiencies. The trend between accretion efficiency and black hole mass would consequently be different for such a sample. This will be visible in later sections when we analyse both local and high redshift sources and locate them in this plane.

In terms of the actual values of accretion parameters, some sources show very low $\epsilon$ and others Eddington ratios above one. The simulated sample discussed above also shows these properties because we did not constrain it further than the DL11 distributions. As discussed in DL11, there are factors not included in their calculations that would raise the values of accretion efficiency obtained, namely the host galaxy contamination and the dust reddening. The host galaxy contribution increases the value of $L_{\rm opt}$, correcting the measured values for this effect would move the red sources to lower $L_{\rm opt}$, which means up and to the left in Fig.~\ref{eff_simulated_dl11_final}. 
The dust correction acts to decrease the observed bolometric luminosity. Correcting for this effect would alter the parameters in the same way that an increase in the bolometric correction would do, which means moving the sources to higher accretion efficiencies as can be seen in the extreme case illustrated in Fig.~\ref{eff_simulated_bolcorr}, upper panel. As mentioned in DL11, it is unlikely that $\epsilon$ is constant for all sources, simply based on the parameter limits plotted in our figures, we can see that at least the Eddington limit will cause some of the regions to be free of sources. The other trends will always depend on the redshift range probed, sensitivity limits and the luminosity and black hole mass functions.
\begin{figure}
\begin{centering}
\includegraphics[width=0.8\columnwidth]{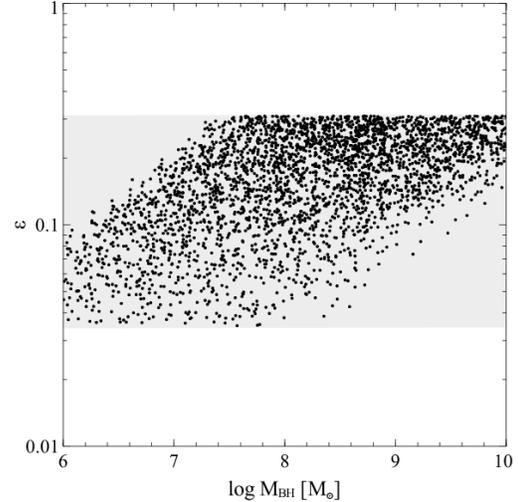}
\caption {Intrinsic distribution of accretion efficiencies and black hole masses for our test sample. Grey region indicates area of efficiencies ranging from the minimum counter-rotating black hole ($\epsilon = 0.035$) to a maximally spinning black hole ($\epsilon = 0.31$). Black dots represent generated uniform sample of AGN selected to have Eddington ratio in the range (0.01 - 1.0).}
\label{eff_intrinsic_and_eddington}
\end{centering}
\end{figure}
\subsection{Uncertainties in inclination and black hole mass}
\label{sec:uncertainties}
In this section we simulate another test sample to investigate the effects of uncertainties in the inclination and black hole mass in the accretion efficiency calculations.\footnote{Please refer to Section 4.2 of DL11 for their discussion on the uncertainties in $M_{\rm BH}$.} We start by simulating a simple set of AGN with well established intrinsic properties: uniform accretion efficiency in the theoretical range $0.035 < \epsilon < 0.31$, black hole mass $10^{6} < M_{\rm BH}/M_{\odot} < 10^{10}$, Eddington ratio $0.01 < \lambda < 1.0$, constant bolometric correction $K_{\rm bol} = 9.1$ and inclination values between 10$^{\circ}$ and 80$^{\circ}$ uniformly distributed in $\cos\thinspace(i)$ to simulate the solid angle weighting. These properties constrain the $L_{\rm opt}$ distribution for these sources. Our aim is then to infer if, starting from this intrinsic $L_{\rm opt}$ distribution and using Eqs. \ref{DL11_mdot} to \ref{DL11_eff2}, we would be able to retrieve the real properties of the sample.

\begin{figure}
\begin{centering}
\includegraphics[width=0.8\columnwidth]{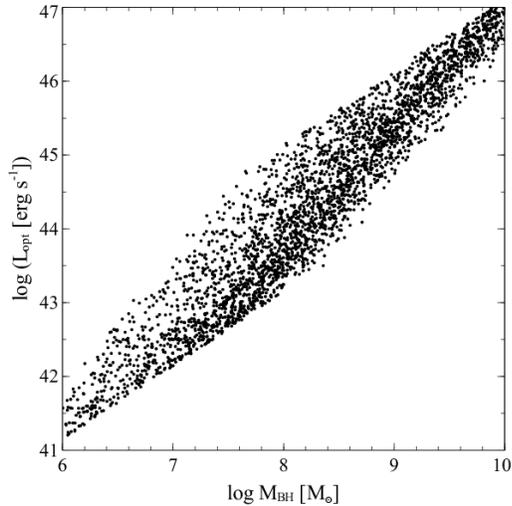}
\caption {Intrinsic optical luminosity ($\nu L_{\nu}$) as a function of black hole mass calculated for a sample with accretion efficiency plotted in black in Fig.~\ref{eff_intrinsic_and_eddington}. See Section~\ref{sec:uncertainties} for calculation details.}
\label{lopt_intrinsic}
\end{centering}
\end{figure}

\begin{figure}
\begin{centering}
\includegraphics[width=0.8\columnwidth]{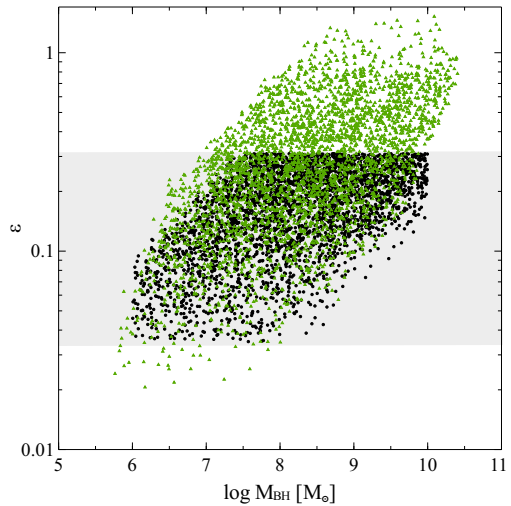}
\caption {Distribution of intrinsic (black dots) and inferred (green triangles) accretion efficiency and black hole masses for a test sample of AGN. The inferred values assume a factor of 3 error in the black hole mass and a fixed inclination. This causes a spread in the diagonal direction creating a difference between what we infer from the data and the intrinsic distribution.}
\label{eff_intrinsic_and_observed}
\end{centering}
\end{figure}

In Fig.~\ref{eff_intrinsic_and_eddington} we show the accretion efficiency uniform distribution before (grey region) and after applying an Eddington ratio cut (black dots). In Fig.~\ref{lopt_intrinsic} we plot the intrinsic $L_{\rm opt}$ distribution calculated using Eq.~\ref{DL11_eff2} and the distribution of $\cos (i)$ (the implicit dependence in this equation is $\epsilon \propto (\cos i)^{1/2}$).
We then do the backward calculation and use this $L_{\rm opt}$ distribution and Eq.~\ref{DL11_eff2} to determine the measured accretion efficiency. The difference this time is that we use an \emph{observed} black hole mass constrained to be within a factor of three of the intrinsic black hole mass ($M_{\rm intr}/3 < M_{\rm obs} < 3\times M_{\rm intr}$) to simulate the typical observational uncertainties in measuring this property. The inclination angle in taken as a constant value of $\cos (i) = 0.8$, although the intrinsic distribution includes angles between 10$^{\circ}$ and 80$^{\circ}$. The inferred accretion efficiency distribution is then calculated using Eq.~\ref{DL11_eff2} and the results are plotted in Fig.~\ref{eff_intrinsic_and_observed}. The diagonal cutoffs depend on the Eddington range assumed, and the diagonal spread to higher and lower values of accretion efficiency are caused by the uncertainties in black hole mass and the effect of assuming a constant angle to a sample that has an intrinsic spread in inclination. With this test we can see that values below or above the predicted accretion efficiency limits can in part be due to uncertainties in the black hole mass and inclination angle. This test also shows that a uniform accretion efficiency distribution, when selected as a function of Eddington ratio, will show a trend with black hole mass. 
Although the correlation found by DL11 is likely due to the parameter space probed, their work has clearly set the method to determine accretion efficiencies for individual AGN. The parameter space covered by their sample would necessarily cause a distribution of sources in a constrained $\epsilon$ vs $M_{\rm BH}$ region. If we consider other samples and their physical properties, based on the DL11 simple arguments, we can now calculate the accretion efficiency for AGN. As referred to in the previous subsections, there are still issues in measuring these properties which can affect our results. We mentioned the effect of host galaxy correction, inclination and black hole mass, but there are more factors that can cause problems in these calculations. In the next sections we will discuss the method of determining mass accretion rates and accretion efficiencies, the possible uncertainties involved in the calculations and apply this to examples of AGN with different characteristics.
\section{Accretion efficiency measurements}
\label{sec:accretion}
In the next sections, we will explore several methods to estimate the accretion efficiency of individual AGN and discuss the uncertainties associated. We will use the standard accretion disc relations described in Section~\ref{sec:model}. Although more complex disc models exist, such as the ones explored in DL11, we use this simple approach because it reflects the main physical dependences in the accretion parameters, while allowing to evaluate analytically the source of uncertainties.
There are two main steps in the accretion efficiency determination: the calculation of the mass accretion rate and the determination of the bolometric luminosity. The former can be calculated from the emission at optical wavelengths, using Eq. \ref{Mdot} and an inclination value. If we assume that this equation reproduces the trend between optical luminosity and accretion rate, the main uncertainties are the disc inclination, the black hole mass, the host galaxy contribution to the emission, and the presence of dust. 
As can be seen in Fig.~\ref{spin}, AGN with similar values of mass accretion rate, can differ in their bolometric luminosities due to the effect of spin: a source with higher spin will have a higher bolometric luminosity.
To constrain the accretion efficiency, and consequently the spin, one needs to have an accurate estimate of the bolometric luminosity. This has always been a complicated issue, AGN emit radiation over a wide spectral range due to different emission mechanisms. Ideally, knowing the spectral energy distribution would allow us to determine the bolometric luminosity, but in practice, this is constrained by the fact that for most sources, the wavelength coverage is not very good, and that there are spectral regions impossible to probe, such as the far-UV regions where emission is obscured by the gas and any intervening dust in our galaxy.
\begin{figure}
\begin{centering}
\includegraphics[width=0.8\columnwidth]{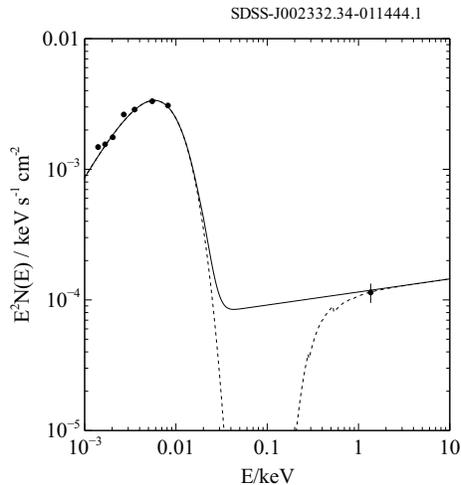}
\caption {Observed frame spectral energy distribution example of a modelled turnover for one of the SDSS sources. Black circles indicate the optical/UV data points corrected for galactic reddening and the X-ray data point. Dashed line is the model for observed emission and solid line is the model for intrinsic emission (before galactic absorption).}
\label{turnover}
\end{centering}
\end{figure}

Here we explore possibilities of determining the bolometric luminosity for different AGN samples. One is, of course, the method adopted by DL11. They gathered all data points and spectral slopes available on the PG quasars and assumed typical quasar slopes for regions without data coverage. Their SEDs are based simply on the data and do not attempt to follow a standard AGN SED model. This allows freedom in the SED shape and variations for individual AGN, but as mentioned in their work, can also result in extreme spectral shapes for a small sub-sample of sources.
Another approach is to select objects that show a turnover in their optical/UV data, assume that it constrains the spin (see Fig.~\ref{turnover}), and integrate the optical to X-ray part of the SED to obtain the intrinsic bolometric luminosity. In the current AGN paradigm, we expect to see reprocessed emission in the infrared (IR), due to the optical/UV disc emission being re-emitted by material within the surrounding parsecs of the AGN. We use this alternative argument to determine the bolometric luminosity from the integrated IR emission and the X-rays, keeping in mind that geometry also plays a role in this case, since not all of the radiation is reprocessed.
\subsection{Optical/UV and X-ray simultaneous observations}
Variability is one of the important factors to take into account when reconstructing an AGN SED. It is well known that most AGN show some sort of flux variability, with timescales that depend on the waveband observed. We first start by determining the mass accretion rate for a sample of objects with simultaneous observations in the optical/UV and X-rays. 

\begin{figure}
\begin{centering}
\includegraphics[width=0.8\columnwidth]{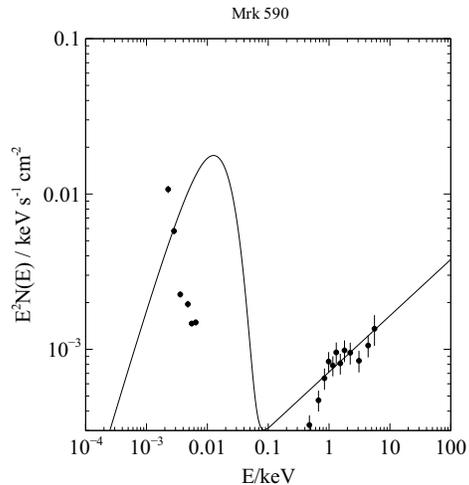}
\caption {Example of the SED of one of the excluded sources from our analysis. Observed data points show a shape that can indicate the presence of dust. The solid line shows the best-fit intrinsic SED for this source when including a fixed value of $E(B-V) = 0.25$ (\citealt{winter10}). Not considering the contribution of dust can underestimate the bolometric luminosity.}
\label{dust_in_SED}
\end{centering}
\end{figure}
The sample consists of 26 local AGN (redshift up to $\sim 0.08$) from the \textit{Swift}/BAT 9-month catalogue with observations in the UV-optical (UVOT) and X-ray telescopes (XRT). These sources were selected due to their low absorption and low spectral complexity and have been analysed in detail in \cite{vasudevan09}, on a study of their physical properties and bolometric corrections. The UVOT flux values we use are from the mentioned work as well, and have been corrected for host galaxy emission and Galactic extinction (see sections 3.1 and 3.2 of \citealt{vasudevan09} for details on this procedure). For most sources there are six data points corresponding to the UVOT filters (see Table 1 of \citealt{vasudevan09}), which we use in this work to determine directly the mass accretion rate. Our final sample consists of 22 sources, since 4 of the 26 sources show clear signs of SEDs affected by dust and were therefore excluded from our analysis: MCG-06-30-15, Mrk 590, Mrk 766 and ESO 490-G026 (see Fig.~\ref{dust_in_SED} as an example of dust in SED).

The optical monochromatic luminosity ($L_{\nu}$) is measured at the rest-frame wavelength of 4392 \AA. For the AGN at low redshift, since the observed and rest-frames do not differ by much, we use the $B$-band flux value at 4392 \AA\thinspace\thinspace observed wavelength and extrapolate it linearly to obtain the luminosity at 4392 \AA \thinspace\thinspace rest-frame of each source. 
The adopted average disc inclination throughout this work is $\cos (i)=0.8$ because we select sources that show low obscuration, and from the AGN standard paradigm are expected to be seen face-on or with low inclination. This is also the value adopted in DL11, which allows an easier comparison of results.
The black hole masses are gathered from different sources in the following order of preference: reverberation mapping (\citealt{denney06, denney10,peterson04}), H$_{\beta}$ virial mass estimates (\citealt{winter10, ho08, grupe10, parisi09}) and $K$-band luminosity (\citealt{vasudevan09}).

\begin{figure}
\begin{centering}
\includegraphics[width=0.8\columnwidth]{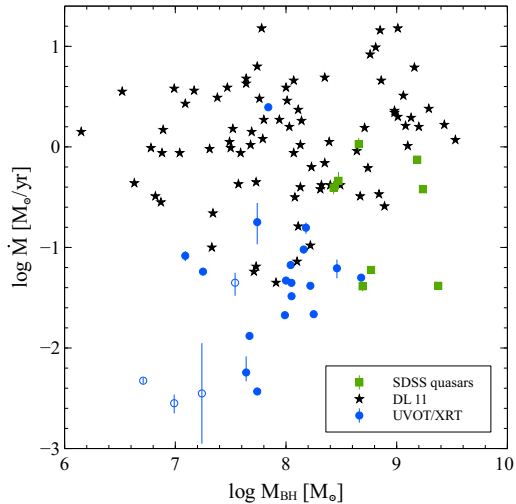}
\caption {Mass accretion rate as a function of black hole mass for our two samples of AGN and the PG quasar sample of DL11 (black stars). Green squares represent higher redshift bright quasars, and blue circles the low redshift \emph{Swift}/BAT AGN sample. Open blue circles indicate sources with more uncertain measurements (the SED shows a strange shape that could be due to uncertain black hole mass or mild presence of dust). Some data points have small error bars which are not visible at the scale of the plot. The data for this figure can be found in Table \ref{mdot_table}.}
\label{mdot_mass}
\end{centering}
\end{figure}

\begin{figure}
\begin{centering}
\includegraphics[width=0.8\columnwidth]{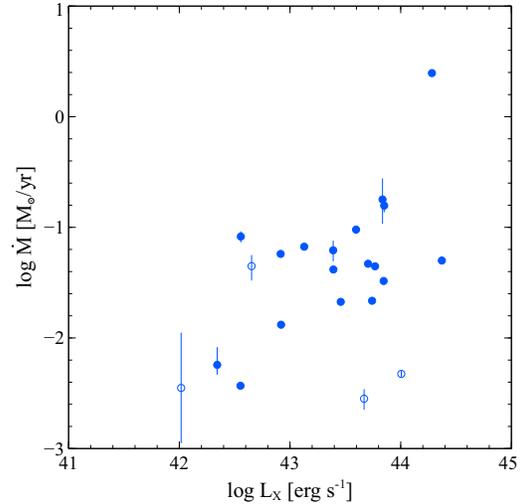}
\caption {Mass accretion rate as a function of X-ray (2-10 keV) luminosity for the \textit{Swift}/BAT local AGN sample. The optical luminosity used to derive $\dot{M}$ and X-ray luminosity are from simultaneous observations. Filled and open circles as in Fig.~\ref{mdot_mass}.}
\label{mdot_vs_lx}
\end{centering}
\end{figure}

The results are shown in Fig.~\ref{mdot_mass} in blue and in Table~\ref{mdot_table}. We also plot the dependence of mass accretion rate with X-ray luminosity in Fig.~\ref{mdot_vs_lx}. The results show a trend of increasing mass accretion rate with X-ray luminosity, which is expected if we assume that higher X-ray luminosities are associated with higher optical luminosities. We stress that for these sources, the X-ray and optical/UV are simultaneous, and that the effect of variability is negligible in this case.
\begin{table*}
\begin{tabular}{|l|l|l|c|c|c|c|c}
\hline
AGN&$z$&log $M_{\rm BH}/M_{\odot}$&log ($\nu L_{\nu}$)&log $\rm{L_{2-10keV}}$&log $\dot{M}$&Mass source&Notes\\
&  (1)&\quad\quad  (2)&   (3)&  (4)&   (5)&   (6)&   (7)\\\hline
1RXS J045205.0+493248&$0.029$&\quad$8.04^{+0.01}_{-0.01}$&	44.08	&	43.13	&	-1.18	&	V09&
\\
2MASX J21140128+8204483&$0.084$&\quad$8.68^{+0.01}_{-0.02}$&	44.43	&	44.37	&	-1.30	&	V09&
\\
3C 120&$0.03301$&\quad$7.74^{+0.2}_{-0.2}$&	44.17	&	43.84	&	-0.75	&	P04&
\\
3C 390.3&$0.0561$&\quad$8.46^{+0.09}_{-0.1}$&	44.34	&	43.39	&	-1.21	&	P04&
\\
Ark 120&$0.032713$&\quad$8.18^{+0.05}_{-0.06}$&	44.42	&	43.85	&	-0.80	&	P04& \\
ESO 548-G081&$0.01448$&\quad$7.74^{+0.01}_{-0.01}$&	43.04	&	42.55	&	-2.43	&	V09	&
\\
IRAS 05589+2828&$0.033$&\quad$8.22^{+0.01}_{-0.01}$&	44.07	&	43.39	&	-1.38	&	W09	&
\\
IRAS 09149-6206&$0.0573$&\quad$7.84$&45.00	&	44.28	&	\thinspace\thinspace0.39	&	P09	&
\\
MCG +04-22-042&$0.032349$&\quad$7.99^{+0.01}_{-0.01}$&43.72	&	43.46	&	-1.67	&	W09	&
\\
Mrk 1018&$0.042436$&\quad$8.25^{+0.02}_{-0.01}$&43.90	&	43.74	&	-1.66	&	W09	&
\\
Mrk 279&$0.030451$&\quad$7.54^{+0.1}_{-0.1}$&43.63	&	42.65	&	-1.35	&	P04	&	*
\\
Mrk 352&$0.014864$&\quad$7.24^{+0.5}_{-0.5}$&42.70	&	42.02	&	-2.45	&	H08	&	*
\\
Mrk 509&$0.034397$&\quad$8.16^{+0.04}_{-0.04}$&44.27	&	43.60	&	-1.02	&	P04	&	
\\
Mrk 841&$0.036422$&\quad$8.05^{+0.02}_{-0.02}$&43.88	&	43.85	&	-1.48	&	W09	&
\\
NGC 4593&$0.009$&\quad$6.99^{+0.09}_{-0.1}$&42.47	&	43.67	&	-2.55	&	D06	&	*
\\
NGC 5548&$0.017175$&\quad$7.64^{+0.16}_{-0.09}$&43.10	&	42.34	&	-2.24	&	D10	&
\\
NGC 7469&$0.016317$&\quad$7.09^{+0.05}_{-0.05}$&43.51	&	42.56	&	-1.08	&	P04	&
\\
NGC 985&$0.043143$&\quad$8.05^{+0.02}_{-0.02}$&43.97	&	43.77	&	-1.35	&	G10	&
\\
SBS 1136+594&$0.0601$&\quad$8.00^{+0.01}_{-0.01}$&43.95	&	43.71	&	-1.33	&	W09	&
\\
SBS 1301+540&$0.0299$&\quad$7.25^{+0.01}_{-0.01}$&43.51	&	42.92	&	-1.24	&	V09	&
\\
UGC 06728&$0.006518$&\quad$6.71^{+0.03}_{-0.03}$&42.43	&	43.99	&	-2.32	&	W09	&	*
\\
WKK 1263&$0.02443$&\quad$7.67^{+0.01}_{-0.01}$&43.37	&	42.92	&	-1.88	&	V09	&
\\
\hline
SDSS J002332.34-011444.1	&	0.483	&	\quad9.24	$\pm$	0.03&	45.38	&	44.48
	&	-0.42	&	S10	\\
SDSS J005709.94+144610.1	&	0.172	&	\quad9.38	$\pm$	0.02&	44.84	&	44.37
	&	-1.38	&	S10	\\
SDSS J092809.43+383000.5	&	0.498	&	\quad9.19	$\pm$	0.02&	45.55	&	44.09
	&	-0.13	&	S10	\\
SDSS J093944.53+392402.8	&	0.455	&	\quad8.66	$\pm$	0.06&	45.30	&	43.49
	&	 \thinspace\thinspace0.03	&	S10	\\
SDSS J111706.39+441333.3	&	0.144	&	\quad8.77	$\pm$	0.02&	44.54	&	43.91
	&	-1.22	&	S10	\\
SDSS J142424.21+595300.4	&	0.135	&	\quad8.69	$\pm$	0.05&	44.38	&	43.73
	&	-1.39	&	S10	\\
SDSS J142455.53+421407.6	&	0.316	&	\quad8.47	$\pm$	0.09&	44.93	&	44.60
	&	-0.34	&	S10	\\
SDSS J163051.74+471118.9	&	0.270	&	\quad8.43	$\pm$	0.07&	44.86	&	44.00
	&	-0.40	&	S10	\\
\hline
\end{tabular}
\caption{Results from the optical/UV and X-ray data on the \textit{Swift}/BAT sample of local AGN and the bright SDSS quasars. (1) Redshift. (2) Log of central black hole mass from different available sources (see column 6), with errors. (3) Optical luminosity at $\lambda = 4392$\thinspace\thinspace\AA \thinspace\thinspace in erg s$^{-1}$. (4) Log of 2--10 keV luminosity in erg s$^{-1}$ from fit to 0.3--10 keV regime for the \textit{Swift}/BAT sample. For the SDSS quasars this value is only an approximation since we assume a powerlaw with index $\Gamma = 1.9$ and no intrinsic absorption. (5) Mass accretion rate in $M_{\odot}$/yr from Eq. \ref{Mdot}. (6) Source for black hole mass value: D06 and D10) Reverberation mapping from \citealt{denney06,denney10} P04) Reverberation mapping from \citealt{peterson04}; W09) $H_{\beta}$ from \citealt{winter10}; H08) $H_{\beta}$ from \citealt{ho08}; G10) $H_{\beta}$ from \citealt{grupe10}; P09) $H_{\beta}$ from \citealt{parisi09}; V09) $K$-band luminosity from \citealt{vasudevan09} and S10) \citealt{shen11}. (7) Symbol * indicates sources with more uncertain values - see text for explanation.}
\label{mdot_table}
\end{table*}
\subsection{Infrared observations}
\label{sec:IR}
One option to remove the need of assuming a full SED shape on the bolometric luminosity calculation, is to measure the emitted reprocessed flux in the IR. With this approach, assuming that the disc emission is re-emitted by the obscuring heated  material, it is possible to integrate the observed IR and the X-ray emission and obtain an estimate of the bolometric luminosity.

\begin{figure}
\begin{centering}
\includegraphics[width=0.8\columnwidth]{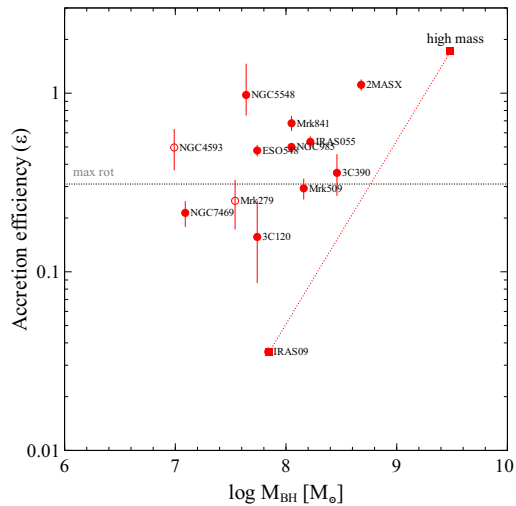}
\caption {Accretion efficiency as a function of black hole mass for \textit{Swift}/BAT sources with IRAS IR measurements. Open circles represent sources with uncertain mass accretion rate measurements (open circles in Fig.~\ref{mdot_mass} as well). Squares indicate the two possible values for IRAS 09149-6206. See discussion in text about extreme sources.}
\label{eff_iras}
\end{centering}
\end{figure}

Some of the sources in the BAT sample for which we determined $\dot{M}$, have been observed by the Infrared Astronomical Satellite (IRAS) and analysed in a thorough way by \cite{vasudevan10} following the method of \cite{pozzi07}. They used the X-ray measured absorbing column density to choose an IR emission template and corrected their values for the disc covering factor, which determines what fraction of the emission will be reprocessed. The host galaxy IR contribution was also corrected using two methods which give similar results: the correlation found between the nuclear 12.3 $\micron$ and 2-10 keV luminosity (\citealt{gandhi09}), and host galaxy IR SED templates. In the following calculations, we adopt the values of integrated $L_{1-1000\thinspace \micron}$ using the latter correction method, and calculate the bolometric luminosity using: $L_{\rm bol}=L_{(1-1000)\thinspace\micron}+L_{\rm X\thinspace (0.5-100) keV}$ (but use the \citealt{gandhi09} method for calculations with high-resolution data below). There are more sophisticated methods of estimating the IR reprocessed bolometric output. For example, \cite{pozzi10} use the torus model from \cite{fritz06} on higher redshift quasars, but recent clumpy torus models such as the one by \cite{nenkova08} are also able to explain detailed features in mid-IR spectra. We nevertheless use the results from the simple method by \cite{pozzi07} employed in \cite{vasudevan10} here, since they are available for the sample of interest. The accretion efficiency is then obtained using these bolometric luminosities and the $\dot{M}$ values calculated in the previous section.
Our results are shown in Table~\ref{eff_iras_table} and in Fig.~\ref{eff_iras}. 
Some of the sources lie above the maximally rotating efficiency limit for a black hole, which lead us to a discussion on possible errors associated with this method. Since these are systematic errors and not statistical, the first obvious question is: are the black hole masses overestimated? This could be true for some of the abnormally high efficiency sources, which have their masses determined from the $K$-band bulge luminosity / black hole mass correlation and from virial methods respectively. These methods have associated uncertainties, specially the $K$-band correlation that depends on the bulge determination (AGN and galaxy disc subtraction). 

One of the sources clearly above the limit is NGC 5548. It has a relatively accurate black hole mass value determined from reverberation mapping, so its high accretion efficiency is most likely caused by an overestimated bolometric luminosity. The archival IRAS data used in this analysis were taken in 1983, while the UVOT data are from a  2007 observation. We are not probing contemporaneous events, which can be important in the presence of flux variability. In fact, this source has been shown to have long term optical and X-ray variability (e.g. 1993 to 1999 - \citealt{merkulova02} and 1996 to 2001 - \citealt{uttley03}). We searched the literature for optical measurements closer in time to the IRAS data campaign and found the specific flux at 4390\thinspace\AA\thinspace\thinspace for this source measured in 1974 (\citealt{debruyn&sargent78}). Simply based on the light travelling times for radiation coming from different parts of the AGN, we expect the IR emission to lag the optical by a factor of months up to years depending on the obscuring gas location (parsecs to tens of parsecs). In any case, the 1974 flux is almost a factor of 1.9 higher than the one from 2007, which means that if we repeat our calculations using the 1974 optical flux and the 1983 IRAS IR flux, we would obtain an accretion efficiency of roughly half the value in our table. This would still be above the limit. It is possible that other systematic uncertainties effect our measurements, such as the constant inclination and accretion disc model used. The more complex models described in DL11 give higher values for the mass accretion rate, and lower values for the accretion efficiency than our simple relations. This could contribute to the higher values observed here.
Going back to the flux variability, NGC 5548 is an example of the variations one can obtain when using non-contemporaneous data to determine the accretion efficiency. 

\begin{figure}
\begin{centering}
\includegraphics[width=0.8\columnwidth]{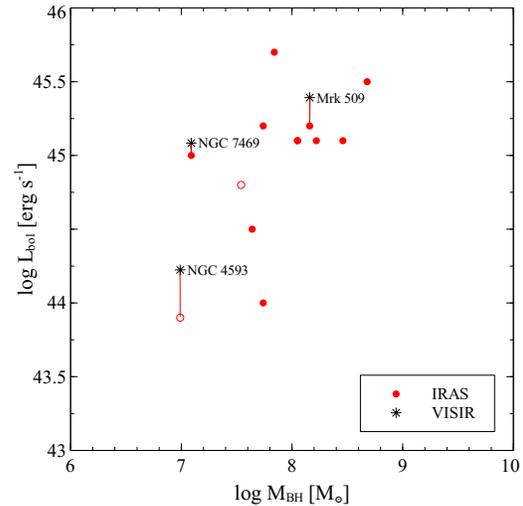}
\caption {Bolometric luminosity as a function of mass for sources with IRAS measurements and for sources with VISIR data from \citealt{gandhi09}. Circles represent results for the IRAS sample in Fig.~\ref{eff_iras}, and asterisks the three sources for which there are high-resolution VISIR data.}
\label{visir_ir}
\end{centering}
\end{figure}

We also note the very low efficiency IRAS 09149-6206. This source was discovered due to its IR emission, but has not been studied in great detail. We used the mass value by \cite{parisi09}, which is $M_{\rm BH} = 7\times10^{7}M_{\odot}$ but from the $K$-band its mass is $M_{\rm BH} = 3\times10^{9} M_{\odot}$ (\citealt{vasudevan09}) which is a very large difference, and moves the source to the upper right region of Fig.~\ref{eff_iras} (the two extreme values are plotted as red squares). The mass value is most likely somewhere in between the two values mentioned above, the lower value adopted gives an extremely high mass accretion rate compared to any other source in our sample.
A set of observations in the mid-IR closer in time to our optical/UV data has been done by \cite{gandhi09}. They used the VISIR instrument on the VLT to determine the luminosity at 12 $\mu$m ($L_{\rm 12\thinspace{\mu}m}$). The high resolution of these observations allowed to measure the flux from the core of a sample of objects with very low contamination from the host galaxy. We use for comparison, their data on three of our sources (Mrk 509, NGC 4593 and NGC 7469). We use their luminosity and bolometric correction (Eq. 5 of their paper) to determine alternative bolometric luminosities. We plot our results of bolometric luminosity as a function of black hole mass in Fig.~\ref{visir_ir} and other properties in Table~\ref{eff_visir_table}. The asterisks in Fig.~\ref{visir_ir} indicate the value of luminosity from the VISIR data and are connected by lines to the correspondent values using the IRAS flux. Red circles represent the same IRAS sample as in Fig.~\ref{eff_iras}. This example illustrates an alternative to obtain the bolometric luminosity. The difference in values can be due to several factors, including the simple bolometric correction used, but most likely has a contribution from flux variability. 
All the AGN in the \textit{Swift}/BAT sample may be affected by variability. The study of individual accretion efficiencies could be extended, providing the other uncertainties are accounted for, by selecting a sample with contemporaneous optical/UV, X-ray and IR data, where the IR lags the optical and UV emission by a few years. There are new surveys for which data is starting to become available, such as AKARI and WISE, that target the IR bands, and could be used to reconstruct contemporaneous SEDs for a study such as the one described above.
\begin{table*}
\begin{tabular}{|l|cl|c|c|c|c}
\hline
AGN&log $\rm L_{bol}$&$\epsilon$&$\lambda_{\rm Edd}$&$1/k_{\rm opt}$&Notes\\
&  (1)&   (2)&   (3)&  (4)  &   (5)\\\hline
Mrk 509                &	45.39	&	0.46	&	1.4e-1	&	7.5e-2
\\
NGC 4593               &	44.22	&	1.04	&	1.4e-1	&	1.7e-2 & *
\\
NGC 7469               &	45.08	&	0.26	&	7.8e-1	&	2.7e-2
\\
\hline
\end{tabular}
\caption{Results for sources with VISIR IR data from \citealt{gandhi09}. (1) Log of observed bolometric luminosity in erg s$^{-1}$ calculated using the X-ray and $L_{12 {\mu}m}$ correlation and the bolometric correction in \citealt{gandhi09}. (2) Accretion efficiency calculated from Eq. \ref{efficiency}. (3) Eddington ratio ($\lambda = L_{\rm bol}/L_{\rm Edd}$). (4) Bolometric correction $L_{\rm opt}$/$L_{\rm bol}$. (5) Symbol * indicates sources with more uncertain values - see text for explanation.}
\label{eff_visir_table}
\end{table*}

\subsection{SDSS quasar sample}
As mentioned in the beginning of Section~\ref{sec:accretion}, for high mass AGN at the right redshift, it is possible to constrain the accretion disc emission peak in the optical/UV wavelengths. In this section, we analyse a sample of 10 bright quasars with SDSS and GALEX data at higher redshifts $z \sim (0.13 - 0.5)$. They have flux measurements in the $u, g, r, i$ and $z$ SDSS bands and non-simultaneous in the Near-UV and Far-UV bands of GALEX, which are corrected for galactic reddening. The host galaxy contribution was not removed for these sources since they are bright quasars and are expected to outshine the host galaxy. There is also a single X-ray measurement, although this is not simultaneous with the UV and optical data. A detailed analysis of their SED and spectral indices will be done in Vasudevan et al. (in prep). Out of the initial sample, two sources were excluded due to signs of dust effects on the SED, which leave us with a sample of 8 quasars.

We determine $L_{\nu}$ at the rest-frame wavelength of 4392 \AA, by modelling the SED in that region, based on the seven bands for which there is data information. The black hole masses used are virial mass estimates from \cite{shen11} (column 139 in their Table 1). Since the redshifts are higher, we do not extrapolate the flux between observed bands but assume a simple model to interpolate the data (see below). This should be spin independent as we already discussed in Fig.~\ref{spin}. The quasars were selected to have their emission peak at the observed bands mentioned above and we determine the mass accretion rate for 8 sources but narrow down the sample for the accretion efficiency measurements. We require to have not only the emission peak at the right energies but also a well constrained turnover by the data, which is only observed for 3 of the sources with similar shapes to Fig.~\ref{turnover} (see list in Table~\ref{eff_sdss_table}).
For the \textit{Swift}/BAT sources, the black hole mass and low redshift set the disc emission peak at higher energies than the ones probed by the UVOT bands. In this case, there is a wide range of accretion efficiencies that are consistent with the measured $\dot{M}$. For the SDSS sources, due to their higher mass values, the data points are distributed around the disc emission peak and we use this to constrain the bolometric luminosity. To obtain an approximate SED shape based on the data, we use {\sc XSPEC} to fit a model to our observations. 
Since we only have one X-ray data point, we cannot constrain the amount of obscuration (hydrogen column density), and the slope of the X-ray powerlaw. To get an approximate shape, we fix the powerlaw index to a typical value of $1.9$, and consider that our source is only affected by galactic absorption. This means that the bolometric luminosity obtained is most likely a lower limit to the real value, due to an underestimated X-ray luminosity. This effect is nevertheless small, because we expect most of the emission to be in the optical/UV region of the SED. For one of our sources, SDSS J092809.43+383000.5, the X-ray flux is an upper limit to the real value. We nevertheless use this limit in our calculations due to the small effect of the X-ray luminosity on the bolometric luminosity and the uncertainties in the photon index mentioned above.
The optical-UV region is modelled by a multi-temperature blackbody accretion disc around a Kerr black hole, to allow blackbody shape variations with spin ({\sc kerrbb}, \citealt{li05}). Our final model is then [{\sc kerrbb + wabs(bknpowerlaw)}], with a redshift factor added manually, and intrinsic reddening taken as zero. The intrinsic SED is then integrated from ($10^{-4} - 100$) keV to obtain the bolometric luminosity. An example of the best-fit model is shown in Fig.~\ref{turnover}. We emphasize that our main objective is to constrain the shape of the SED with the data points to be able to obtain $L_{\rm bol}$, and do not use the model output for the spin parameter. 
One of the important factors when modelling the accretion disc emission is the spectral hardening factor $f_{\rm col}$ (\citealt{shimura&takahara95}), which gives the ratio between the colour temperature of the blackbody used to model the emission and the disc effective temperature. A change in this parameter will not only shift the modelled emission peak but change the spectral shape (\citealt{ross92}), causing degeneracies with the spin determination. Here we assumed a constant typical value of $f_{\rm col} = 1.7$, but note that previous work has shown that $f_{\rm col}$ could also vary with other physical parameters (\citealt{merloni00}). In the model used, a change in $f_{\rm col}$ corresponds to a simple energy shift. This adds uncertainties since, as mentioned above, we should be including a spectral shape change as well. The value we obtain is just an approximation, valid for our assumptions, but we wanted to stress the importance of considering more complex accretion disc models that can tackle this issue (examples of some of the developments in this field are given in DL11 and references therein).
The results are shown in Fig.~\ref{eff_all} (green circles) and in Table~\ref{eff_sdss_table}. The quasars probe the highest black hole mass range, but show a spread in accretion efficiency from the slowly rotating to highly rotating black holes. They could also be affected by the accretion disc model used to determine the mass accretion rate as mentioned in the previous section.

With this method, we can only measure the accretion efficiency for sources of relatively high masses, since our data has to sample the peak of emission to be able to constrain the bolometric luminosity. 
\begin{table*}
\begin{tabular}{|l|c|c|c|c|c}
\hline
AGN&log $\rm L_{bol}$&$\epsilon$&$\lambda_{\rm Edd}$&$1/k_{\rm opt}$&Notes\\
&  (1)&   (2)&   (3)&  (4)  &   (5)\\\hline
2MASX J21140128+8204483&	45.50	&	1.11	&	5.3e-2	&	8.4e-2
\\
3C 120                 &	45.20	&	0.16	&	2.3e-1	&	9.3e-2
\\
3C 390.3               &	45.10	&	0.36	&	3.5e-2	&	1.7e-1
\\
ESO 548-G081           &	44.00	&	0.48	&	1.5e-2	&	1.1e-1
\\
IRAS 05589+2828        &	45.10	&	0.53	&	6.1e-2	&	9.2e-2
\\
IRAS 09149-6206        &	45.70	&	0.04	&	2.0e-1	&	2.0e-1
\\
Mrk 279                &	44.80	&	0.25	&	1.5e-1	&	6.8e-2 & *
\\
Mrk 509                &	45.20	&	0.29	&	8.8e-2	&	1.2e-1
\\
Mrk 841                &	45.10	&	0.68	&	9.0e-2	&	6.1e-2
\\
NGC 4593               &	43.90	&	0.50	&	6.5e-2	&	3.7e-2 & *
\\
NGC 5548               &	44.50	&	0.98	&	5.8e-2	&	4.0e-2
\\
NGC 7469               &	45.00	&	0.21	&	6.5e-1	&	3.2e-2
\\
NGC 985                &	45.10	&	0.50	&	9.0e-2	&	7.4e-2
\\
\hline
\end{tabular}
\caption{Results for sources with IRAS IR data available. (1) Log of observed bolometric luminosity in erg\thinspace s$^{-1}$ calculated from $L_{\rm IR}(1 - 1000 \micron)$ + $L_{\rm X}$ (0.5 - 100 keV) (\citealt{vasudevan09}). (2) Accretion efficiency calculated from Eq. \ref{efficiency}. (3) Eddington ratio ($\lambda = L_{\rm bol}/L_{\rm Edd}$). (4) Bolometric correction $L_{\rm opt}$/$L_{\rm bol}$. (5) Symbol * indicates sources with more uncertain values - see text for explanation.}
\label{eff_iras_table}
\end{table*}
\subsection{Constraining the SED observationally}
\label{sec:sed}
In the previous sections we referred to different methods of determining the bolometric luminosity. Ideally, an accurate luminosity should be the result of a fully observationally probed SED. The observed SED includes the obscuration effects and dust reddening, together with a full wavelength coverage, one would also have to model the obscuration signatures imprinted on an intrinsic emitted SED.

From theoretical arguments, we would expect to have the energy peak of the disc emission depending on the black hole mass, with lower mass black holes peaking at higher energies. For the highest mass black holes ($M_{\rm BH} \sim 10^{9} M_{\odot}$) such as our SDSS quasars, the optical and UV data can be used to constrain the peak turnover, but the Wien type fall-over slope is not possible to determine because it is expected to be at the energies where radiation is absorbed by obscuring gas (intrinsic and/or the Milky-Way line of sight material). Without these data, it is difficult to verify the deviations from a black-body type decreasing slope, which are expected from more complex models of accretions discs (see for example Fig.~2 of DL11). When we go to lower masses, the optical/UV data probes the Rayleigh-Jeans type part of the disc emission. This allows to determine the mass accretion rate but does not break the degeneracies in terms of black hole spin (such as the case for our \textit{Swift}/BAT sources).

One other option, if we probe even smaller black hole masses, would be to look at the low X-ray energies ($E < 2$ keV), to sample the Wien tail of the accretion disc emission. \cite{yuan10} analysed the data on RX J1633+4718, an AGN with black hole mass of the order $(2-4)\times$10$^{6} M_{\odot}$. In this work, they claim that the X-ray data are probing the higher energy wing of the accretion disc emission. Although there would always be regions more difficult to probe observationally, this approach could be used to constrain the highest energy range for low mass black hole disc emission. 
\section{Discussion}
\begin{figure}
\begin{centering}
\includegraphics[width=0.8\columnwidth]{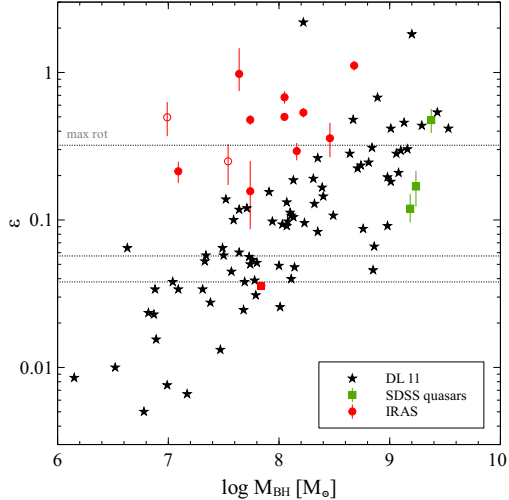}
\caption {Summary plot of our results. Accretion efficiency for all the AGN in this work compared with the distribution found by DL11 (black stars), as a function of their black hole mass. The red circles represent the \textit{Swift}/BAT sources with bolometric luminosity determined from IR measurements, and the green squares are the SDSS sources with bolometric luminosity calculated from a modelled accretion disc turnover. The horizontal dotted lines are from top to bottom, the maximum spin, non-rotating and maximum counter-rotating spin for a black hole.}
\label{eff_all}
\end{centering}
\end{figure}
\begin{figure}
\begin{centering}
\includegraphics[width=0.8\columnwidth]{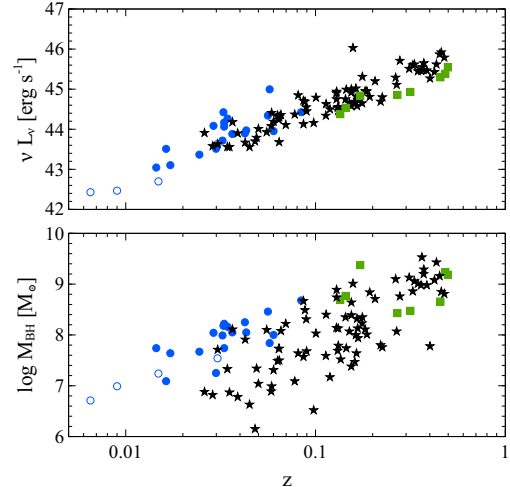}
\caption {Black hole mass and optical luminosity distribution for all sources. DL11 PG quasar sample (black stars), green squares for SDSS quasars and blue circles for \textit{Swift}/BAT AGN. Open circles indicate sources with more uncertain $L_{\rm opt}$ and are indicated with a different symbol in Table~\ref{mdot_table}.}
\label{mass_opt_z_all}
\end{centering}
\end{figure}

We have measured the mass accretion rates for 22 local \textit{Swift}/BAT AGN and 8 SDSS quasars. The results in Fig.~\ref{mdot_mass} show that with these two samples, we are, as expected, probing different black hole mass ranges. The $\dot{M}$ distribution shows a large spread for any of the samples, but the local, less luminous AGN tend to extend to lower values of mass accretion rate than the SDSS quasars and the PG quasar sample of DL11.
The distribution of properties for our samples is shown in Fig.~\ref{mass_opt_z_all}, where blue circles represent the \textit{Swift}/BAT sources and green squares the SDSS quasars. For comparison we over-plot the DL11 distribution in black.

\begin{figure}
\begin{centering}
\includegraphics[width=0.8\columnwidth]{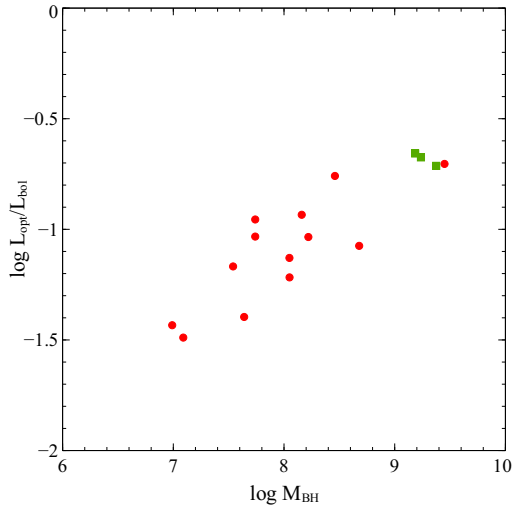}
\caption {Bolometric corrections for optical luminosity at 4392\thinspace\thinspace\AA. Red circles represent local AGN sample detected with \textit{Swift}/BAT and green squares SDSS quasars. Sources show less than a factor of ten spread in bolometric corrections, as one would expect (\citealt{marconi04}).}
\label{lopt_lbol_all}
\end{centering}
\end{figure}

\begin{figure}
\begin{centering}
\includegraphics[width=0.7\columnwidth]{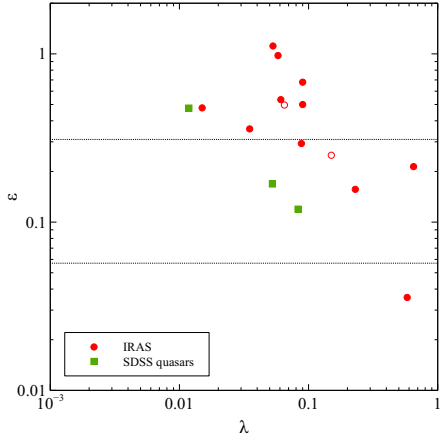}
\caption {Accretion efficiency as a function of Eddington ratio for local and high redshift AGN (colour coded as in Fig.~\ref{eff_all}). Our AGN lie in the region of unobscured sources of Fig.~3 in \citealt{raimundo&fabian09}, as expected due to our selection criteria of low obscuration sources.}
\label{edd_vs_eff}
\end{centering}
\end{figure}

To determine the accretion efficiencies, we selected a sub-sample of these 30 AGN for which we could obtain better estimates of the bolometric luminosity. Two different methods of determining the bolometric luminosity were used: constraining the turnover of the quasi-blackbody accretion disc and considering the reprocessed emission in the IR, using different sets of available archival data.
The final results are shown in Fig.~\ref{eff_all}. The accretion efficiencies show a spread between the minimum and maximum spin case ($0.057 < \epsilon < 0.31$, \citealt{novikov&thorne73}), with the exception of 4 sources which were discussed in Section~\ref{sec:IR}. They do not show a trend with black hole mass, which is what is expected from the discussion in Section~\ref{sec:correlations}. Since we are using different samples with different selection effects which extend to lower $L_{\rm opt}$, we expect to see a different distribution in the accretion efficiency vs black hole mass plane. 
The SDSS sources have lower accretion efficiencies, and from Fig.~\ref{eff_simulated} could appear to be above their Eddington ratio. In fact, from Fig.~\ref{lopt_lbol_all}, where we plot the distribution of bolometric corrections, we can see that they are in the group of highest corrections. This means that if we simulate their distribution as in Section~\ref{sec:correlations}, the constant $L_{\rm opt}$ limits will correspond to a lower $L_{\rm bol}$ and a lower Eddington ratio as well, as can from the Eddington ratio values in Table \ref{eff_sdss_table}.
In Fig.~\ref{edd_vs_eff} we plot the distribution of accretion efficiencies as a function of their Eddington ratio ($\lambda = L_{\rm bol}/L_{\rm Edd}$) for the sources with IRAS measurements and the SDSS quasars. We see that in general they tend to have moderate to high Eddington ratios ($\lambda > 0.01$) and high accretion efficiencies. This will locate them in the unobscured (low absorption) AGN region predicted from AGN evolution models (\citealt{raimundo&fabian09}), which is what we expect since these sources, local AGN and bright quasars, were selected to have low obscuration.\\

As we have been mentioning throughout the paper there are major uncertainties in the determination of accretion efficiencies. Here we summarise the main factors:\\
- \textit{Selection effects}: Our data coverage is limited, and selecting sources from different surveys will introduce different flux and volume limitations. Targeting large samples allow us to use the same method on a large scale and have all sources affected by approximately the same uncertainties, but will only provide access to a limited parameter space. This can cause apparent correlations between black hole accretion parameters. In this work we analysed different types of sources, but are still far from being able to draw conclusions on the general behaviour of AGN.\\
- \textit{Dust and host galaxy}: In our analysis, and also in DL11 previous work, we assumed $E(B-V) = 0$ for all our sources. This introduces errors which mainly affect the bolometric luminosity measurement, underestimating its value. Although we have used very bright quasars and low absorption \textit{Swift}/BAT AGN and excluded sources which have signs of dust in their SED, it is probable that every source will have some dust contamination. One option would be to use $E(B-V)$ measured spectroscopically, but since dust and spin are degenerate in simple accretion disc models, errors on the dust estimate could affect the bolometric luminosity by a significant factor. In the case of the IR bolometric luminosities this factor is less important, since we are integrating the emission, but there is still the host galaxy contamination. The host galaxy will not only contribute to the IR SED part but also when measuring the optical luminosity. Ignoring the host galaxy emission will cause an overestimated $L_{\rm opt}$ and $\dot{M}$ and an underestimated accretion efficiency\\
- \textit{Variability}: We addressed this issue in Section~\ref{sec:IR}, when we determined the bolometric luminosity by adding the intrinsic X-ray emission and the reprocessed optical/UV emission in the IR. Our mass accretion rates were determined using recent observations in the optical/UV, but the IR data were gathered $\sim$20 years before. For a variable source, such as some of the ones we analysed and discussed, the flux variation in time can be very large, which means that we are measuring the mass accretion rate and the bolometric luminosity at different AGN states. One approach to remove this effect would be to get contemporaneous data on the IR and optical/UV, as we exemplified with the VISIR data. The IR is expected to lag the optical emission by a few months up to years, which means that the observations would not have to be simultaneous but close in time.\\
- \textit{Black hole mass}: This is always an uncertainty in AGN studies, in our case it will contribute to every type of measurement we do. In our sample there are some black holes with reverberation mapping masses, which are the most accurate, but for some AGN we only have values from correlation with $K$-band host galaxy luminosity, which has larger uncertainties. 

Two other assumptions in this work are the accretion disc inclination (applied in every mass accretion rate determination) and the assumed accretion disc model (applied in every mass accretion rate determination and accretion efficiency for SDSS quasars). The inclination has a factor of $(\cos i)^{-3/2}$ in the mass accretion rate. We used $\cos (i) = 0.8$ to compare our results with DL11 but also because in the standard AGN paradigm, sources with low obscuration (such as ours) have most likely a smaller angle between the disc normal and our line of sight. A face-on disc establishes a lower limit for $\dot{M}$, and an upper limit for the accretion efficiency. It is probable that AGN show a distribution of different inclinations and an average value approach is not valid. The values presented in Table \ref{mdot_table} can be easily corrected for different inclination angles. The disc model we used is a simplification and previous work have shown that accretion discs are likely more complex than our simple approximation. This method, described in Section~\ref{sec:model} is useful to establish the main physical relations, and show almost the same trends as the best-fit relations found for more complex disc models in DL11.  

The methods explored in this paper can be used to target different types of AGN using a multi-wavelength data coverage. As described in Section~\ref{sec:sed}, black holes of different masses and redshifts will have their emission peak at different wavelength ranges, making it possible to constrain some characteristics of their spectra in limited wavebands. These observations can be used to complement the information we already have, providing the sources of uncertainties mentioned above are considered. 
\begin{table*}
\begin{tabular}{|l|c|c|c|c|}
\hline
AGN&log $\rm L_{bol}$&$\epsilon$&$\lambda_{\rm Edd}$&$1/k_{\rm opt}$\\
&  (1)&   (2)&   (3)&  (4)\\\hline
SDSS J002332.34-011444.1	&	46.06	&	0.17	&	5.2e-2
 & 0.21	\\
SDSS J005709.94+144610.1	&	45.55	&	0.48	&	1.2e-2
 & 0.19	\\
SDSS J092809.43+383000.5	&	46.20	&	0.12	&	8.3e-2
 & 0.22		\\
\hline
\end{tabular}
\caption{Results from SED fitting for sources with observed turnover. (1) Log of observed bolometric (0.0001--100 keV) luminosity in erg s$^{-1}$ measured by integrating the SED which best fits the UV/optical region. (2) Accretion efficiency calculated from Eq. \ref{efficiency}. (3) Eddington ratio ($\lambda = L_{\rm bol}/L_{\rm Edd}$). (4) Bolometric correction $L_{\rm opt}$/$L_{\rm bol}$.}
\label{eff_sdss_table}
\end{table*}
\section{Conclusions}
Motivated by the interesting work of \cite{davis&laor11}, who determined the accretion efficiency distribution for a large sample of quasars, we investigated the methods and uncertainties in calculating the accretion properties of individual AGN. We explored the correlation between the accretion efficiency and black hole mass ($\epsilon \propto M^{\sim 0.5}$) found by \cite{davis&laor11}, and conclude that it is mainly caused by the selection effects in their sample. In terms of the uncertainties, we find that the distribution of calculated efficiencies will show an artificial spread, due to possible errors in the black hole mass and due to assuming a constant inclination for sources which intrinsically have a broad distribution of inclination angles.
We also exemplify how samples at different luminosities lie in the efficiency black hole mass plane, and show that an Eddington limited sample (such as $\lambda < 1$) adds a mass dependent boundary. 

To test the method of determining the mass accretion rate and accretion efficiency, we select a low redshift sample of AGN with available multi-wavelength data and a sample of SDSS quasars with higher mass and luminosity. Our final results show that, with simplifying assumptions, it is possible to calculate the accretion efficiencies for individual AGN, although the values are largely sensitive to the unknown parameters and uncertainties. From the calculations, some AGN give accretion efficiency values above that for the maximally spinning case, which is not expected. The efficiencies are affected by uncertainties in the bolometric luminosity calculation, variability, accretion disc model, the black hole mass, inclination, dust and host galaxy contamination. The systematic errors dominate over the statistical errors, which makes it difficult to understand the general behaviour of AGN, and to determine correlations without further observational data on the sources.
The methods explored in this paper can be applied to other samples in the future, providing we are able to determine the parameters more accurately and constrain the errors involved. We discuss how such a study can be improved by observing sources with different properties, in particular the black hole mass, since their accretion disc emission will be expected to peak at different wavelengths. In the future, sources with good wavelength coverage, and especially with contemporaneous observations to account for variability, would be a good starting sample for such a study.

The determination of mass accretion rate and accretion efficiency for individual AGN is of great importance in understanding the growth of black holes and their relation with the host galaxies. The accretion efficiency is related to the black hole spin parameter, which has been studied theoretically as a method to learn the mechanisms of black hole growth (\citealt{volonteri05}, \citealt{king&pringle06}, \citealt{berti&volonteri08}). If determined for large samples, these parameters would not only shed light on why black holes show different physical properties but would also be able to constrain models of AGN populations and evolution.
\section{Acknowledgements}
SR thanks W.~N. Brandt for useful comments on the paper and for providing the data on the SDSS quasars, A. Zoghbi for kindly helping to update XSPEC models and R. Canning for useful discussions.
SR acknowledges financial support from FCT - Funda\c c\~ao para a Ci\^encia e a Tecnologia (Portugal). ACF thanks the Royal Society for support.
\bibliographystyle{mnras}
\bibliography{AGN}

\begin{thebibliography}{}

\bibitem[\protect\citeauthoryear{{Agol} \& {Krolik}}{{Agol} \&
  {Krolik}}{2000}]{agol&krolik00}
{Agol} E.,  {Krolik} J.~H., 2000, \apj, 528, 161

\bibitem[\protect\citeauthoryear{{Berti} \& {Volonteri}}{{Berti} \&
  {Volonteri}}{2008}]{berti&volonteri08}
{Berti} E.,  {Volonteri} M., 2008, \apj, 684, 822

\bibitem[\protect\citeauthoryear{{Bian} \& {Zhao}}{{Bian} \&
  {Zhao}}{2002}]{bian&zhao02}
{Bian} W.,  {Zhao} Y., 2002, A\&A, 395, 465

\bibitem[\protect\citeauthoryear{{Bian} \& {Zhao}}{{Bian} \&
  {Zhao}}{2003}]{bian&zhao03}
{Bian} W.,  {Zhao} Y., 2003, \pasj, 55, 599

\bibitem[\protect\citeauthoryear{{Boroson} \& {Green}}{{Boroson} \&
  {Green}}{1992}]{boroson&green92}
{Boroson} T.~A.,  {Green} R.~F., 1992, \apjs, 80, 109

\bibitem[\protect\citeauthoryear{{Boyle} et~al.}{{Boyle}
  et~al.}{2000}]{boyle00}
{Boyle} B.~J., {Shanks} T., {Croom} S.~M., {Smith} R.~J., {Miller} L.,
  {Loaring} N.,  {Heymans} C., 2000, \mnras, 317, 1014

\bibitem[\protect\citeauthoryear{{Brenneman} \& {Reynolds}}{{Brenneman} \&
  {Reynolds}}{2006}]{brenneman&reynolds06}
{Brenneman} L.~W.,  {Reynolds} C.~S., 2006, \apj, 652, 1028

\bibitem[\protect\citeauthoryear{{Cao}}{{Cao}}{2007}]{cao07}
{Cao} X., 2007, \apj, 659, 950

\bibitem[\protect\citeauthoryear{{Collin} et~al.}{{Collin}
  et~al.}{2002}]{collin02}
{Collin} S., {Boisson} C., {Mouchet} M., {Dumont} A., {Coup{\'e}} S., {Porquet}
  D.,  {Rokaki} E., 2002, A\&A, 388, 771

\bibitem[\protect\citeauthoryear{{Davis} \& {Laor}}{{Davis} \&
  {Laor}}{2011}]{davis&laor11}
{Davis} S.~W.,  {Laor} A., 2011, \apj, 728, 98 (DL11)

\bibitem[\protect\citeauthoryear{{de Bruyn} \& {Sargent}}{{de Bruyn} \&
  {Sargent}}{1978}]{debruyn&sargent78}
{de Bruyn} A.~G.,  {Sargent} W.~L.~W., 1978, \aj, 83, 1257

\bibitem[\protect\citeauthoryear{{Denney} et~al.}{{Denney}
  et~al.}{2006}]{denney06}
{Denney} K.~D. et~al., 2006, \apj, 653, 152

\bibitem[\protect\citeauthoryear{{Denney} et~al.}{{Denney}
  et~al.}{2010}]{denney10}
{Denney} K.~D. et~al., 2010, \apj, 721, 715

\bibitem[\protect\citeauthoryear{{Elvis}, {Risaliti}, \& {Zamorani}}{{Elvis}
  et~al.}{2002}]{elvis02}
{Elvis} M., {Risaliti} G.,  {Zamorani} G., 2002, \apjl, 565, L75

\bibitem[\protect\citeauthoryear{{Fabian} \& {Iwasawa}}{{Fabian} \&
  {Iwasawa}}{1999}]{fabian&iwasawa99}
{Fabian} A.~C.,  {Iwasawa} K., 1999, \mnras, 303, L34

\bibitem[\protect\citeauthoryear{{Fabian} et~al.}{{Fabian}
  et~al.}{1989}]{fabian89}
{Fabian} A.~C., {Rees} M.~J., {Stella} L.,  {White} N.~E., 1989, \mnras, 238,
  729

\bibitem[\protect\citeauthoryear{{Fritz}, {Franceschini}, \&
  {Hatziminaoglou}}{{Fritz} et~al.}{2006}]{fritz06}
{Fritz} J., {Franceschini} A.,  {Hatziminaoglou} E., 2006, \mnras, 366, 767

\bibitem[\protect\citeauthoryear{{Gammie}}{{Gammie}}{1999}]{gammie99}
{Gammie} C.~F., 1999, \apjl, 522, L57

\bibitem[\protect\citeauthoryear{{Gandhi} et~al.}{{Gandhi}
  et~al.}{2009}]{gandhi09}
{Gandhi} P., {Horst} H., {Smette} A., {H{\"o}nig} S., {Comastri} A., {Gilli}
  R., {Vignali} C.,  {Duschl} W., 2009, A\&A, 502, 457

\bibitem[\protect\citeauthoryear{{Grupe} et~al.}{{Grupe}
  et~al.}{2010}]{grupe10}
{Grupe} D., {Komossa} S., {Leighly} K.~M.,  {Page} K.~L., 2010, \apjs, 187, 64

\bibitem[\protect\citeauthoryear{{Ho}, {Darling}, \& {Greene}}{{Ho}
  et~al.}{2008}]{ho08}
{Ho} L.~C., {Darling} J.,  {Greene} J.~E., 2008, \apjs, 177, 103

\bibitem[\protect\citeauthoryear{{King} \& {Pringle}}{{King} \&
  {Pringle}}{2006}]{king&pringle06}
{King} A.~R.,  {Pringle} J.~E., 2006, \mnras, 373, L90

\bibitem[\protect\citeauthoryear{{Koratkar} \& {Blaes}}{{Koratkar} \&
  {Blaes}}{1999}]{koratkar&blaes99}
{Koratkar} A.,  {Blaes} O., 1999, \pasp, 111, 1

\bibitem[\protect\citeauthoryear{{Laor}}{{Laor}}{1991}]{laor91}
{Laor} A., 1991, \apj, 376, 90

\bibitem[\protect\citeauthoryear{{Laor} \& {Davis}}{{Laor} \&
  {Davis}}{2011}]{Laor&Davis11}
{Laor} A.,  {Davis} S.~W., 2011, ArXiv e-prints 1106.4969

\bibitem[\protect\citeauthoryear{{Li} et~al.}{{Li} et~al.}{2005}]{li05}
{Li} L., {Zimmerman} E.~R., {Narayan} R.,  {McClintock} J.~E., 2005, \apjs,
  157, 335

\bibitem[\protect\citeauthoryear{{Lynden-Bell}}{{Lynden-Bell}}{1969}]{lynden-b%
ell69}
{Lynden-Bell} D., 1969, \nat, 223, 690

\bibitem[\protect\citeauthoryear{{Marconi} et~al.}{{Marconi}
  et~al.}{2004}]{marconi04}
{Marconi} A., {Risaliti} G., {Gilli} R., {Hunt} L.~K., {Maiolino} R.,
  {Salvati} M., 2004, \mnras, 351, 169

\bibitem[\protect\citeauthoryear{{Merkulova}}{{Merkulova}}{2002}]{merkulova02}
{Merkulova} N.~I., 2002, A\&A, 387, 40

\bibitem[\protect\citeauthoryear{{Merloni}}{{Merloni}}{2004}]{merloni04}
{Merloni} A., 2004, \mnras, 353, 1035

\bibitem[\protect\citeauthoryear{{Merloni}, {Fabian}, \& {Ross}}{{Merloni}
  et~al.}{2000}]{merloni00}
{Merloni} A., {Fabian} A.~C.,  {Ross} R.~R., 2000, \mnras, 313, 193

\bibitem[\protect\citeauthoryear{{Nenkova} et~al.}{{Nenkova}
  et~al.}{2008}]{nenkova08}
{Nenkova} M., {Sirocky} M.~M., \v{Z}. {Ivezi{\'c}} ,  {Elitzur} M., 2008, \apj,
  685, 147

\bibitem[\protect\citeauthoryear{{Novikov} \& {Thorne}}{{Novikov} \&
  {Thorne}}{1973}]{novikov&thorne73}
{Novikov} I.~D.,  {Thorne} K.~S., 1973, in {C.~Dewitt \& B.~S.~Dewitt} , ed,
  Black Holes (Les Astres Occlus), p. 343

\bibitem[\protect\citeauthoryear{{Parisi} et~al.}{{Parisi}
  et~al.}{2009}]{parisi09}
{Parisi} P. et~al., 2009, A\&A, 507, 1345

\bibitem[\protect\citeauthoryear{{Peterson} et~al.}{{Peterson}
  et~al.}{2004}]{peterson04}
{Peterson} B.~M. et~al., 2004, \apj, 613, 682

\bibitem[\protect\citeauthoryear{{Pozzi} et~al.}{{Pozzi}
  et~al.}{2010}]{pozzi10}
{Pozzi} F. et~al., 2010, \aap, 517, A11

\bibitem[\protect\citeauthoryear{{Pozzi} et~al.}{{Pozzi}
  et~al.}{2007}]{pozzi07}
{Pozzi} F. et~al., 2007, \aap, 468, 603

\bibitem[\protect\citeauthoryear{{Raimundo} \& {Fabian}}{{Raimundo} \&
  {Fabian}}{2009}]{raimundo&fabian09}
{Raimundo} S.~I.,  {Fabian} A.~C., 2009, \mnras, 396, 1217

\bibitem[\protect\citeauthoryear{{Reynolds} \& {Fabian}}{{Reynolds} \&
  {Fabian}}{2008}]{reynolds&fabian08}
{Reynolds} C.~S.,  {Fabian} A.~C., 2008, \apj, 675, 1048

\bibitem[\protect\citeauthoryear{{Ross}, {Fabian}, \& {Mineshige}}{{Ross}
  et~al.}{1992}]{ross92}
{Ross} R.~R., {Fabian} A.~C.,  {Mineshige} S., 1992, \mnras, 258, 189

\bibitem[\protect\citeauthoryear{{Salpeter}}{{Salpeter}}{1964}]{salpeter64}
{Salpeter} E.~E., 1964, \apj, 140, 796

\bibitem[\protect\citeauthoryear{{Salucci} et~al.}{{Salucci}
  et~al.}{1999}]{salucci99}
{Salucci} P., {Szuszkiewicz} E., {Monaco} P.,  {Danese} L., 1999, \mnras, 307,
  637

\bibitem[\protect\citeauthoryear{{Schmidt} \& {Green}}{{Schmidt} \&
  {Green}}{1983}]{schmidt&green83}
{Schmidt} M.,  {Green} R.~F., 1983, \apj, 269, 352

\bibitem[\protect\citeauthoryear{{Shakura} \& {Sunyaev}}{{Shakura} \&
  {Sunyaev}}{1973}]{shakura&sunyaev73}
{Shakura} N.~I.,  {Sunyaev} R.~A., 1973, A\&A, 24, 337

\bibitem[\protect\citeauthoryear{{Shankar}, {Weinberg}, \&
  {Miralda-Escud{\'e}}}{{Shankar} et~al.}{2009}]{shankar09}
{Shankar} F., {Weinberg} D.~H.,  {Miralda-Escud{\'e}} J., 2009, \apj, 690, 20

\bibitem[\protect\citeauthoryear{{Shen} et~al.}{{Shen} et~al.}{2011}]{shen11}
{Shen} Y. et~al., 2011, \apjs, 194, 45

\bibitem[\protect\citeauthoryear{{Shields}}{{Shields}}{1978}]{shields78}
{Shields} G.~A., 1978, \nat, 272, 706

\bibitem[\protect\citeauthoryear{{Shimura} \& {Takahara}}{{Shimura} \&
  {Takahara}}{1995}]{shimura&takahara95}
{Shimura} T.,  {Takahara} F., 1995, \apj, 445, 780

\bibitem[\protect\citeauthoryear{{Small} \& {Blandford}}{{Small} \&
  {Blandford}}{1992}]{small&blandford92}
{Small} T.~A.,  {Blandford} R.~D., 1992, \mnras, 259, 725

\bibitem[\protect\citeauthoryear{{Soltan}}{{Soltan}}{1982}]{soltan82}
{Soltan} A., 1982, \mnras, 200, 115

\bibitem[\protect\citeauthoryear{{Uttley} et~al.}{{Uttley}
  et~al.}{2003}]{uttley03}
{Uttley} P., {Edelson} R., {McHardy} I.~M., {Peterson} B.~M.,  {Markowitz} A.,
  2003, \apjl, 584, L53

\bibitem[\protect\citeauthoryear{{Vasudevan} et~al.}{{Vasudevan}
  et~al.}{2010}]{vasudevan10}
{Vasudevan} R.~V., {Fabian} A.~C., {Gandhi} P., {Winter} L.~M.,  {Mushotzky}
  R.~F., 2010, \mnras, 402, 1081

\bibitem[\protect\citeauthoryear{{Vasudevan} et~al.}{{Vasudevan}
  et~al.}{2009}]{vasudevan09}
{Vasudevan} R.~V., {Mushotzky} R.~F., {Winter} L.~M.,  {Fabian} A.~C., 2009,
  \mnras, 399, 1553

\bibitem[\protect\citeauthoryear{{Volonteri} et~al.}{{Volonteri}
  et~al.}{2005}]{volonteri05}
{Volonteri} M., {Madau} P., {Quataert} E.,  {Rees} M.~J., 2005, \apj, 620, 69

\bibitem[\protect\citeauthoryear{{Winter} et~al.}{{Winter}
  et~al.}{2010}]{winter10}
{Winter} L.~M., {Lewis} K.~T., {Koss} M., {Veilleux} S., {Keeney} B.,
  {Mushotzky} R.~F., 2010, \apj, 710, 503

\bibitem[\protect\citeauthoryear{{Yuan} et~al.}{{Yuan} et~al.}{2010}]{yuan10}
{Yuan} W., {Liu} B.~F., {Zhou} H.,  {Wang} T.~G., 2010, \apj, 723, 508

\end{thebibliography}

\begin{appendix}
\appendix
\section{Spectral energy distributions}


\begin{figure*}
\centering
\begin{tabular}{cccc}
\epsfig{file=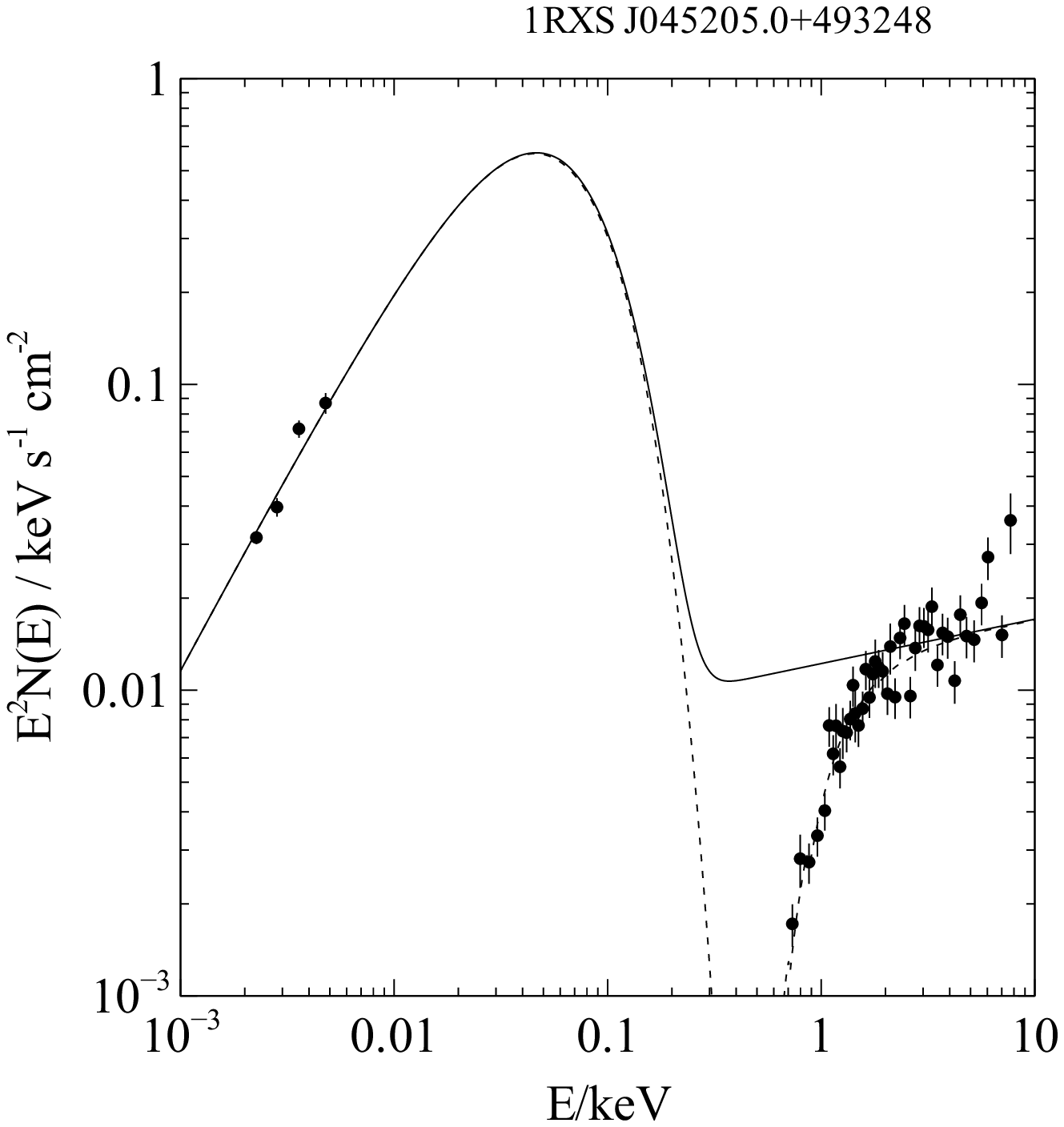,width=0.5\columnwidth,clip=} 
\epsfig{file=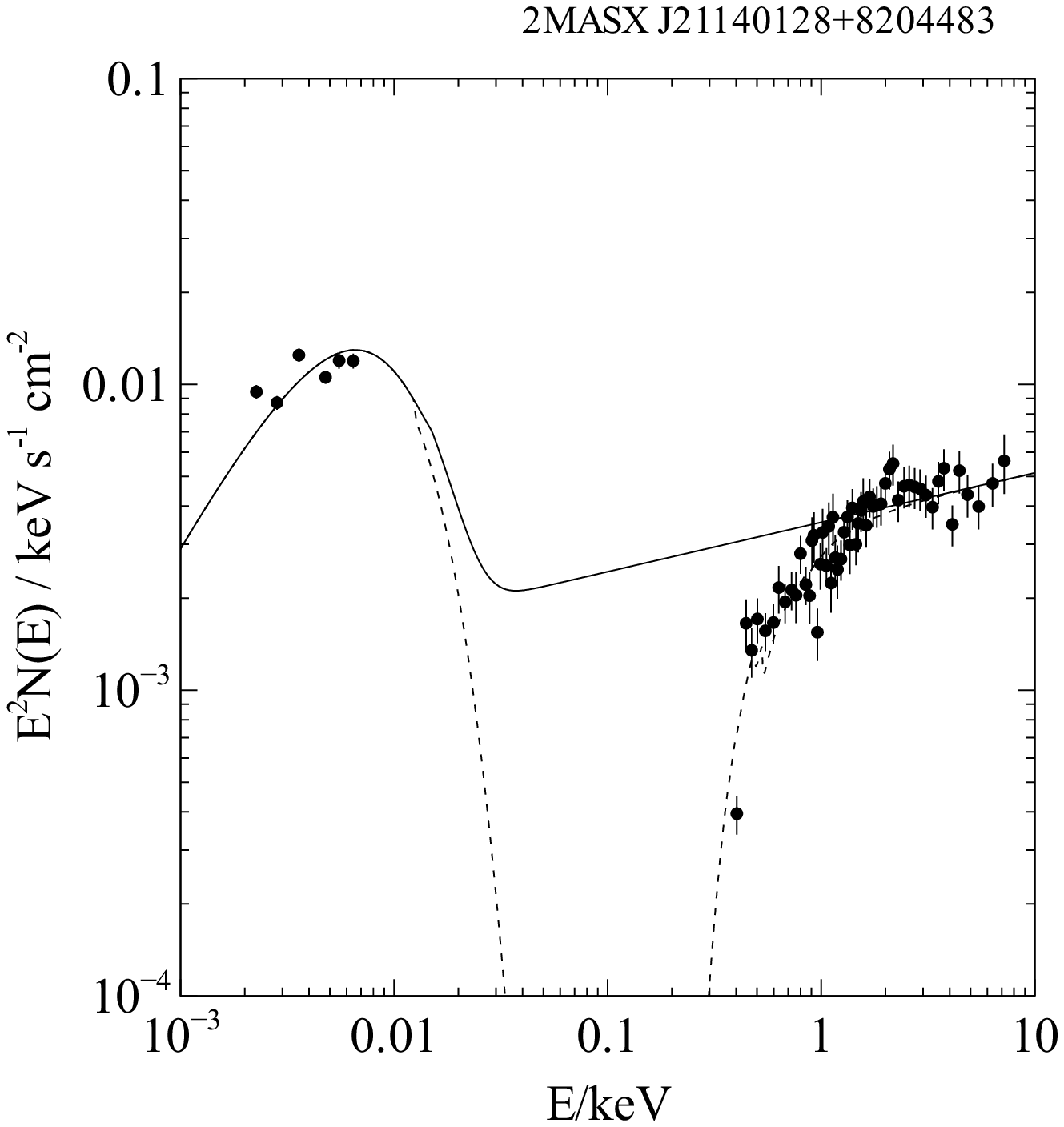,width=0.5\columnwidth,clip=} 
\epsfig{file=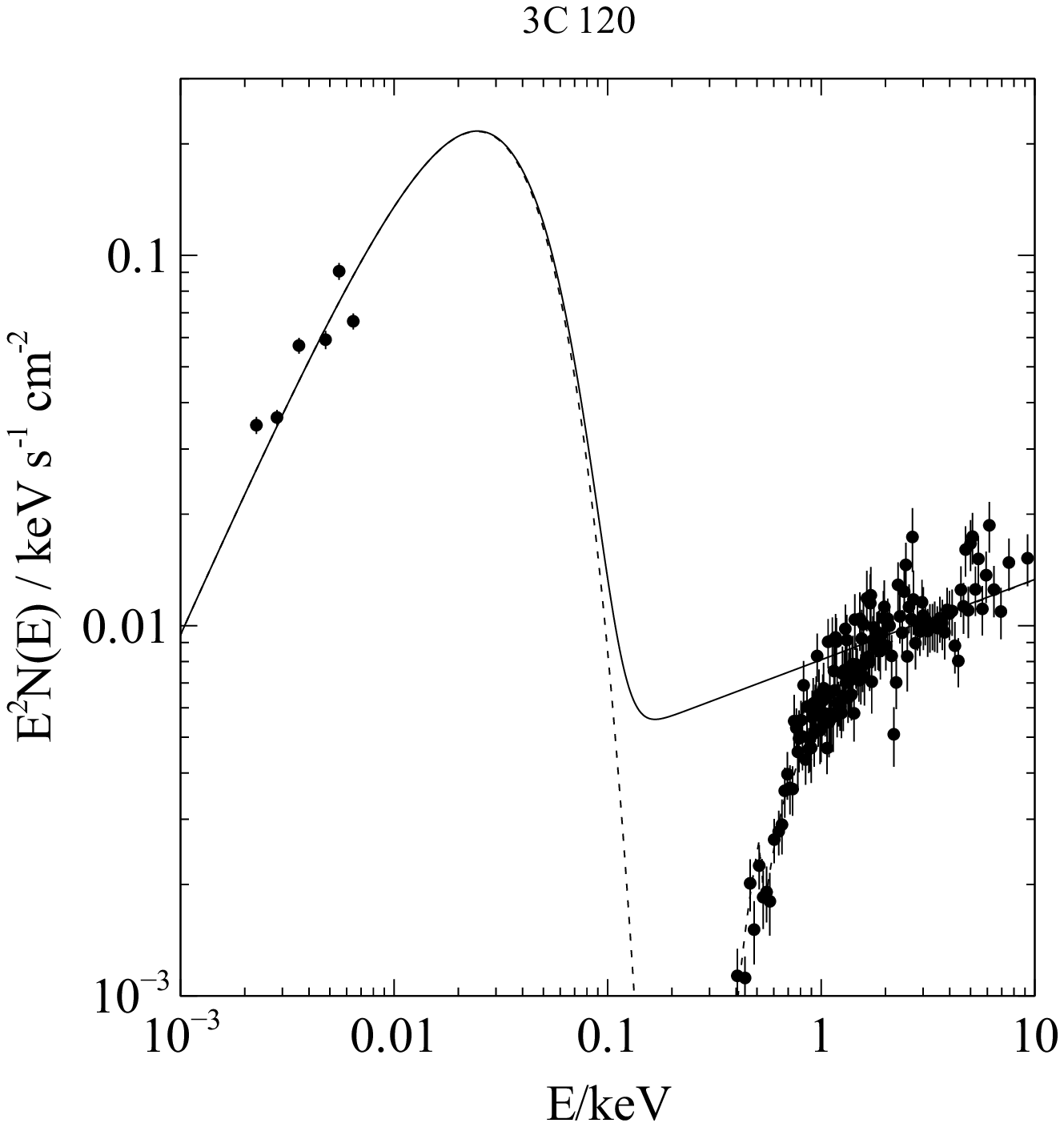,width=0.5\columnwidth,clip=}
\epsfig{file=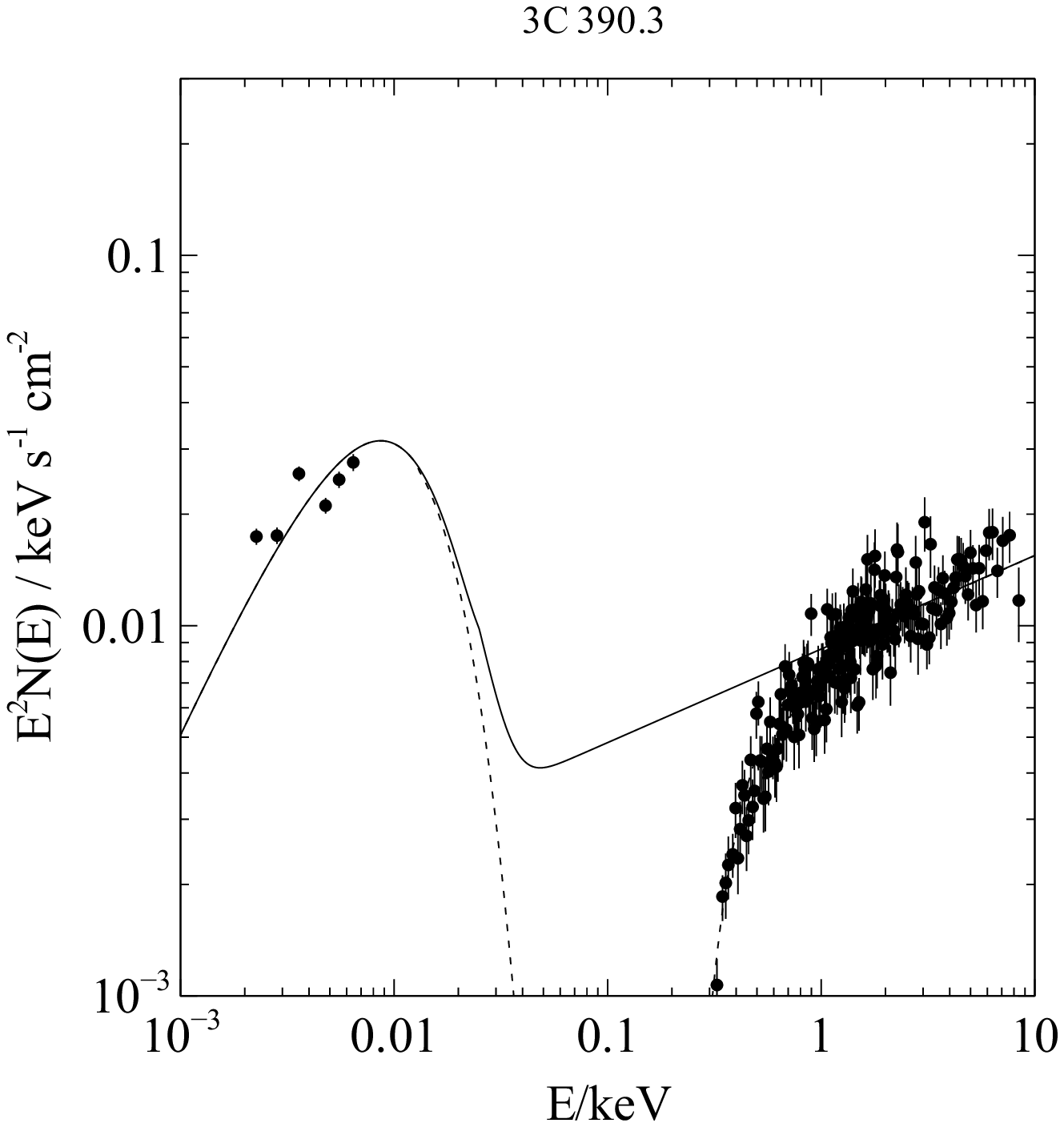,width=0.5\columnwidth,clip=}\\ 
\epsfig{file=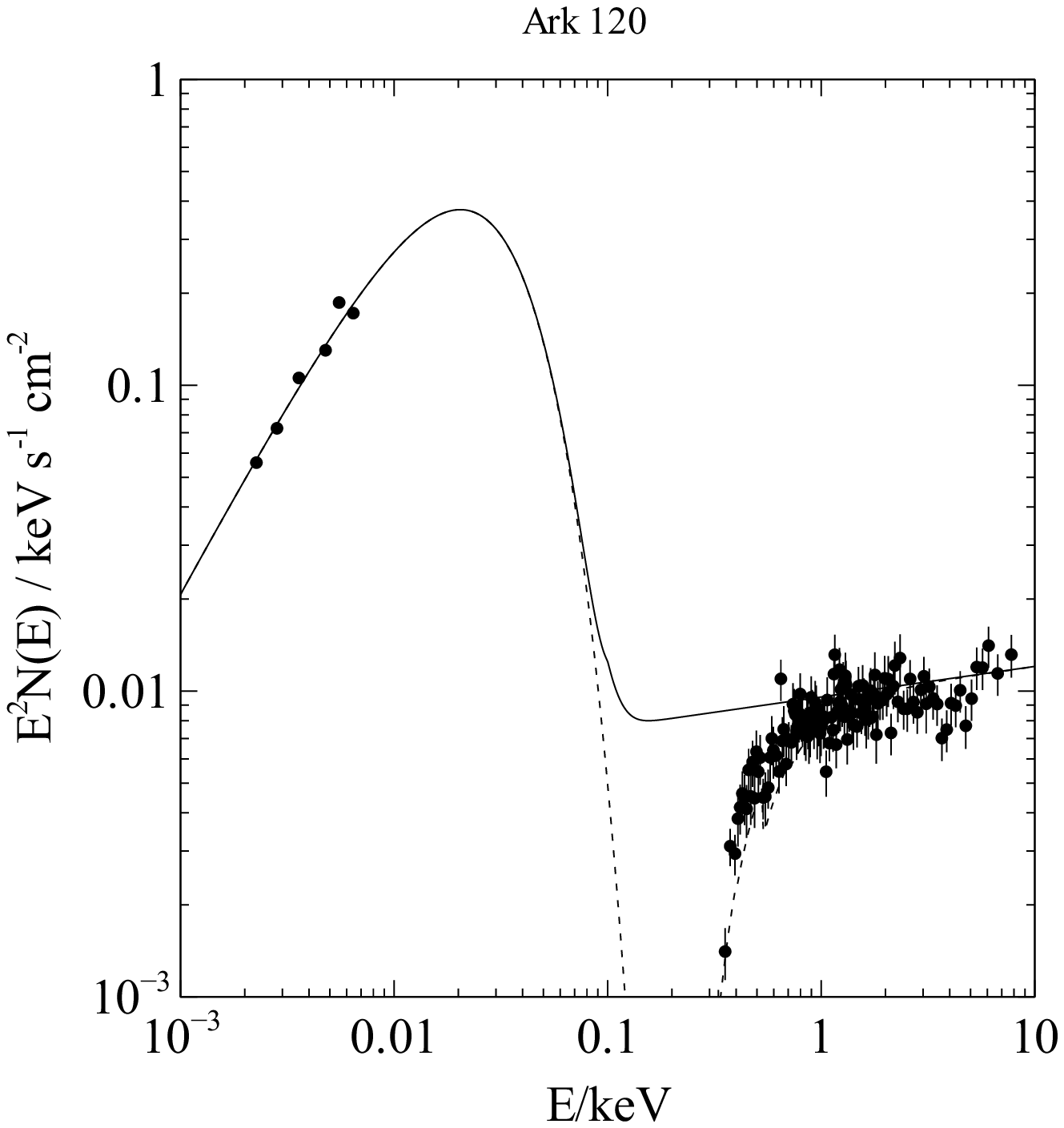,width=0.5\columnwidth,clip=}
\epsfig{file=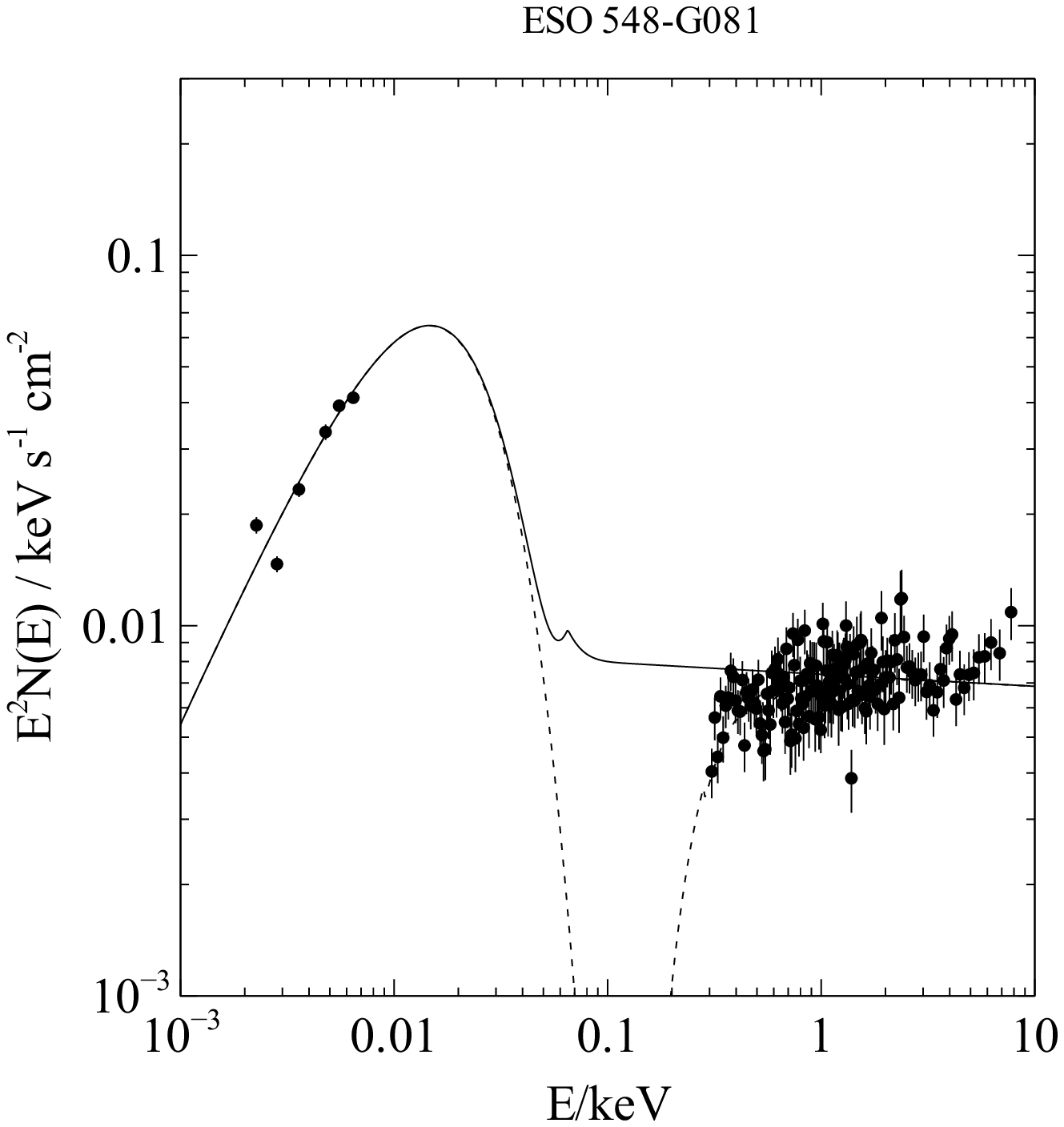,width=0.5\columnwidth,clip=} 
\epsfig{file=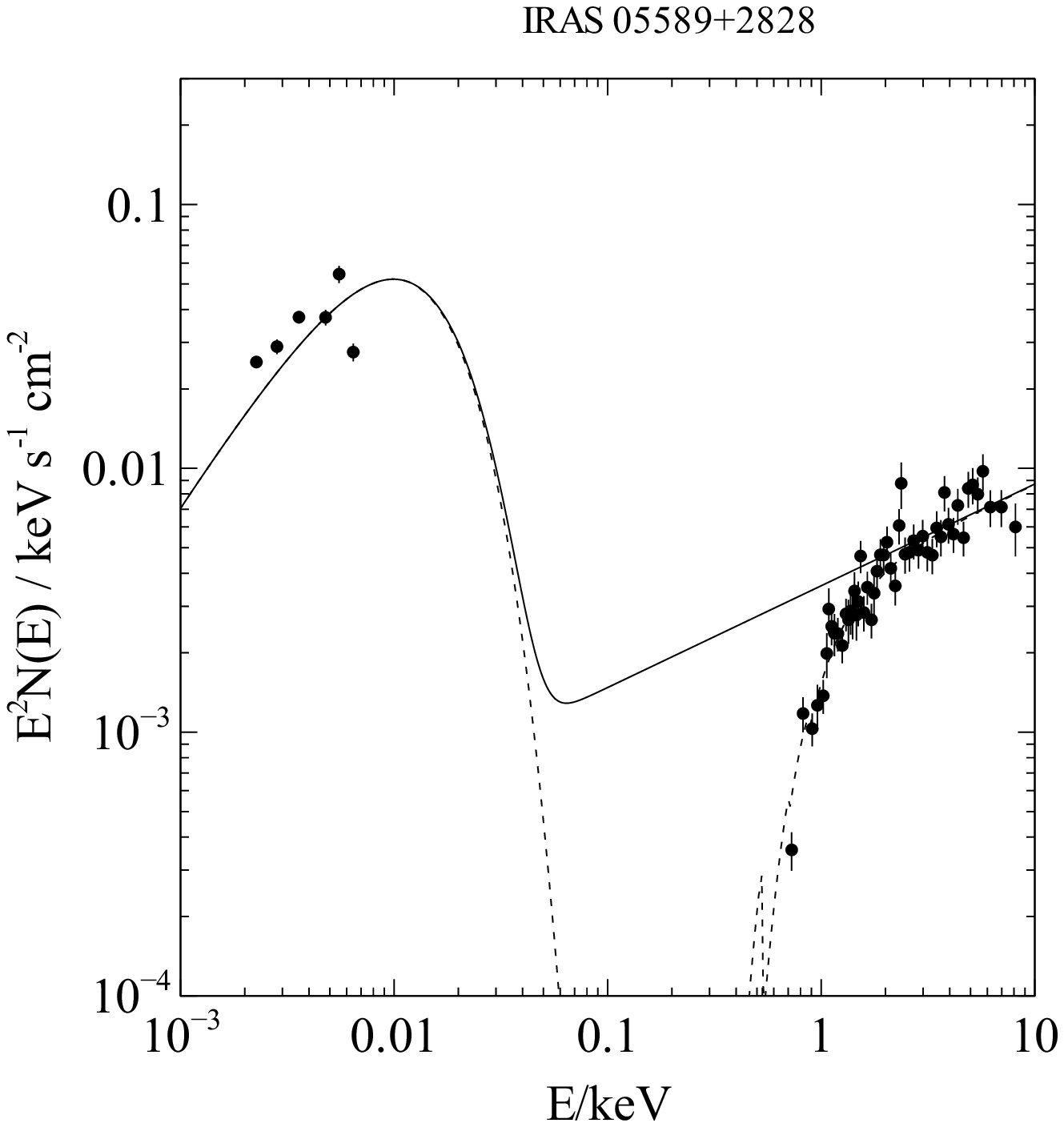,width=0.5\columnwidth,clip=}
\epsfig{file=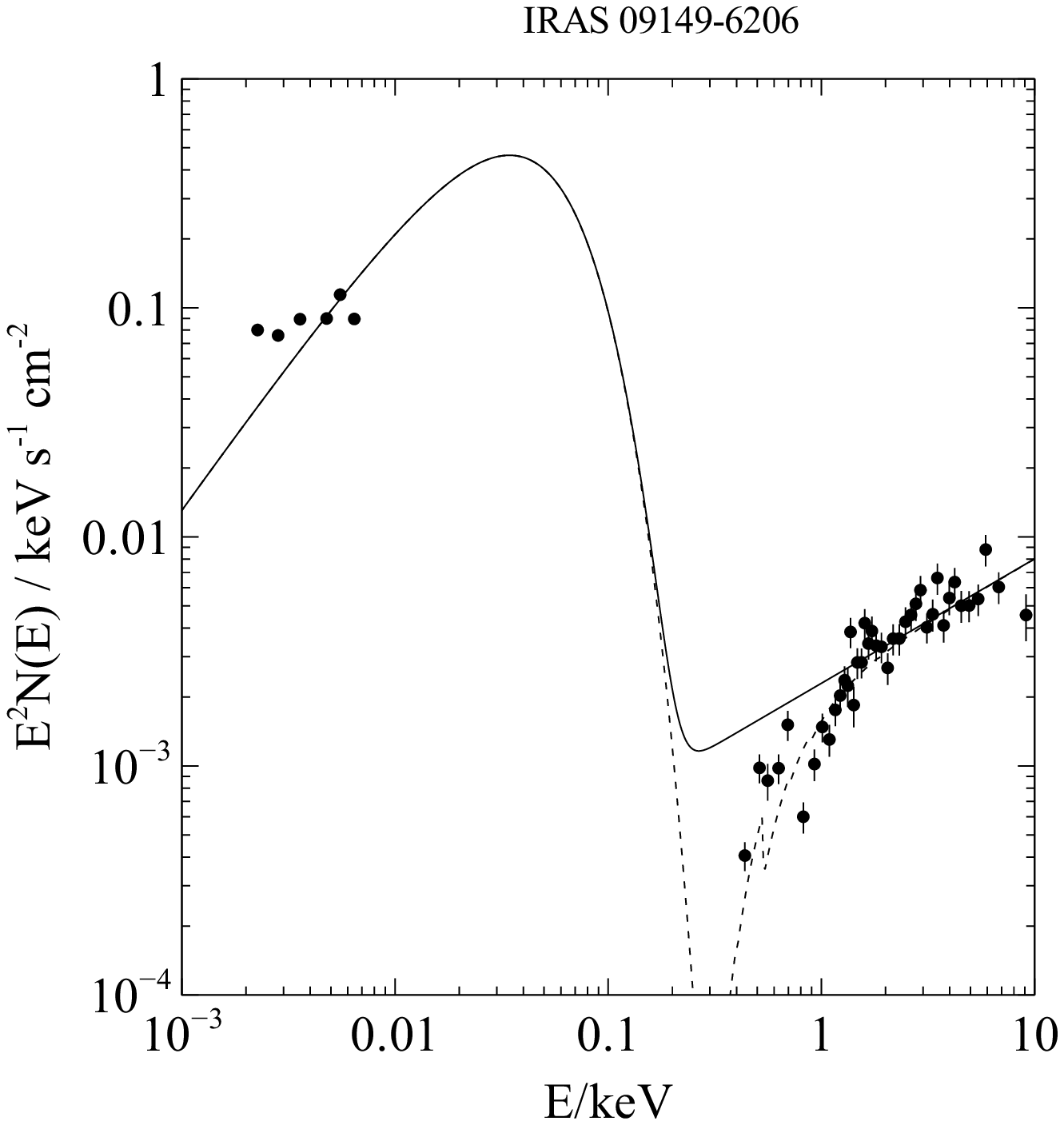,width=0.5\columnwidth,clip=} \\
\epsfig{file=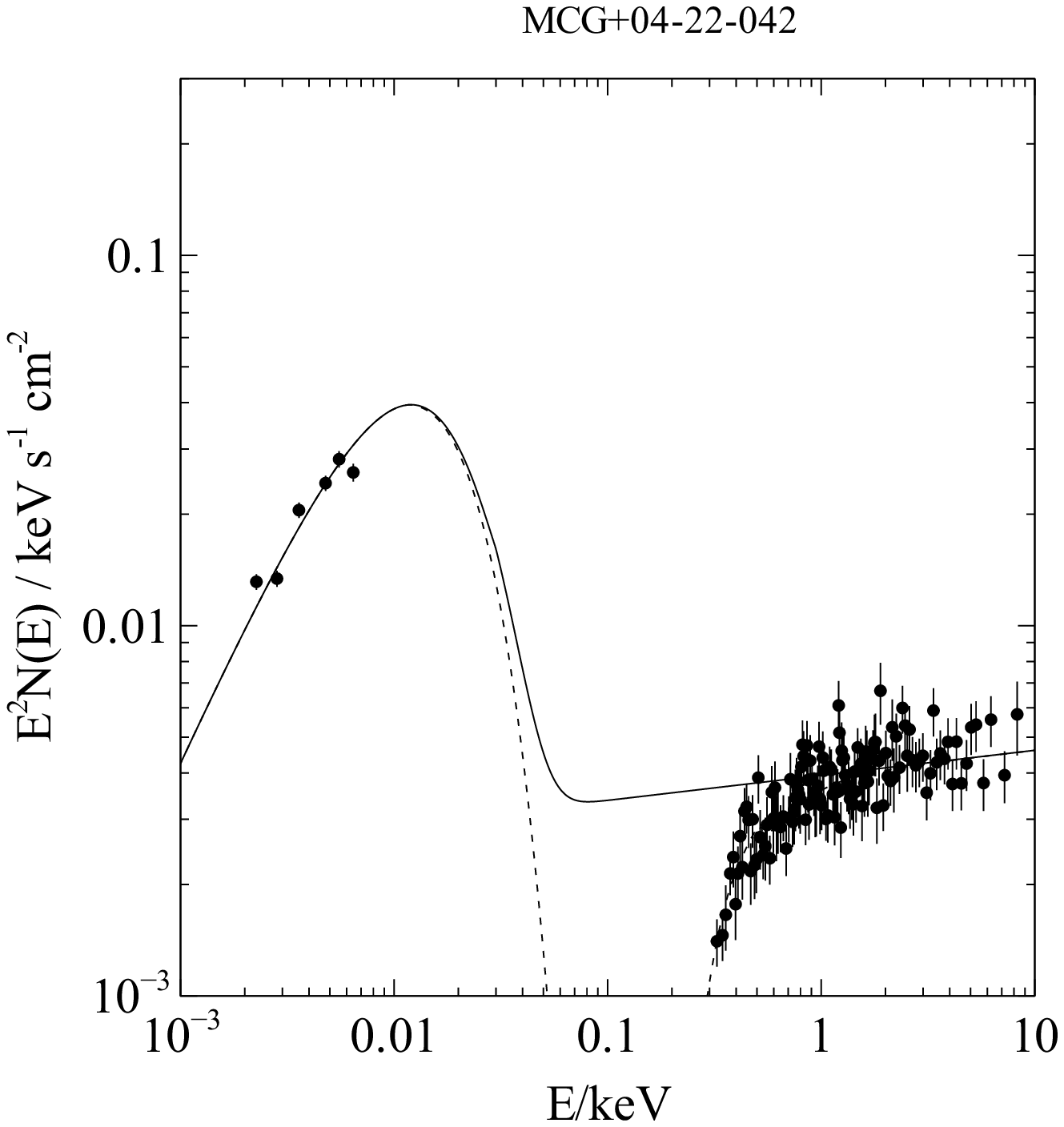,width=0.5\columnwidth,clip=} 
\epsfig{file=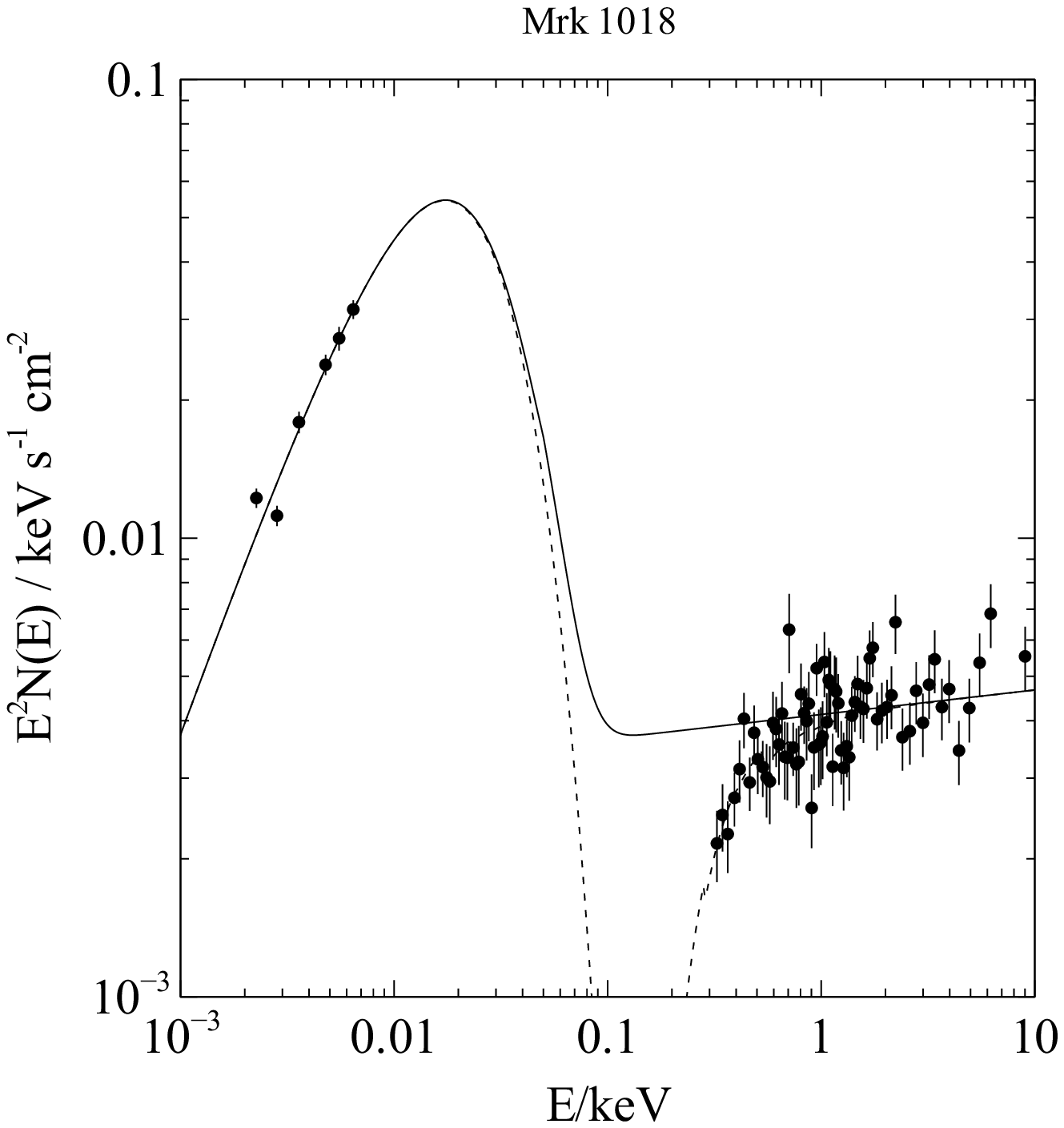,width=0.5\columnwidth,clip=} 
\epsfig{file=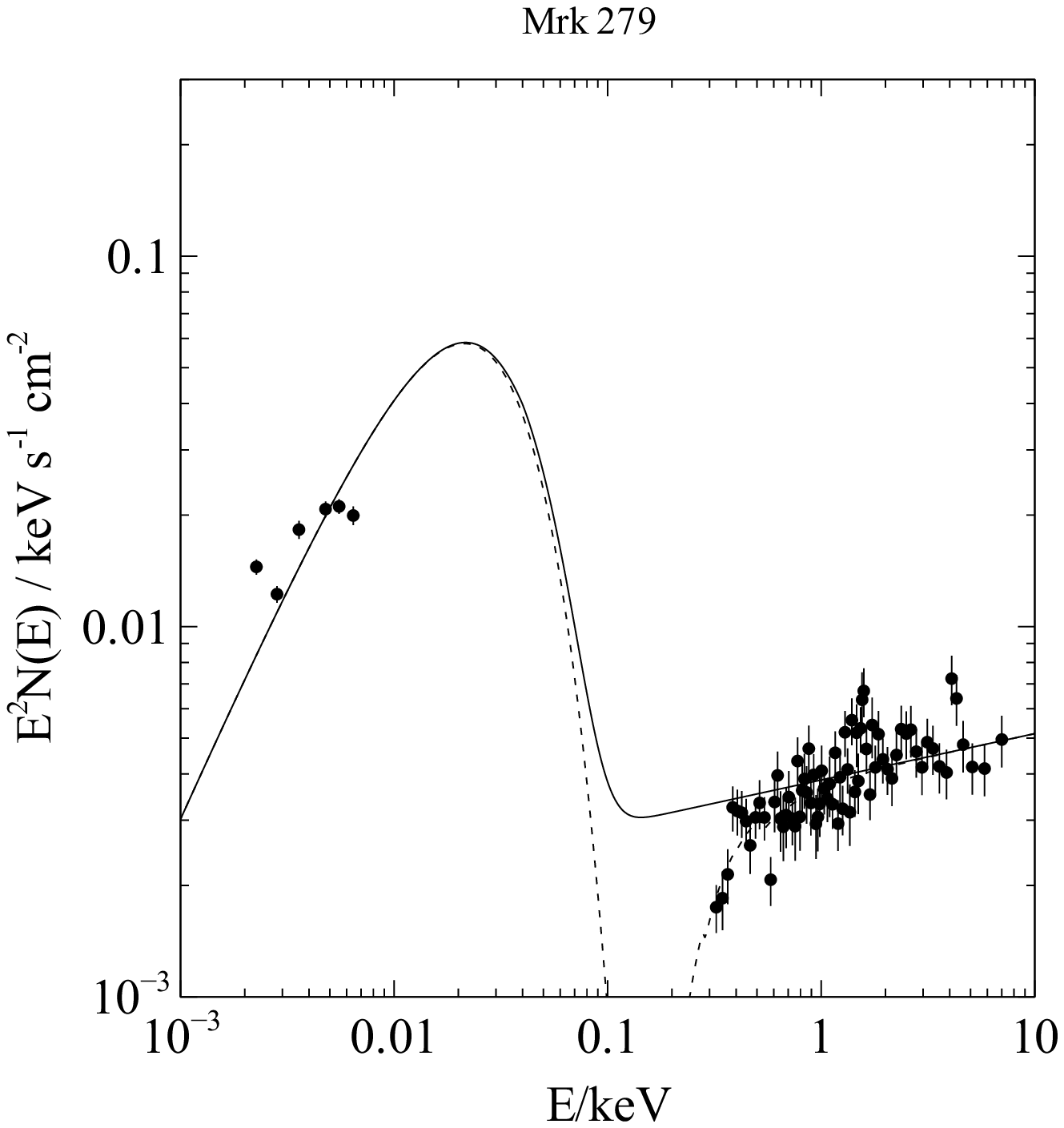,width=0.5\columnwidth,clip=} 
\epsfig{file=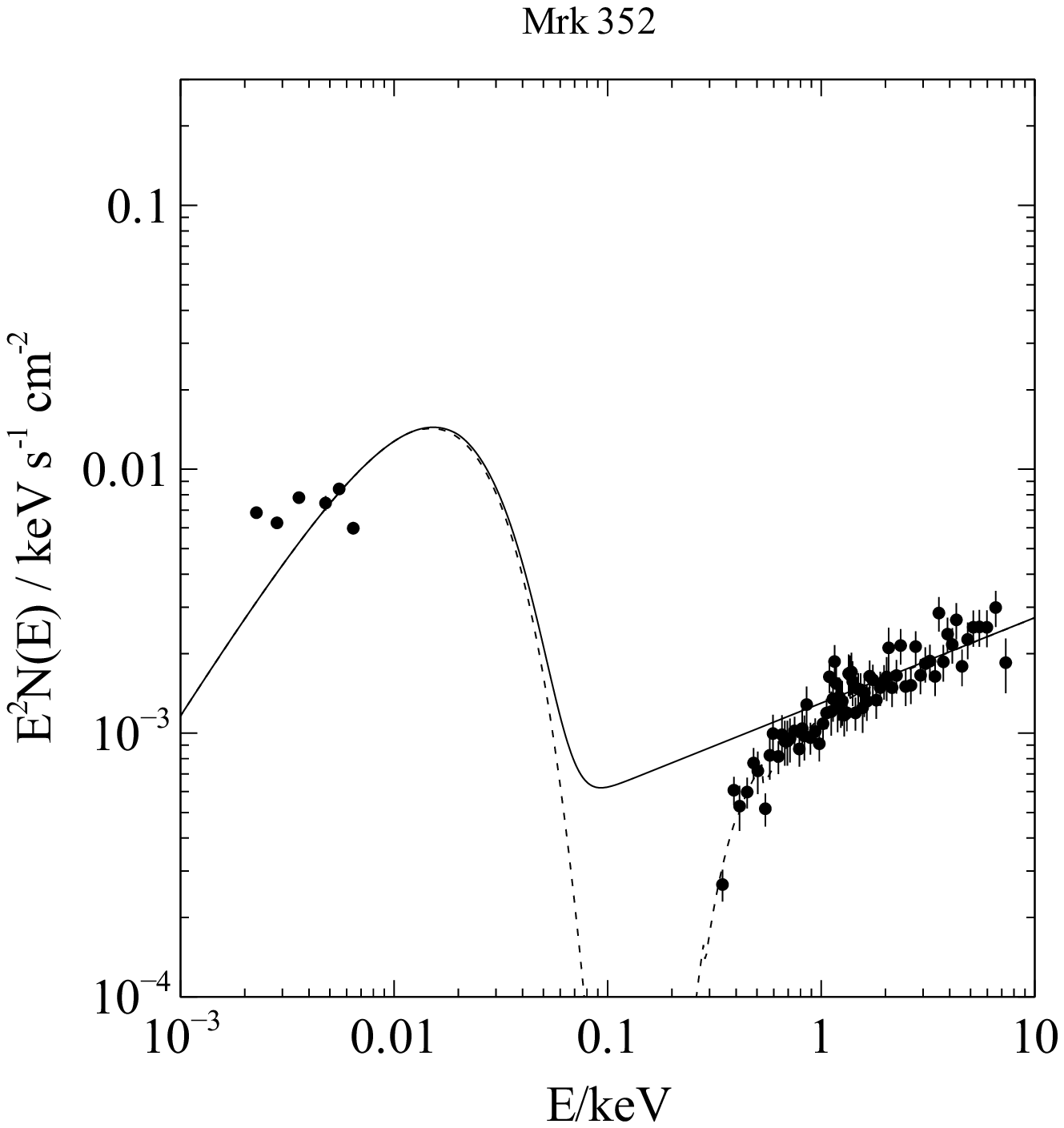,width=0.5\columnwidth,clip=} \\
\epsfig{file=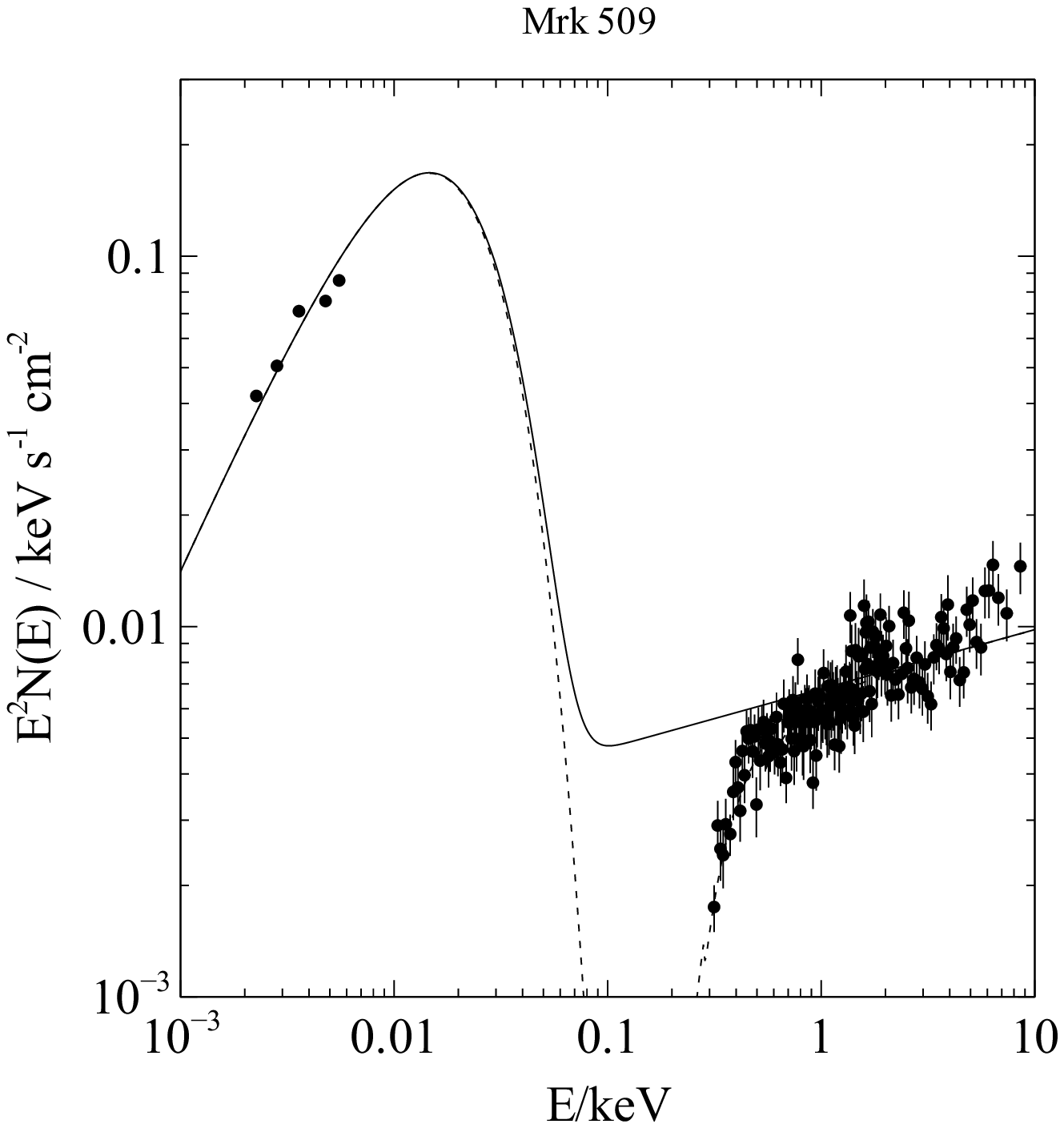,width=0.5\columnwidth,clip=} 
\epsfig{file=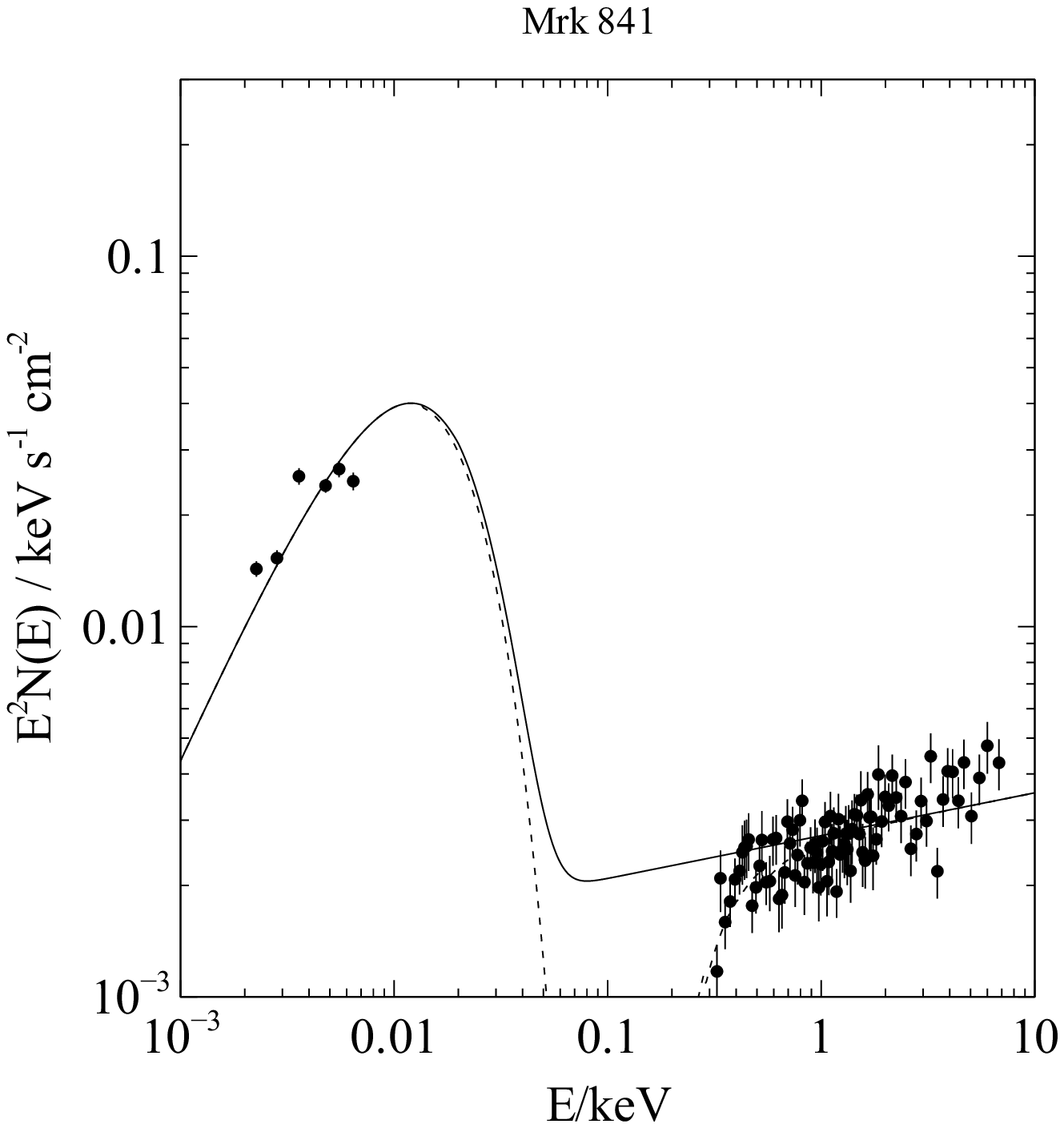,width=0.5\columnwidth,clip=} 
\epsfig{file=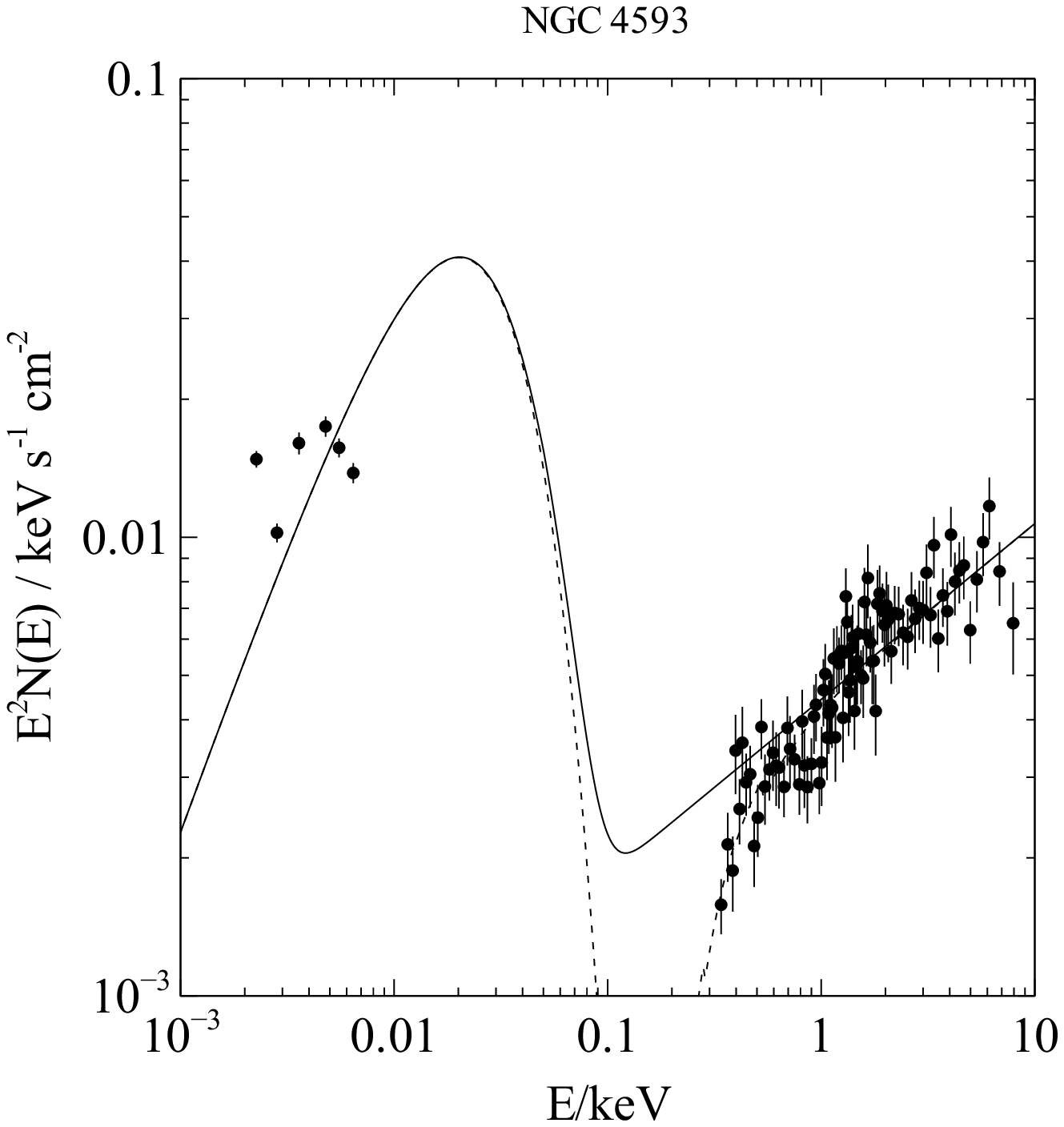,width=0.5\columnwidth,clip=} 
\epsfig{file=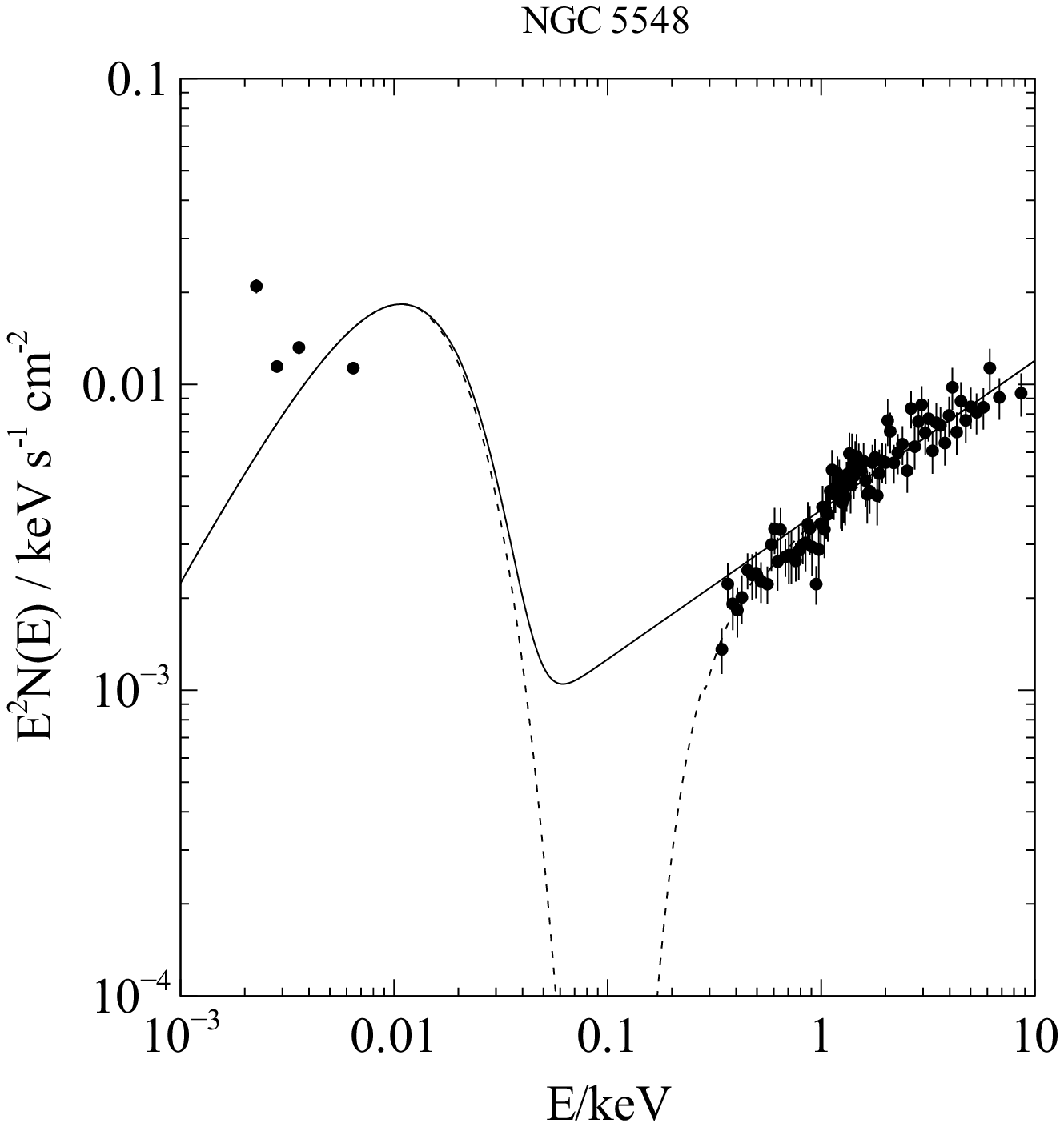,width=0.5\columnwidth,clip=}
\end{tabular}
\begin{center}
\emph{Fig.~\ref{sed_list} (continued on next page)}
\end{center}
\end{figure*}
\begin{figure*}
\centering
\begin{center}
\emph{Fig.~\ref{sed_list} (continued)}\\
\end{center}
\begin{tabular}{cccc}
\epsfig{file=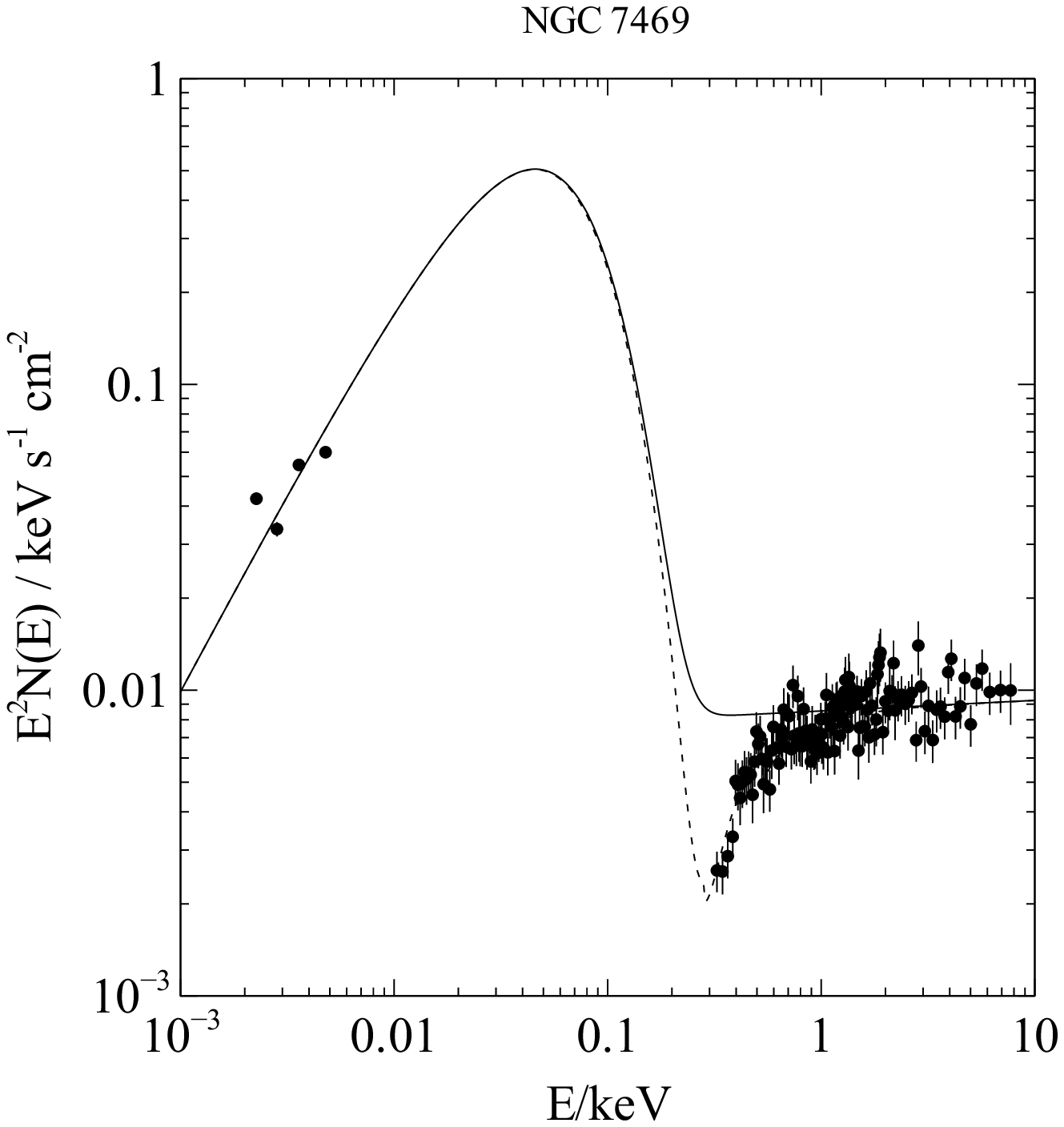,width=0.5\columnwidth,clip=} 
\epsfig{file=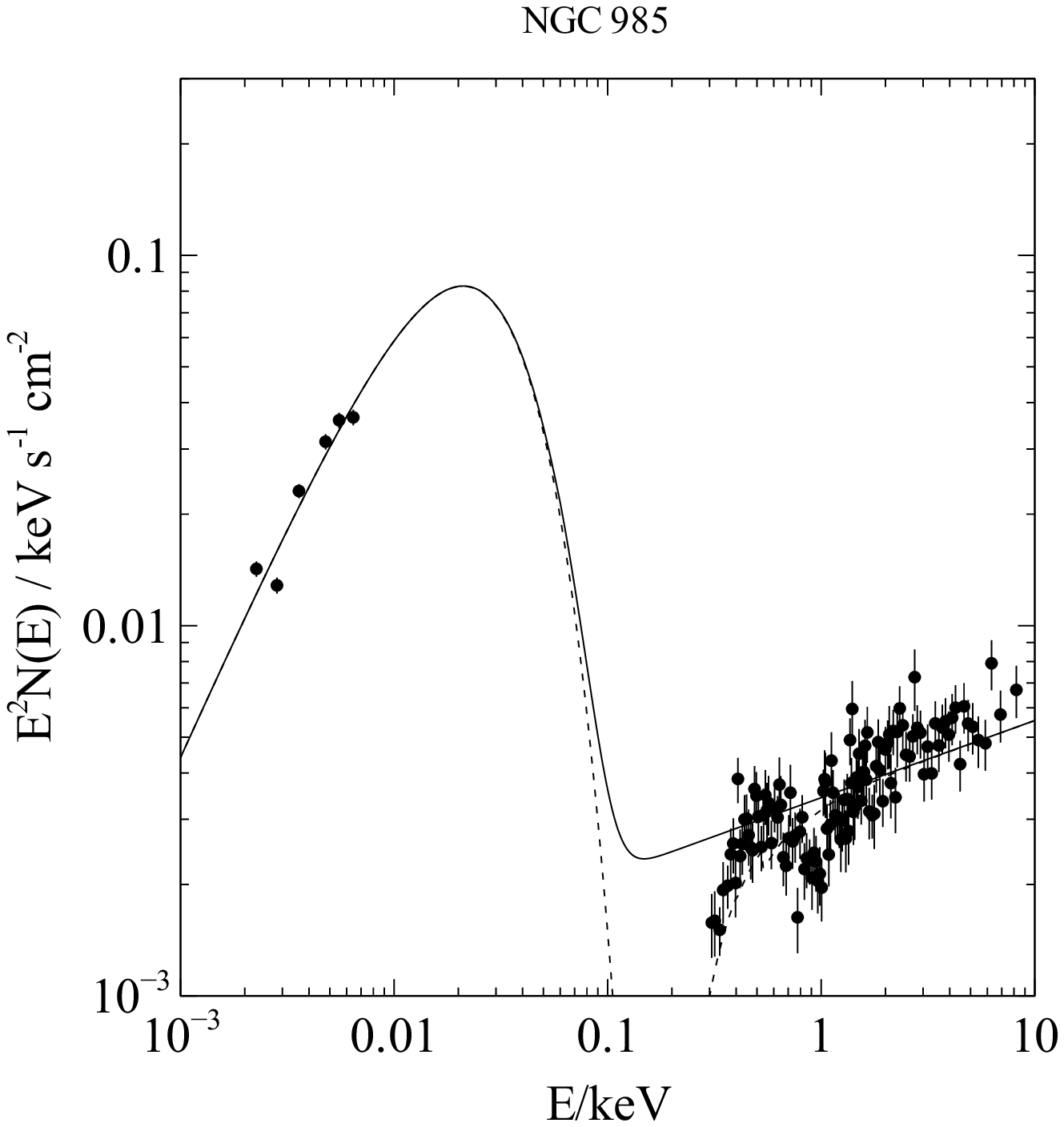,width=0.5\columnwidth,clip=} 
\epsfig{file=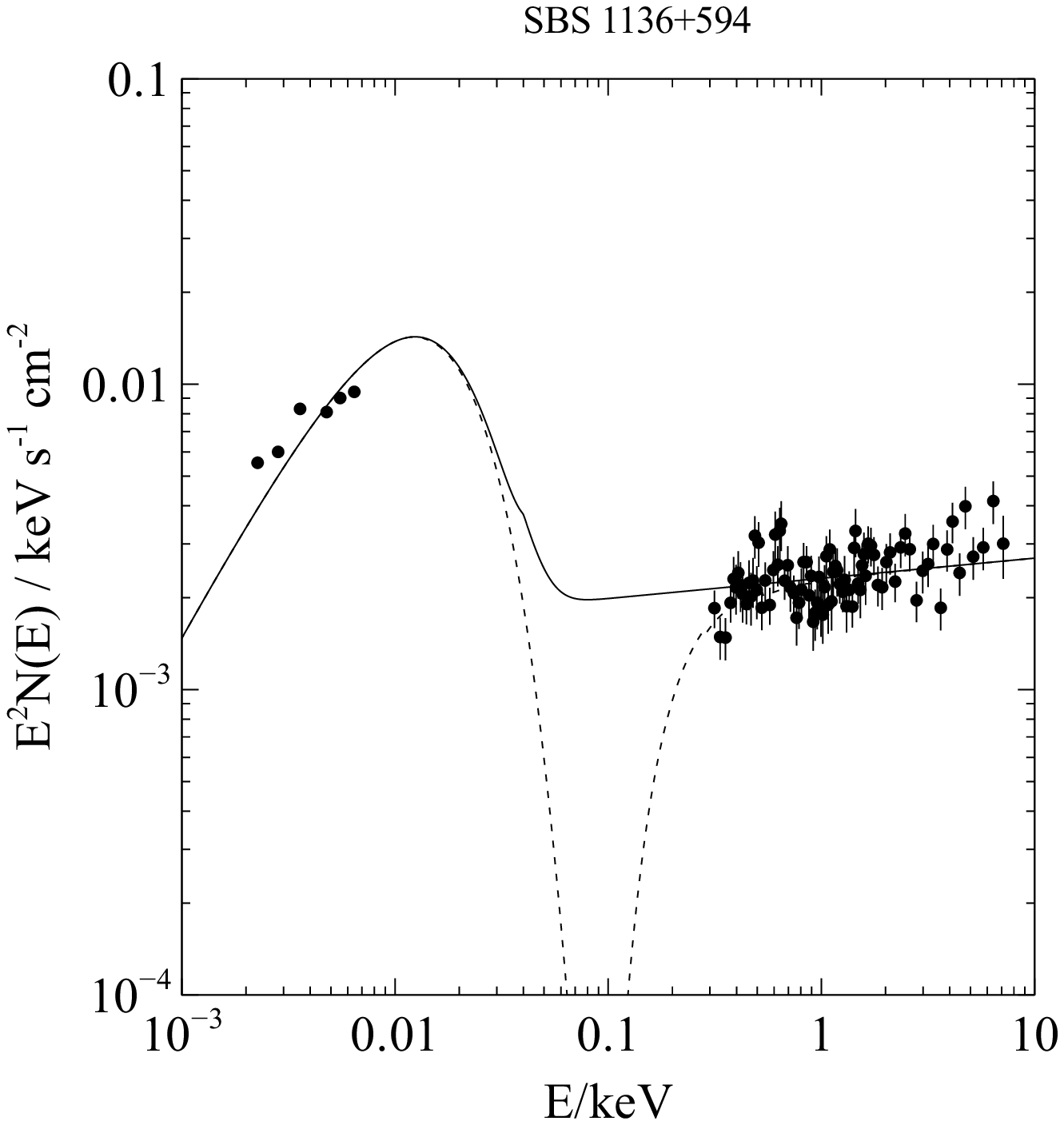,width=0.5\columnwidth,clip=} 
\epsfig{file=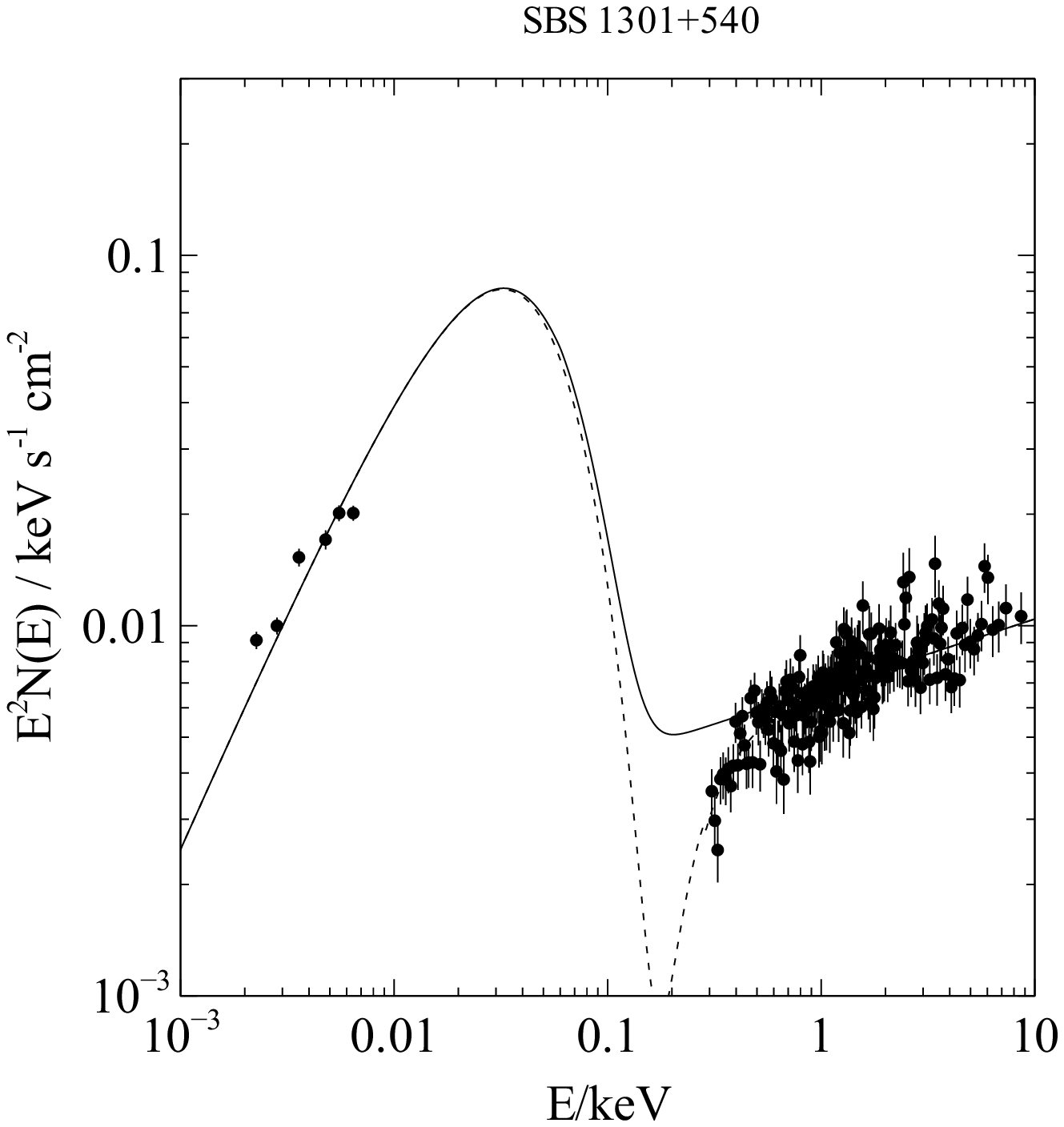,width=0.5\columnwidth,clip=} \\
\epsfig{file=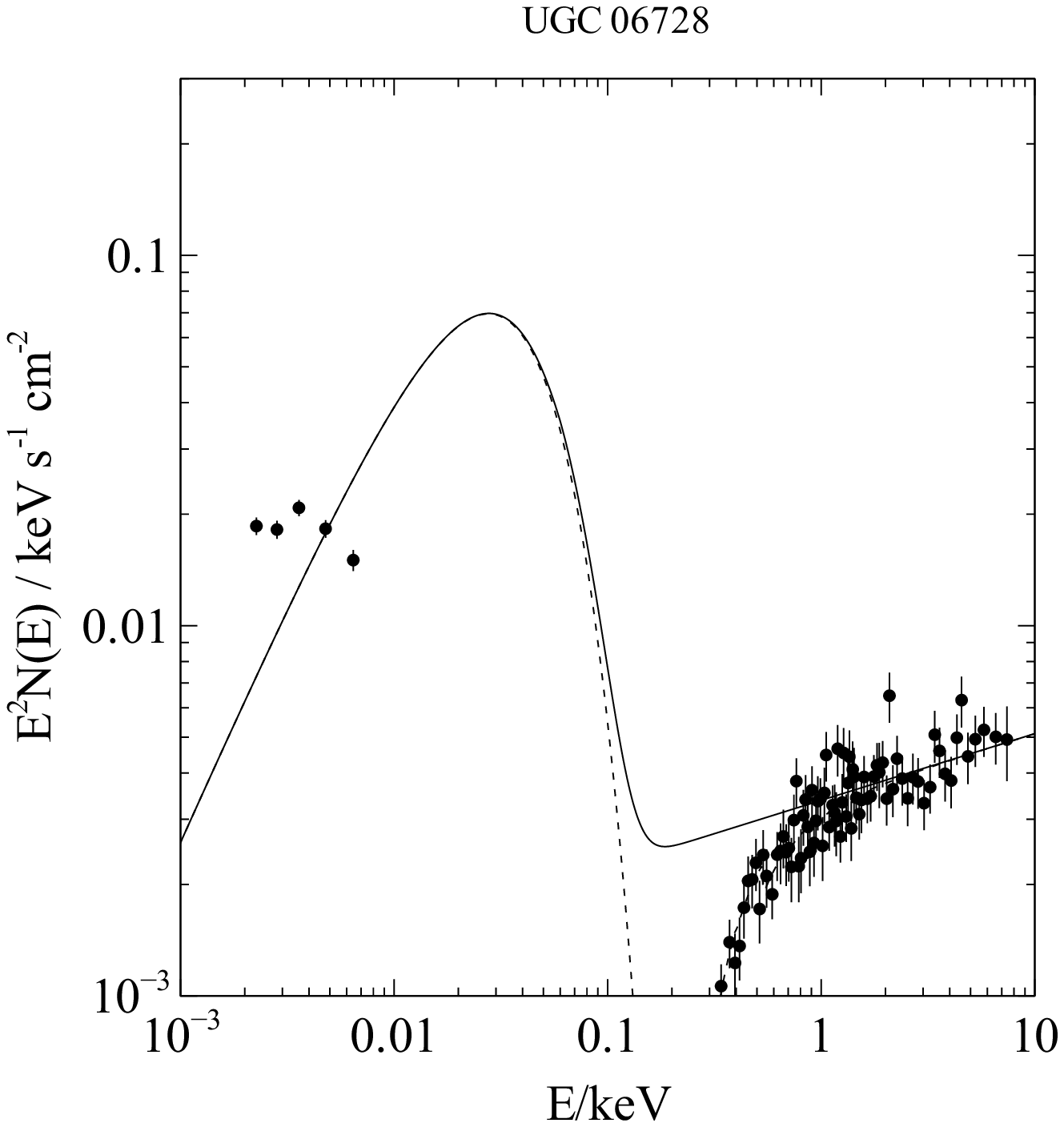,width=0.5\columnwidth,clip=} 
\epsfig{file=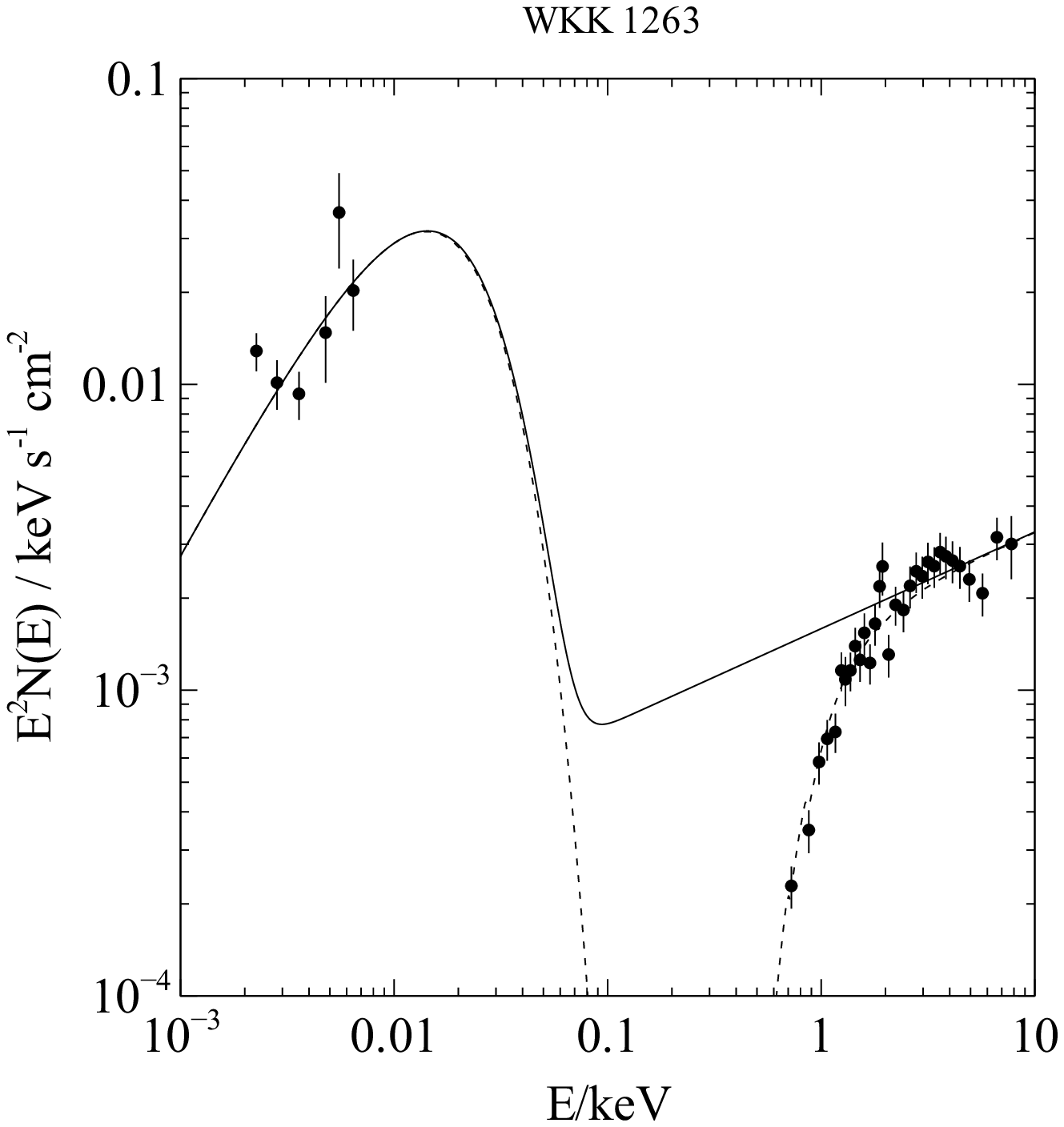,width=0.5\columnwidth,clip=} 
\epsfig{file=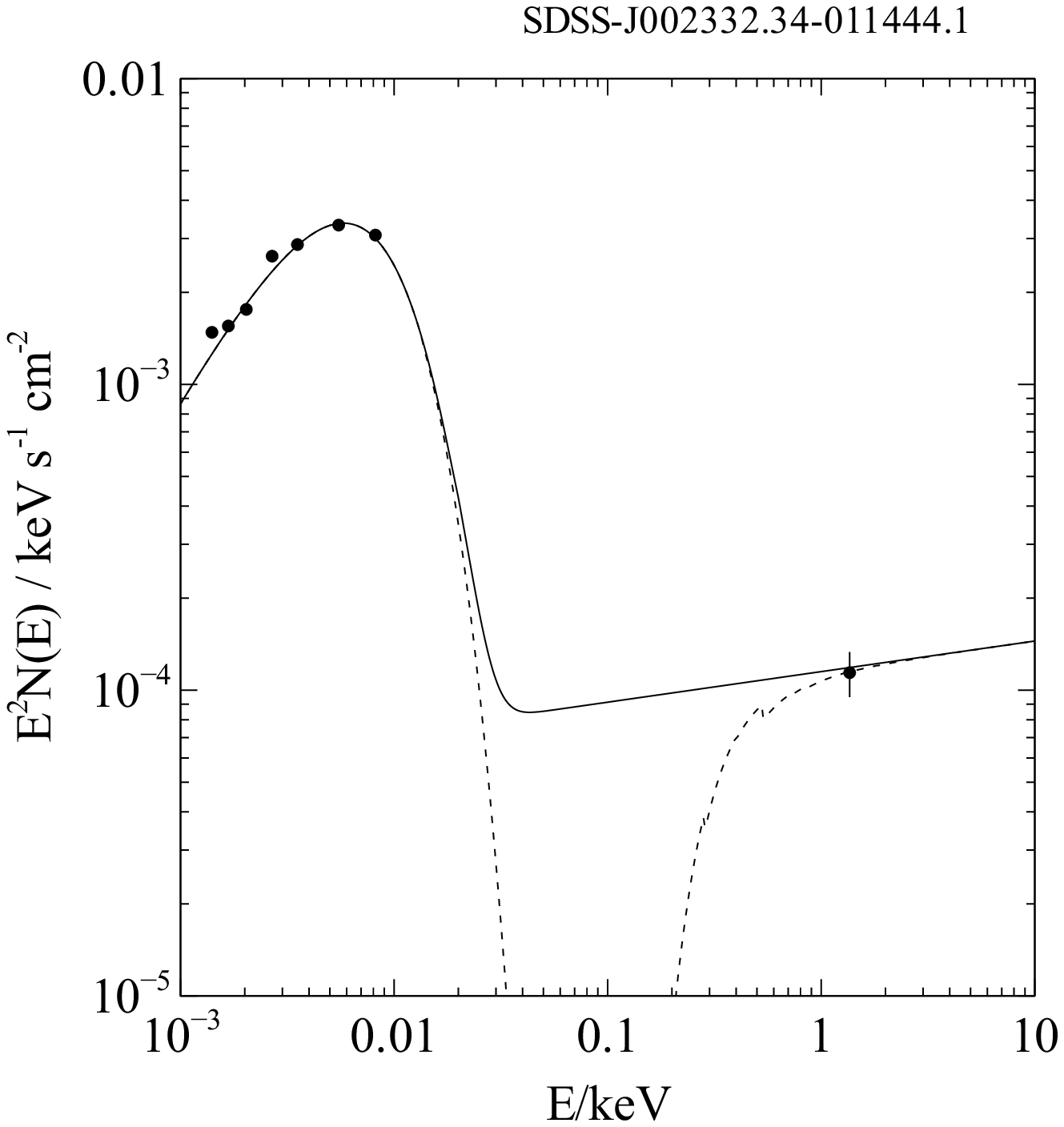,width=0.5\columnwidth,clip=} 
\epsfig{file=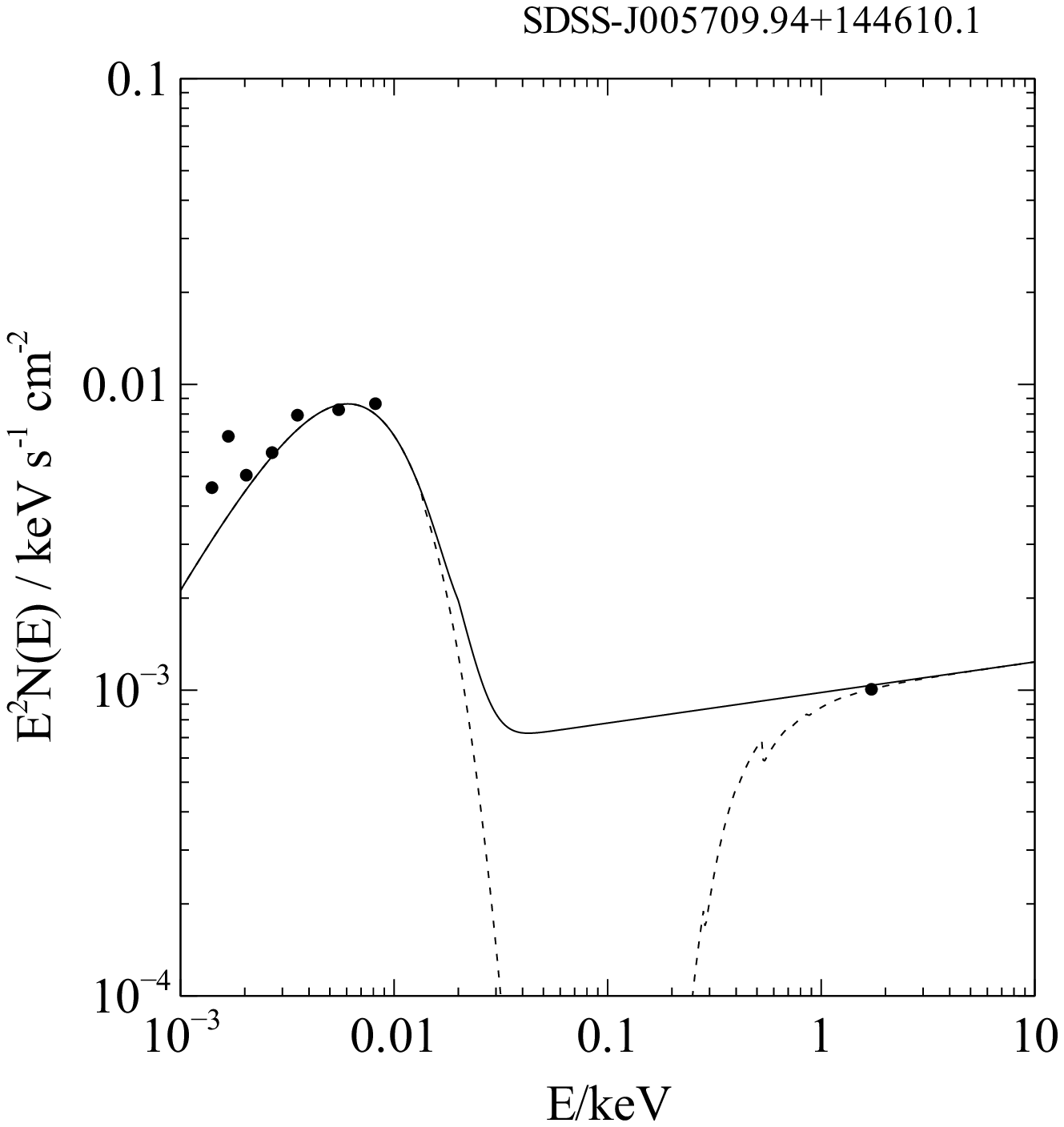,width=0.5\columnwidth,clip=} \\
\epsfig{file=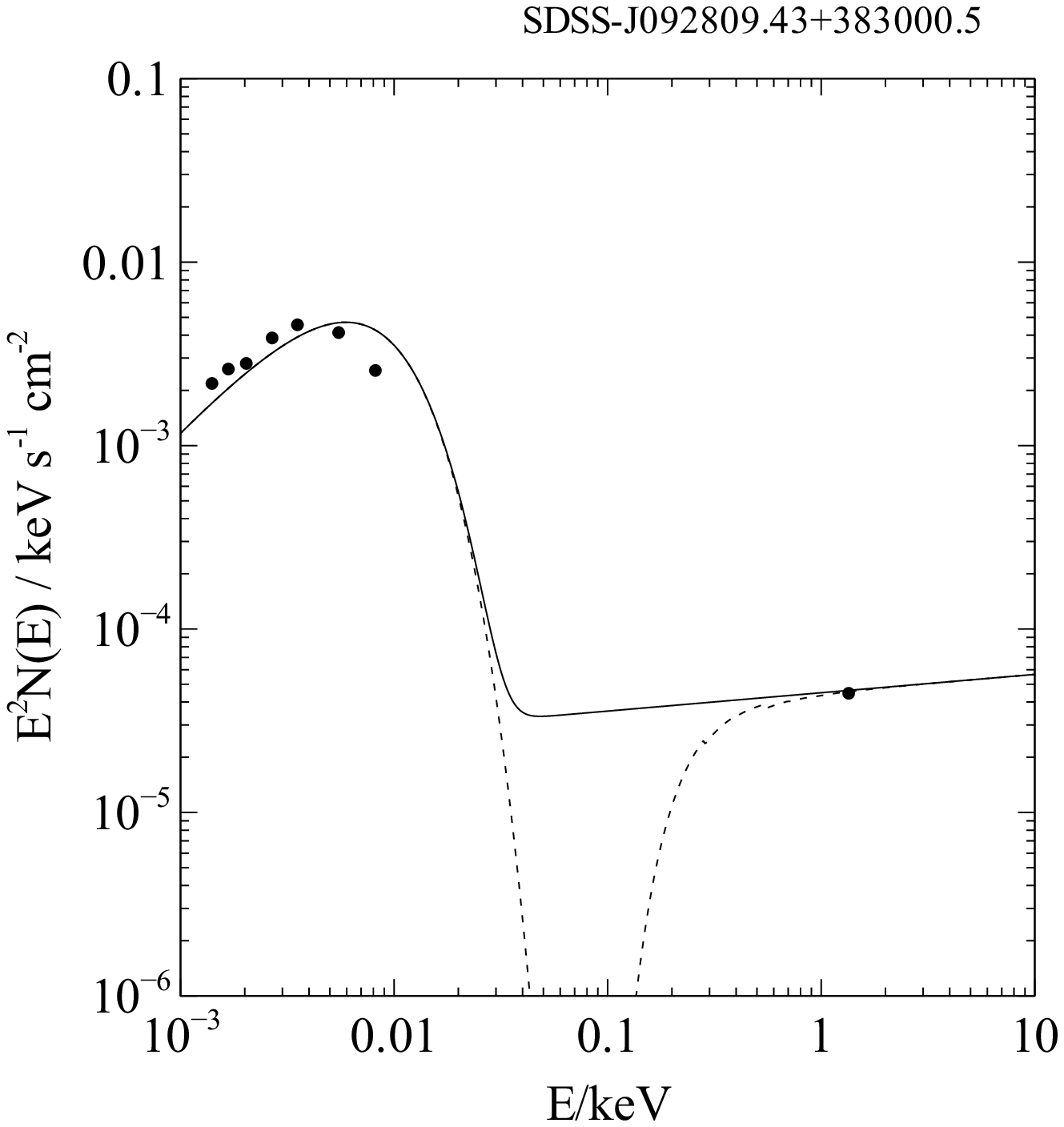,width=0.5\columnwidth,clip=} 
\epsfig{file=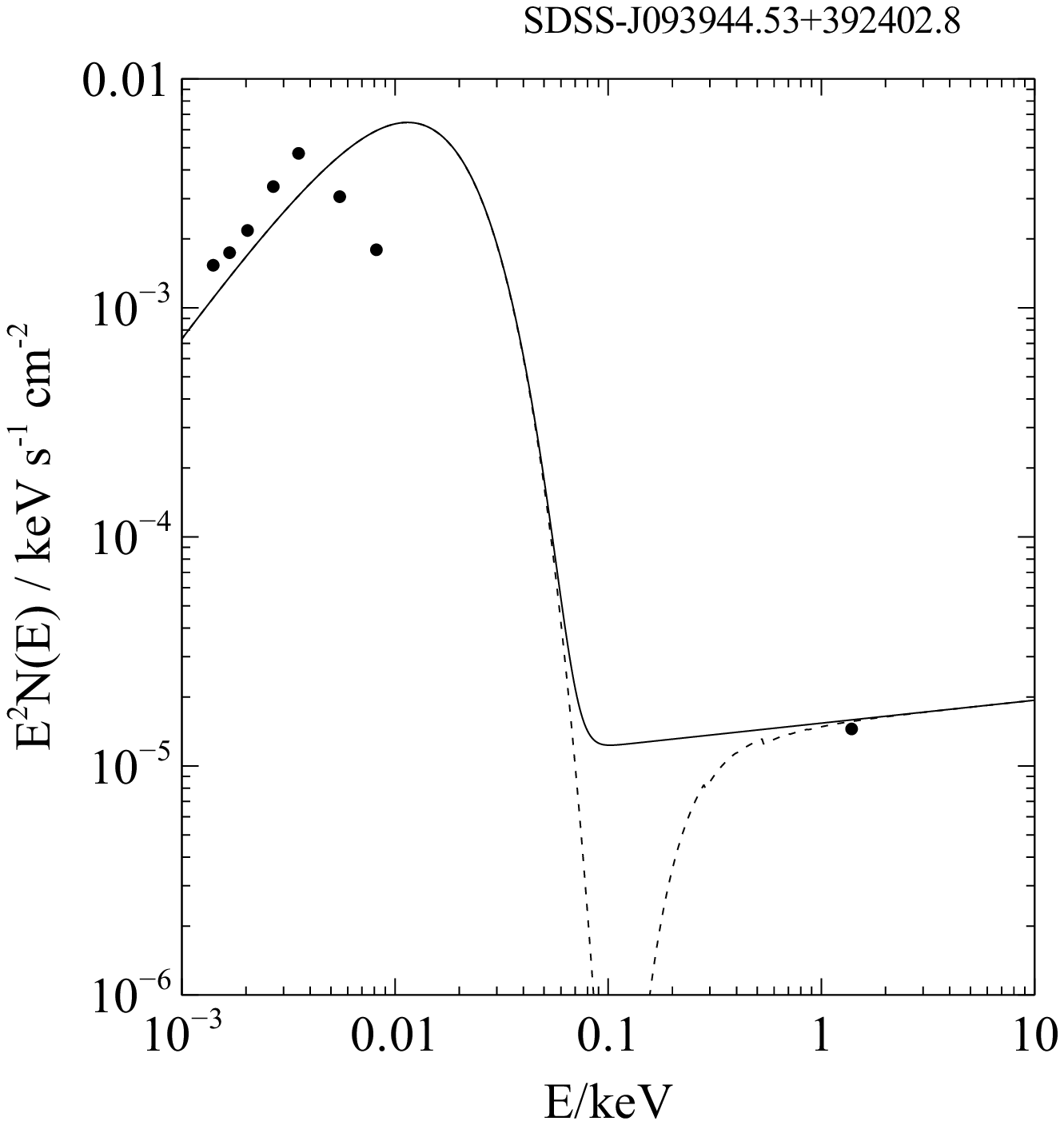,width=0.5\columnwidth,clip=} 
\epsfig{file=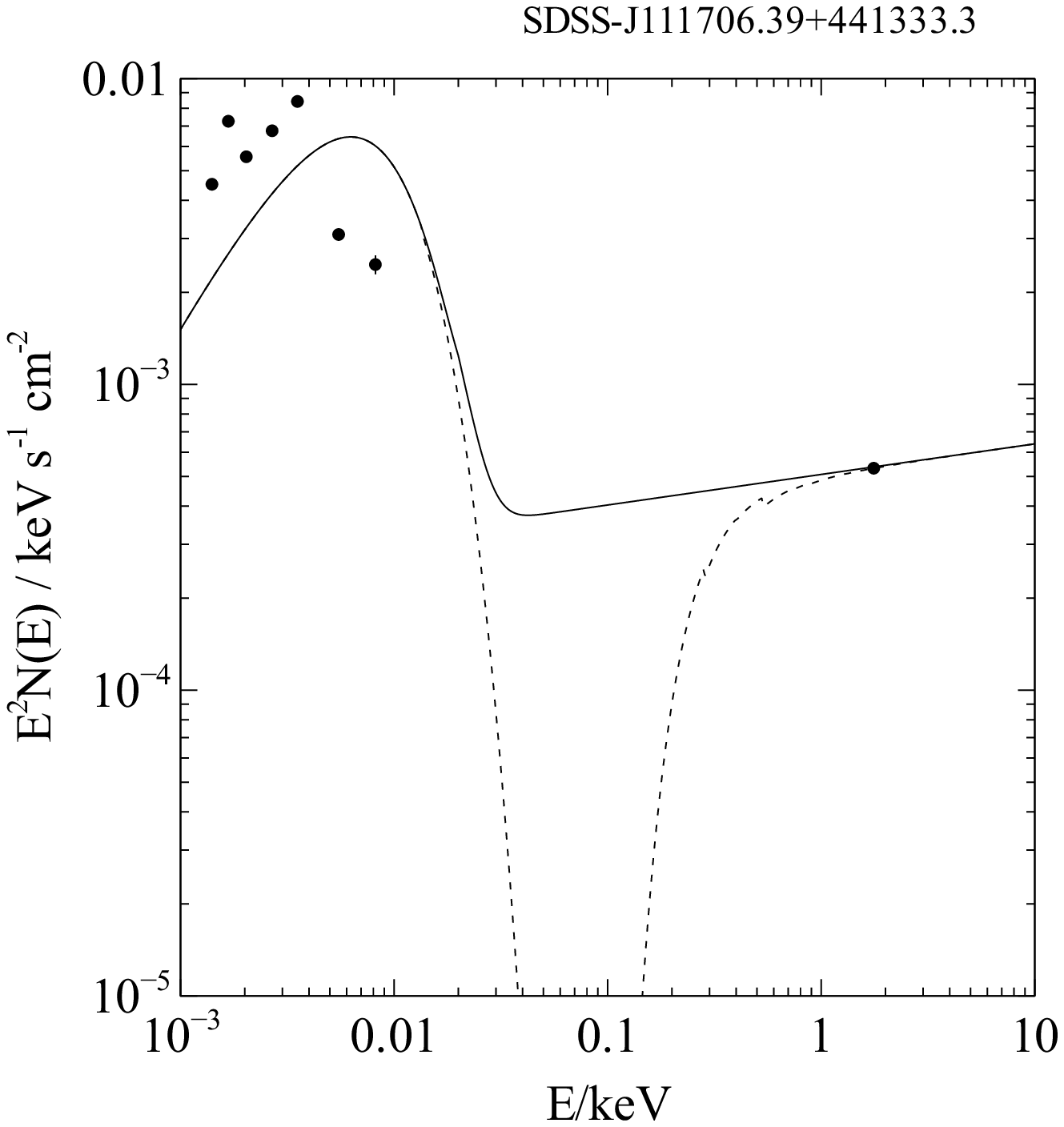,width=0.5\columnwidth,clip=} 
\epsfig{file=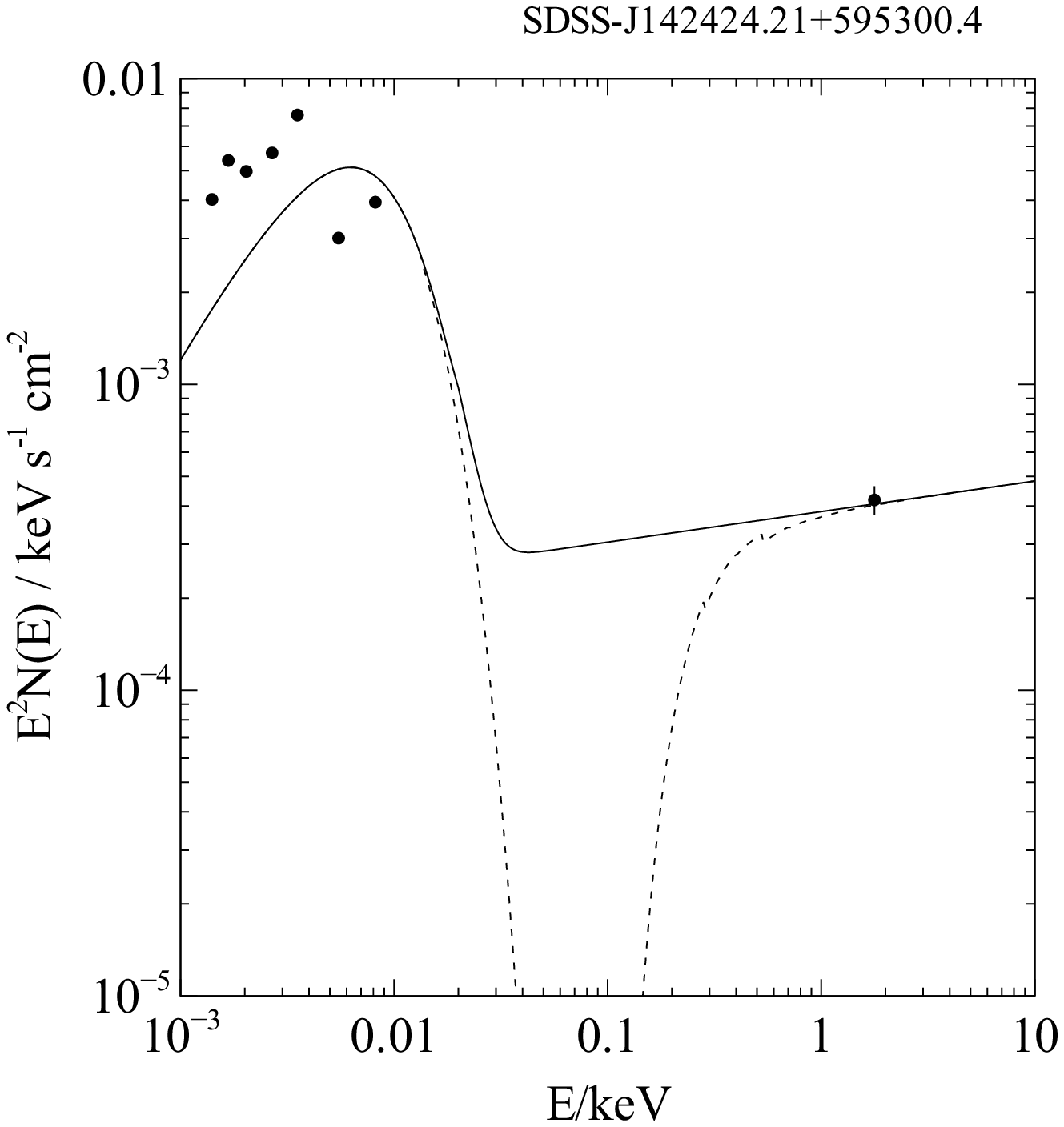,width=0.5\columnwidth,clip=} \\
\epsfig{file=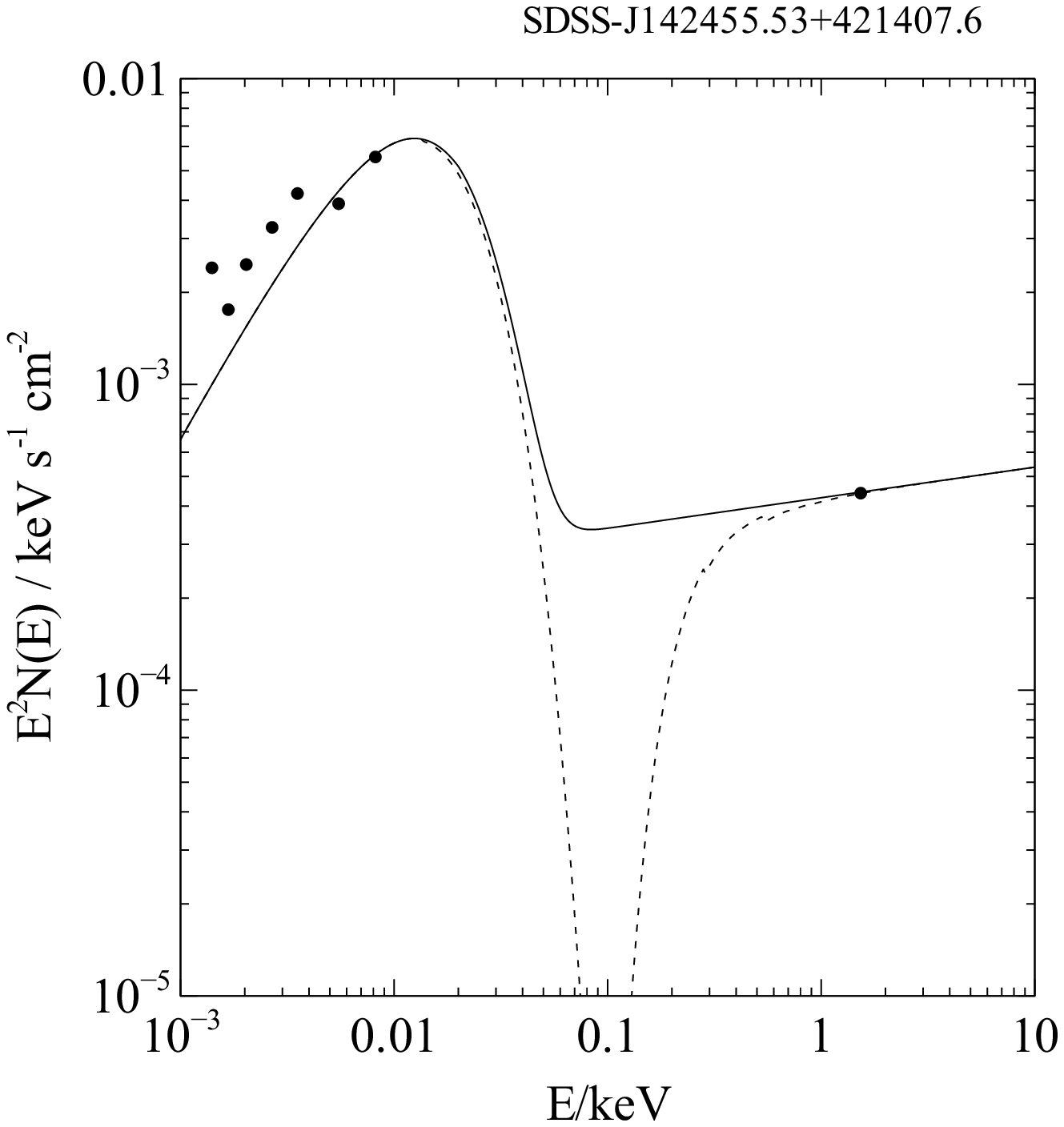,width=0.5\columnwidth,clip=} 
\epsfig{file=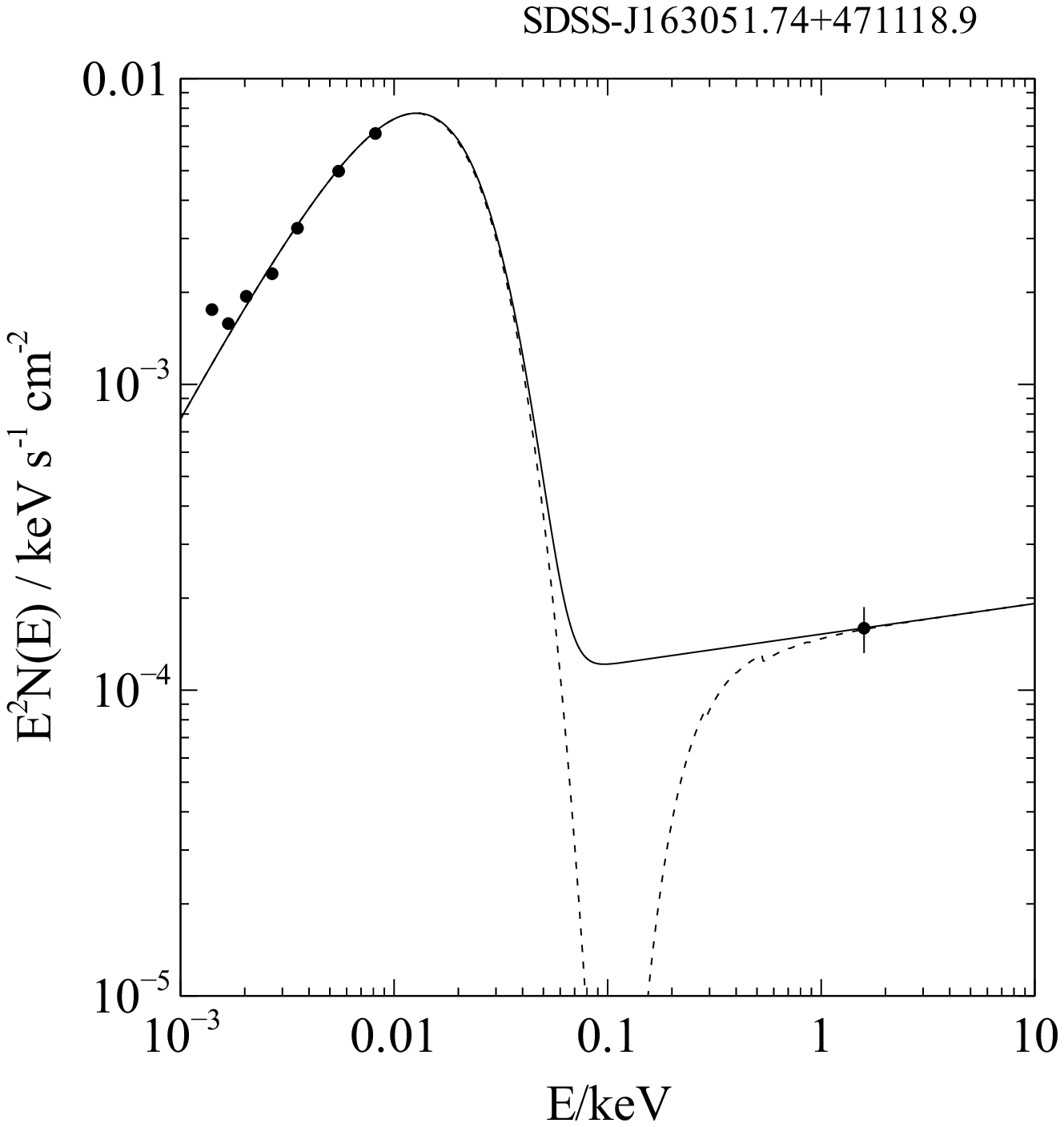,width=0.5\columnwidth,clip=} 
\end{tabular}
\caption {Observed-frame spectral energy distributions for sources in Table~\ref{mdot_table}. Lines are plotted to indicate possible shape of SED (dashed lines for observed and solid lines for intrinsic before absorption) assuming $E(B-V) = 0$. These are not used in the calculations except for SDSS J002332.34-011444.1, SDSS J005709.94+144610.1 and SDSS J092809.43+383000.5. Mass accretion rate is determined directly from data points and bolometric luminosity from IR measurements for the majority of sources (see description in Section~\ref{sec:accretion} for details). For most of the sources plotted the optical/UV turnover is not constrained by the data. We assume a value of spin for plotting purposes.}
\label{sed_list}
\end{figure*}

\end{appendix}

\clearpage

\end{document}